\documentclass[12pt]{article}

\usepackage[dvips]{graphics}

\usepackage{epsfig}

\DeclareFontEncoding{OT2}{}{}

\usepackage{graphicx}

\usepackage{amsfonts}
\usepackage{amsmath, amsthm, amssymb,ulem}

\begin{document}



\newcommand{\dbar}{d\mkern-6mu\mathchar'26}

\def\cM{\mathcal M}
\def\cA{\mathcal A}
\def\cC{\mathcal C}
\def\cB{\mathcal B}
\def\cO{\mathcal O}
\def\cT {\mathcal T}
\def\cL {\mathcal L}
\def\cS {\mathcal S}
\def\cP{\mathcal P}
\def\cK{\mathcal K}
\def\cH{\mathcal H}
\def\cQc{\mathcal Q_\circ}
\def\cQ{\mathcal Q}
\def\cD{\mathcal D}
\def\cR{\mathcal R}
\def\bR{\mathbb R}
\def\bZ{\mathbb Z}
\def\ia{\mathit A}
\def\bT{\mathbb T}
\def \da {\partial^{\alpha}}
\def \db {\partial^{\beta}}
\def \ulx {\underline{x}}

\pagenumbering{roman}

\def\bcw{\mathbin{\bigcirc\mkern-15mu\wedge}}

\def\r#1{{\mathop{#1}\limits^\circ}}
\def\ring{\r}

\def\=x{\stackrel{=}{x}}

\def\nint{\mathbin{\int\mkern-18mu\diagup \;} }

\newcommand{\namelistlabel}[1] {\mbox{#1}\hfil}
\newenvironment{namelist}[1]{%
\begin{list}{}
{\let\makelabel\namelistlabel
\settowidth{\labelwidth}{#1}
\setlength{\leftmargin}{1.1\labelwidth}}
}{%
\end{list}}
\newcommand{\norm}[1]{\left\Vert#1\right\Vert}
\newcommand{\abs}[1]{\left\vert#1\right\vert}

\newtheorem{lem}{Lemma}
\newtheorem{thm}{Theorem}
\newtheorem{cor}{Corollary}

\newtheorem{Lem}{Lemma}[section]
\newtheorem{Thm}{Theorem}[section]

\newtheorem{Cor}[Lem]{Corollary}

\newtheorem{Prop}[Lem]{Proposition}

\newtheorem{prop}{Proposition}[section]

\newtheorem{Pro}[Thm]{Proposition}

\newtheorem{co}[lem]{Corollary}

\newtheorem{corr}[thm]{Corollary}

\newtheorem{C}[Lem]{Claim}

\newtheorem{Def}{Definition}[section]
\newtheorem{remark}{Remark}[section]


\parskip=10pt

\pagenumbering{arabic}
\begin{titlepage}
\baselineskip=15pt

\thispagestyle{empty}
\hfill

\vspace{.25in}
\begin{center}
{\Large \bf \sf Blow Ups of Complex Solutions of the
3$\mathcal{D}$-Navier-Stokes System} \\
{\Large \bf \sf and} \\
{\Large \bf \sf Renormalization Group Method}

\bigskip
{\large \sf by}

\bigskip

{\large \sf Dong Li}\footnote{Program in Applied and Computational Mathematics, Princeton University,
Princeton, New Jersey, U.S.A.} {\large \sf and}
\addtocounter{footnote}{-1}{\large \sf Ya. G. Sinai\footnote{Mathematics
Department, Princeton University, Princeton, New Jersey, U.S.A. \&}
\footnote{Landau Institute of Theoretical Physics, Moscow, Russia}}

\end{center}

\medskip

\vfill
\begin{quote}
\noindent
\underline{\large \sf Abstract:}
We consider complex-valued solutions of the three-dimensional Navier-Stokes system
without external forcing on $R^3$. We show that there exists an open set
in the space of $10$-parameter families of initial conditions such that
for each family from this set there are values of parameters for which
the solution develops blow up in finite time.
\medskip
\noindent
\normalsize
\end{quote}

\vfill
\begin{quote}
\noindent
\underline{\large \sf Keywords: }
Navier-Stokes system, renormalization group theory, fixed point,
the linearization near the fixed point, spectrum of the linearized
group, Hermite polynomials
\end{quote}
\vfill

\end{titlepage}

\newpage
\baselineskip=18pt
\thispagestyle{empty}

\renewcommand{\thefootnote}{\fnsymbol{footnote}}
\addtocounter{footnote}{-2}

\begin{center}
{\large \S1. \sf Introduction.}
\end{center}

There are many phenomena in nature which can be considered as some
manifestation of blow ups, like hurricanes, tornadoes, sandstorms,
etc. If we believe that Navier-Stokes system describes well enough
the motions of real gases and fluids under normal conditions, then
it gives some reasons to expect that blow ups in solutions of this
system also exist.

We consider in this paper the 3$\cD$-Navier-Stokes system for
incompressible fluids moving without external forcing on $R^3$
with viscosity equal to 1.  After Fourier transform it becomes a
non-local, non-linear equation for a non-known function $v ( k , t
)$ with values in $C^3$, $k \in R^3$, $t > 0$.  The
incompressibility condition takes the form\footnote{Since $k \in
R^3$, $v ( k , t ) \in C^3$, the order in the inner product is
important.} $\langle v ( k , t )$, $k \rangle = 0$ and
\begin{equation}
\begin{split}
v ( k , t ) = \exp \, \{ - t | k |^2 \} \, v ( k , 0 ) &+ i
\displaystyle{\int\limits_{0}^{t}} \, \exp \, \{ - ( t - s ) | k
|^2 \} \, d s\cdot
\\
&\displaystyle{\int\limits_{R^3}} \, \left\langle \, v ( k - k^\prime , s ) ,
\, k \right\rangle \, \cdot \, P_k \, v ( k^\prime, s ) \, d k^\prime
\end{split}
\end{equation}
In this expression $v ( k , 0)$ is an initial condition and $P_k$ is the
orthogonal projection to the subspace orthogonal to $k$, i.e.
$P_k v = v
- \, \frac{\langle v, k \rangle \, \cdot \, k}{\langle k , k \rangle}$.
The formula (1) shows
  that the Navier-Stokes system is genuinely infinite-dimensional
dynamical system: the value $v ( k , t )$ is determined by the
integration over all ``degrees of freedom'' and previous moments of
time.

The problem of blow ups in solutions of the Navier-Stokes System(NSS)
appeared after classical works of J. Leray (see [Le 1]) where he
proved the existence of the weak solutions of NSS. O. Ladyzenskaya
proved the existence of strong solutions of 2-dim NSS in bounded
domains (see [La 1]). Many important contributions to the modern
understanding of the 2-dim fluid dynamics were done by C. Foias and
R. Temam (see [FT]), V. Yudovich (see [Y1]), Giga ([G1]) and
others. However, the situation with the 3-dim NSS remained unclear. The
Clay mathematical institute announced the problem of existence of strong
solutions of the 3-dimensional NSS as one of the most important problem in
mathematics of the XXI-century (see [Cl]).

In this paper we omit the condition that $v ( k , t )$ is the Fourier
transform of a real-valued vector field in the $x$-space and consider
(1) in the space of all possible complex-valued functions with values in
$C^3$.  In this situation the energy inequality does not hold. Detailed
assumptions concerning the initial condition $v ( k , 0)$
will be discussed later (see \S 7).  In all cases $v ( k , 0)$ will be
bounded functions whose support is a neighborhood of some
point $(0,0, k^{(0)})$. The incompressibility condition implies
that the components $v_1 ( k , 0)$,$v_2 (k , 0)$ of $v ( k , 0)$
are arbitrary functions of $k$ while $v_3 ( k , 0 )$ can be found
from the incompressibility condition $\langle v,k \rangle = 0$.

Various methods (see, for example, [K], [C], [S1]) allow to prove
in such cases the existence and uniqueness of classical solutions
of (1) on finite intervals of time.  For these solutions (see, for
example [S2])
\begin{equation}
|v ( k , t ) | \, \leq \, {\sf const} \, \exp \, \{ - \, {\sf const}
\, \sqrt{t} \, \cdot \, | k | \} \, , \, 0 \leq t \leq t_0 \, .
\end{equation}
Presumably, $v ( k , t )$ has an asymptotics of this type but this
requires more work.  According to a conventional wisdom, possible blow
ups are connected with the violation of (2).

In this paper we fix $t$ and consider one-parameter families of
initial conditions $v_A ( k , t ) = A v ( k , 0)$.  We show that
for some special $v ( k , 0 )$ one can find critical values
$A_{cr} = A_{cr} ( t )$ such that the solution $v_{A_{cr}} ( k , s
)$ blows up at $t$ so that for $t^\prime < t$ both the energy and the
enstrophy are finite while at $t^\prime = t$ they both become infinite.
Even more, for $t^\prime < t$ the solution decays exponentially outside
some region depending on $t$.  As $t^\prime \uparrow t$ this region
expands to an unbounded domain in $R^3$.

Our main approach is based on the renormalization group method
which is so useful in probability theory, statistical physics and
the theory of dynamical systems.  It is rather difficult to give
the exact formulation of our result in the introduction because it
uses some notions, parameters, etc., which will appear in the
later sections.  Loosely speaking, we show that in
$\ell$-parameter families of initial conditions, for $\ell = 10$,
one can find values of parameters for which the solutions develop
blow ups of the type we already described.
The meaning of $\ell$ is explained in \S4, \S5, \S6.

We thank C. Fefferman, W.E, K. Khanin and V. Yakhot for many
useful discussions.  A big part of the text was prepared during
the visit of the second author of the Mathematics Department of
California Institute of Technology and we thank the Department for
its very warm hospitality.  We also thank G. Pecht for her
excellent typing of the manuscript. The financial support from NSF Grant
DMS 0600996 given to the second author is highly acknowledged.

\pagebreak
\begin{center}
{\large \S2. \ Power Series for Solutions of the
3$\cD$-Navier-Stokes-Systems and Preliminary Changes of Variables}
\end{center}

Our general approach is based upon the method of power series which were
introduced in [S1], [S2].  We write down the solution of (1) in the
form:
\begin{equation}
v_A ( k , t ) \, = \, \exp \, \{ - t | k |^2 \} \, \cdot \, A \, v ( k ,
0 ) \, + \,
\displaystyle{\int\limits_{0}^{t}} \, \exp \, \{ - ( t - s ) | k
|^2 \} \, \cdot \, \displaystyle{\sum\limits_{p > 1}} \, A^p \, h_p ( k
, s ) \, d s
\end{equation}

\noindent
The substitution of (3) into (1) gives the system of recurrent equations
connecting the functions $h_p$:
\begin{equation}
h_1 ( k , s ) \, = \, \exp \, \{ - s | k |^2 \} \, v ( k , 0 ),
\end{equation}
\begin{equation}
h_2 ( k , s ) \,  = \,   i \displaystyle{\int\limits_{R^3}} \, \langle v
( k - k^\prime, 0 ) \, , k \, \rangle \, P_k \, v ( k^\prime, 0 ) \,
\cdot \, \exp \, \{ - s | k - k^\prime |^2 \, - \, s | k^\prime
|^2 \} \, d^3 k^\prime,
\end{equation}
\(
h_p ( k , s ) \, = \, i \displaystyle{\int_{0}^{s}} \, d s_2 \,
\displaystyle{\int\limits_{R^3}} \, \left\langle \, v ( k - k^\prime, 0) , k \, \right\rangle
\, P_k h_{p-1} ( k^\prime, s_2 ) \, \cdot \, \\ \nonumber
\exp \, \{ - s | k - k^\prime |^2 - ( s - s_2 ) | k^\prime|^2 \} \, d^3
k^\prime \, +
\, i \displaystyle{\sum\limits_{p_1 + p_2 = p\atop{p_1, p_2 > 1}}} \,
\displaystyle{\int_{0}^{s}} \, ds_1 \,
\displaystyle{\int_{0}^{s}} \, ds_2 \,
\displaystyle{\int\limits_{R^3}} \, \left\langle \, h_{p_1} ( k - k^\prime , s_1 )
, k  \, \right\rangle \, \cdot
\)

\vspace{.9em}
\hspace{-.30in}
\(
P_k h_{p_2} ( k^\prime , s_2 ) \, \cdot \, \exp \{ - ( s - s_1 ) | k -
k^\prime |^2 - ( s - s_2 ) | k^\prime |^2 \} \, d^3 k^\prime
\; +  \)
\begin{equation}
\, i \displaystyle{\int_{0}^{s}} \, ds_1 \,
\displaystyle{\int\limits_{R^3}} \,
\left\langle \, h_{p-1} ( k - k^\prime , s_1 )
, \, k \, \right\rangle \, P_k \, v ( k^\prime, 0 ) \, \cdot
\, \exp \, \{ - ( s - s_1 ) | k - k^\prime |^2 - s | k^\prime |^2 \} \, d^3
k^\prime \, .
\end{equation}
Clearly, $h_p ( k , s ) \, \perp k$ for every $p \geq 1$, $k \in R^3$.

It follows from the results of [S2] that the series (3) converges for
sufficiently small $s$ and gives a classical solution of (1).  Make the
following change of variables which simplifies (4), (5), (6).  Put
$\tilde{k} = k \sqrt{s}$, $\tilde{k}^\prime = k^\prime \sqrt{s}$,
introduce relative times $\tilde{s}_1 , \tilde{s}_2, s_1 = \tilde{s}_1 s
, s_2 = \tilde{s}_2 s$ and denote $g_r ( \tilde{k}, s ) = h_r \left(
\frac{\tilde{k}}{\sqrt{s}}, s \right), r \geq 1$.  Then
\[
\hspace{1.5in} g_1 ( \tilde{k}, s ) \, = \,
\exp \, \{ - | \tilde{k} |^2 \} \,  \cdot \, v
\left(
\frac{\tilde{k}}{\sqrt{s}}, 0 \right) \, , \hspace{2.0in} \mbox{(4$^\prime$)}
\]

\[
\hspace{.75in} g_2 ( \tilde{k}, s) \, = \, h_2 \left(
\frac{\tilde{k}}{\sqrt{s}} , s \right) \, = \,
\frac{i}{s^2} \, \displaystyle{\int\limits_{R^3}} \, \langle \, v \left(
\frac{\tilde{k} - \tilde{k}^\prime}{\sqrt{s}} , 0 \right), \tilde{k} \,
\rangle \, \cdot \]
\[
\hspace{1.40in}P_{\tilde{k}} v \, \left(
\frac{\tilde{k}^\prime}{\sqrt{s}} , 0 \right) \, \exp \,
\{ - | \tilde{k} - \tilde{k}^\prime |^2 - | \tilde{k}^\prime|^2 \} \,
d^3 \tilde{k}^\prime \, , \hspace{1.65in} \mbox{(5$^\prime$)}
\]

\vspace{.5in}
\[
\hspace{-.75in} g_p ( \tilde{k} , s ) \, = \, \frac{i}{s} \,
\displaystyle{\int\limits_{0}^1} \, d \tilde{s}_2 \,
\displaystyle{\int\limits_{R^3}} \, \langle \, v \left(
\frac{\tilde{k} - \tilde{k}^\prime}{\sqrt{s}}, 0 \right) , \tilde{k} \,
\rangle \, \cdot P_{\tilde{k}} g_{p-1} ( \tilde{k} \sqrt{\tilde{s}_2} \,,
\tilde{s}_2 s ) \]
\[\exp \, \{ - | \tilde{k} - \tilde{k}^\prime |^2 -
( 1 - \tilde{s}_2 ) | \tilde{k}^\prime |^2 \} d^3 \tilde{k}^\prime \,
+ \]

\[
+ \, i \,
\displaystyle{\sum\limits_{p_1 + p_2 = p \atop{p_1 > 1 , p_2 > 1}}} \,
\displaystyle{\int\limits_{0}^{1}} \, d \tilde{s}_1 \,
\displaystyle{\int\limits_{0}^1} \,
d \tilde{s}_2 \,
\displaystyle{\int\limits_{R^3}} \, \langle \, g_{p_1} ( (
\tilde{k} - \tilde{k}^\prime ) \,
\sqrt{\tilde{s}_1},
\tilde{s}_1 s ) , \tilde{k} \, \rangle \, \cdot \]

\[
P_{\tilde{k}} g_{p_2} ( \tilde{k}^\prime \sqrt{\tilde{s}_2} ,
\tilde{s}_2 s ) \, \exp \,
\{
- ( 1 - \tilde{s}_1 ) \, | \tilde{k} - \tilde{k}^\prime |^2 - ( 1 -
  \tilde{s}_2 ) | \tilde{k}^\prime |^2 \} \, d^3 \tilde{k}^\prime
\]

\[ + \,
\frac{i}{s} \,
\displaystyle{\int\limits_{0}^1} \, d \tilde{s}_1 \,
\displaystyle{\int\limits_{R^3}} \, \langle \,
g_{p-1} ( ( \tilde{k} - \tilde{k}^\prime ) ) \,
\sqrt{\tilde{s}_1} \, , \, \tilde{s}_1
s) , \tilde{k} \rangle P_{\tilde{k}} \, v \left(
\frac{\tilde{k}^\prime}{\sqrt{s}} , 0 \right) \, \cdot
\]

\[
\hspace{1.0in} \exp \, \{
- ( 1 - \tilde{s}_1 ) | \tilde{k} - \tilde{k}^\prime |^2 \, - \, |
  \tilde{k}^\prime |^2 \} \, d^3 \tilde{k}^\prime
\hspace{2.25in} \mbox{(6$^\prime$)}
\]

The function $g_2 ( \tilde{k}, s )$ has a singularity at $s = 0$
even in the case of functions with compact support: for small $s$
its values are of order $\dfrac{1}{\sqrt{s}}$.  This singularity
is integrable and all $g_p ( k , s )$, $p > 2$, are bounded.  The
singularity is connected with our choice of the coordinates
$\tilde{k}, \tilde{k}^\prime$.

The formulas (4)-(6) or (4$^\prime$)-(6$^\prime$) resemble convolutions
in probability theory.  For example, if $C = {\sf supp} \, v ( k , 0)$ then
${\sf supp} \, h_p = \underbrace{C + C + \cdots + C}_{p \ \mbox{times}}$.
Therefore it is natural to expect that $h_p$ and $g_p$ satisfy some form
of the limit theorem of probability theory.  This question will be
discussed in more detail in the next sections.

Make another change of variables.  Assume that we have some $p$.
The terms in (6$^\prime$) with $p_1 \leq p^{1/2}$ and $p_2 \leq
p^{1/2}$ will be called boundary terms.  They will be treated as
remainder terms and will be estimated later.  Suppose that we have
some number $\tilde{k}^{(0)}$ which later will be assumed to be
sufficiently large.  Introduce the vector
$\widetilde{\cK}^{(r)} = ( 0 , 0 , r \tilde{k}^{(0)}  )$. These
will be the points near which all $g_r$ will be
concentrated,\break $p^{1/2} \leq r \leq p - p^{1/2}$.  We write
$\tilde{k} = \widetilde{\cK}^{(r)} + \sqrt{r} \, \cdot \, Y , Y \,
\in \, R^3$. Thus instead of $\tilde{k}$ we have the new variable
$Y = ( Y_1 , Y_2 , Y_3 )$ which typically will take values $O(1)$.
Put $\tilde{\kappa}^{(0)} = ( 0 , 0 ,
\tilde{k}^{(0)} )$.

In all integrals over $\tilde{s}_1 , \tilde{s}_2$ in (6$^\prime$)
make another change of variables $1 - \tilde{s}_j =
\frac{\theta_j}{p^2_j}$, $j = 1,2$.  Instead of the variable of
integration $\tilde{k}^\prime$ introduce $Y^\prime$ where
$\tilde{k}^\prime = \widetilde{\cK}^{(p_2)} + \sqrt{p} Y^\prime$.
We write $\tilde{g}_r ( Y , s ) = g_r ( \widetilde{\cK}^{(r)} +
\sqrt{r} \, Y, s )$, $\gamma = \frac{p_1}{p}$, $\frac{p_2}{p} = 1
- \gamma$.  Then from (6$^\prime$)

\(
\tilde{g}_p ( Y , s ) \, = \, g_p ( \tilde{\cK}^{(p)} \, + \, \sqrt{p}
\, Y , s ) \, = \,
p^{5/2}
\left[
i \; \displaystyle{\sum\limits_{p_1, p_2 > \sqrt{p}\atop{p_1 + p_2 =
p}}} \,
\displaystyle{\int\limits_{0}^{p^2_1}} \, d \theta_1 \,
\displaystyle{\int\limits_{0}^{p^2_2}} \, d \theta_2 \, \cdot \,
\frac{1}{p^2_1 \, \cdot \, p^2_2} \, \cdot \,
\right. \)

\( \displaystyle{\int\limits_{R^3}} \, \left\langle \, \tilde{g}_{p_1} \,
\left( \frac{Y - Y^\prime}{\sqrt{\gamma}} \, , \left( 1 -
\frac{\theta_1}{p^2_1} \right) s \right) \,, \tilde{\kappa}^{(0)}
\, + \, \frac{Y}{\sqrt{p}} \, \right\rangle \, \cdot \,
P_{\tilde{\kappa}^{(0)} \, + \, \frac{Y}{\sqrt{p}}} \,
\tilde{g}_{p_2} \left( \frac{Y^\prime}{\sqrt{1-\gamma}} \, ,
\left( 1 - \, \frac{\theta_2}{p^2_2} \right) s \right) \, \cdot \,
\)
\begin{equation}
\exp \, \left\{ - \theta_1 \abs{ \tilde{\kappa}^{(0)} \, + \, \frac{Y -
Y^\prime}{\sqrt{p} \cdot \gamma} }^2 - \theta_2  \abs{
\tilde{\kappa}^{(0)} \, + \, \frac{Y^\prime}{\sqrt{p} \,
(1-\gamma)} }^2 \right\} \, d^3 \, Y^\prime \Biggr] \, . \hspace{1.75in}
\end{equation}

\noindent
This is the main recurrent relation which we shall study in the next
sections.  It is of some importance that in front of (7) we have the
factor $p^{5/2}$ and inside the sum the factor $\frac{1}{p^2_1} \, \cdot
\, \frac{1}{p^2_2}$.  Both are connected with the new scaling inherent
to the Navier-Stokes system.

\begin{center}
{\large \S3. \sf The Renormalization Group Equation}
\end{center}

As $p \longrightarrow \infty$ the recurrent equation (7) takes
some limiting form which will be derived in this section.  All
remainders which appear in this way are listed and estimated in
\S8.

The main contribution to (7) comes from $p_1 , p_2$ of order $p$.
If $Y , Y^\prime \, = \, O ( 1)$ then $\frac{Y -
Y^\prime}{\sqrt{p}}$, $\frac{Y^\prime}{\sqrt{p}}$ are small
compared to $\tilde{\kappa}^{(0)} = ( 0 , 0 , \tilde{k}^{(0)}$).
Therefore the Gaussian term in
(7) can be replaced by $\exp \, \{ - ( \theta_1 + \theta_2) |
\tilde{k}^{(0)} |^2 \}$, $\tilde{s}_1$ and $\tilde{s}_2$ can be
replaced by 1 and the integrations over $\theta_1$, $\theta_2$ and
$Y^\prime$ can be done separately. Thus instead of (7) we get
a simpler recurrent relation:

\( \tilde{g}_p ( Y , s ) \, = \, \dfrac{i}{| \tilde{k}^{(0)} |^4}
\, p^{5/2} \; \displaystyle{\sum\limits_{p_1, p_2> p^{1/2}
\atop{p_1 + p_2 = p}}} \: \frac{1}{p^2_1 \, \cdot \, p^2_2} \,
\cdot \, \displaystyle{\int\limits_{R^3}} \, \left\langle \, \tilde{g}_{p_1}
\, \left( \frac{Y - Y^\prime}{\sqrt{\gamma}} \, , s \right) \, ,
\tilde{\kappa}^{(0)} \, + \, \frac{Y}{\sqrt{p}} \, \right\rangle \, \cdot \,
\)
\begin{equation}
\cdot \, P_{\tilde{\kappa}^{(0)} + \frac{Y}{\sqrt{p}}} \,
\tilde{g}_{p_2} \, \left( \frac{Y^\prime}{\sqrt{(1-\gamma)}} , s
\right) \, d^3 \, Y^\prime \, .
\end{equation}
In view of incompressibility

\begin{equation*}
\begin{array}{ll}
&  \left\langle \tilde{g}_{p_1} \, \left( \frac{Y - Y^\prime}{\sqrt{\gamma}},
s \right) \, , \ \tilde{\kappa}^{(0)} \, +
\, \frac{Y}{\sqrt{p}} \, \right\rangle \, = \nonumber \\
\nonumber \\
= & \frac{1}{p_1} \, \left\langle \, g_{p_1} \left( \kappa^{(0)} \, \cdot \,
p_1 \, + \, \frac{Y - Y^\prime}{\sqrt{\gamma}} \, \cdot \,
\sqrt{p_1}, s \right), \, \kappa^{(0)}
p_1 \, + \, Y \, \cdot \, \gamma \sqrt{p} \, \right\rangle \nonumber \\
\nonumber \\
=  & \frac{1}{p_1} \, < \, g_{p_1} \left( \kappa^{(0)} p_1 \, + \,
\frac{Y - Y^\prime}{\sqrt{\gamma}} \, \cdot \, \sqrt{p_1} \, , s
\right), \, \kappa^{(0)} \, p_1 \, + \, \frac{Y -
Y'}{\sqrt{\gamma}}
\, \sqrt{p_1} > \, + \nonumber \\
\nonumber \\
+  & \frac{1}{p_1} \, < \, g_{p_1} \left( \kappa^{(0)} \, p_1 \, +
\, \frac{Y - Y^\prime}{\sqrt{\gamma}} \, \cdot \, \sqrt{p_1} , s
\right),  \, Y \gamma \, \cdot \, \sqrt{p} - ( Y - Y^\prime ) \,
\sqrt{p} \, > \, =
\nonumber \\
\nonumber \\
=  &  \frac{1}{\sqrt{p_1}} \, \cdot \, < \, \tilde{g}_{p_1} \left(
\frac{Y - Y^\prime}{\sqrt{\gamma}} , s \right) \, , \, \frac{Y -
Y^\prime}{\sqrt{\gamma}} \, > \, \cdot ( \gamma - 1) \, +
\end{array}
\end{equation*}
\begin{equation}
\begin{array}{ll}
{\hspace{-2.5in}} +  & \frac{1}{\sqrt{p_2}} \, < \,
\tilde{g}_{p_1} \left( \frac{Y - Y^\prime}{\sqrt{\gamma}} , s
\right) \, , \, Y^\prime \sqrt{(1-\gamma)} \, >
\end{array}
\end{equation}

Write $\tilde{g}_p$ in the form
\begin{equation}
\tilde{g}_p ( Y , s ) \, = \, ( G^{(p)}_1 ( Y , s ) \, , \, G^{(p)}_2 (
Y , s ) \, ,
\frac{1}{\sqrt{p}} \, F^{(p)} ( Y , s ) ) \, .
\end{equation}

\noindent
Since $\tilde{k} = \tilde{\kappa}^{(0)} \cdot p \, + \, Y \sqrt{p}$, the
incompressibility implies
\begin{equation}
< g_r (
\tilde{k} , s ) \, , \, \tilde{k} \, > \, = \, < \,
g_r ( \tilde{k} , s ) \, , \,
\frac{\tilde{k}}{r} \, > \, = \, 0
\end{equation}

\noindent
and for $Y \, = \, O ( 1)$
\begin{equation}
\frac{Y_1}{\sqrt{r}} \, \cdot \,
G^{(r)}_1 ( Y , s ) \, + \,
\frac{Y_2}{\sqrt{r}} \, G^{(r)}_2 ( Y , s ) \, + \,
\frac{\tilde{k}^{(r)}}{\sqrt{r}} \, \cdot \, F^{(r)} ( Y , s ) \, = \, O
\left( \frac{1}{r} \right) \, .
\end{equation}

\noindent
In our approximation we replace (12) by

\begin{equation}
Y_1 G^{(r)}_1 ( Y , s ) \, + \, Y_2 G^{(r)}_2 ( Y , s ) \, + \, F^{(r)} ( Y
, s ) \, = \, 0  \, .
\end{equation}

\noindent
Thus for given $Y_1 , Y_2, Y_3$ the component $F_r$ can be
expressed through $G^{(r)}_1 , G^{(r)}_2$.  This remains to be true even
if we do not neglect the $rhs$ of (13).  Return back to (9).  From (13)

\begin{equation*}
\begin{array}{ll}
< & \tilde{g}_{p_1} \left( \frac{Y - Y^\prime}{\sqrt{\gamma}} , s
\right), \tilde{\kappa}^{(0)} \, + \, \frac{Y}{\sqrt{p}} > = \,
\frac{1}{\sqrt{p}} \left[ \frac{\gamma - 1}{\sqrt{\gamma}} \, < \,
\tilde{g}_{p_1} \, \left( \frac{Y - Y^\prime}{\sqrt{\gamma}} , s
\right) , \, \frac{Y - Y^\prime}{\sqrt{\gamma}} \, >
\right. \\
\\
+ & \sqrt{1 - \gamma} \, < \, \tilde{g}_{p_1} \left( \frac{Y -
Y^\prime}{\sqrt{\gamma}} , s \right)  \, , \frac{Y^\prime}{\sqrt{1
- \gamma}} >
\left. \right]  \, = \\
\\
= & \frac{1}{\sqrt{p}} \left[ \frac{\gamma - 1}{\sqrt{\gamma}} \,
\left( \frac{Y_1 - Y^\prime_1}{\sqrt{\gamma}} \, G^{(p_1)}_{1} \,
\left( \frac{Y - Y^\prime}{\sqrt{\gamma}} , s \right) \right.
\right. + \,  \frac{Y_2 - Y^\prime_2}{\sqrt{\gamma}} \,
G^{(p_1)}_{2} \, \left( \frac{Y - Y^\prime}{\sqrt{\gamma}} , \, s
\right) \\
\\
+ & \frac{Y_3 - Y^\prime_3}{\sqrt{\gamma}} \, \cdot \,
\frac{1}{\sqrt{p_1}} \, \cdot \, F^{(p_1)} \, \left( \frac{Y -
Y^\prime}{\sqrt{\gamma}} , s \right) \, + \, \sqrt{(1-\gamma)} \,
\left( \frac{Y^\prime_1}{\sqrt{1 - \gamma}} \, G^{(p_1)}_{1} \,
\left( \frac{Y - Y^\prime}{\sqrt{\gamma}} , s \right) \right.
\end{array}
\end{equation*}

\begin{equation}
\begin{array}{ll}
+ & \frac{Y^\prime_2}{\sqrt{1 - \gamma}} \, G^{(p_1)}_{2} \,
\left( \frac{Y - Y^\prime}{\sqrt{\gamma}} , s \right) \, + \,
\frac{1}{\sqrt{p_2}} \, \frac{Y^\prime_3}{\sqrt{1 - \gamma}} \,
F^{(p_1)} \, \left. \left. \left( \frac{Y - Y^\prime}{\sqrt{\gamma}}
, s \right) \right) \right] \,.
\end{array}
\end{equation}

\noindent
In our approximation the inner product in (14) can be replaced by

\( \frac{1}{\sqrt{p}} \left[ \frac{\gamma - 1}{\sqrt{\gamma}} \,
\left( \frac{Y_1 - Y^\prime_1}{\sqrt{\gamma}} \, G^{(p_1)}_{1} \,
\left( \frac{Y - Y^\prime}{\sqrt{\gamma}}, s \right) \, + \,
\frac{Y_2 - Y^\prime_2}{\sqrt{\gamma}} \, G^{(p_1)}_{2} \,
 \left(
\frac{Y - Y^\prime}{\sqrt{\gamma}} , s \right) \right) \right. \)
\begin{equation}
\left. + \, \sqrt{1 - \gamma} \left( \frac{Y^\prime_1}{\sqrt{1 -
\gamma}} \, G^{(p_1)}_{1} \, \left( \frac{Y -
Y^\prime}{\sqrt{\gamma}}, s \right) \, + \, \frac{Y_2}{\sqrt{1 -
\gamma}} \, G^{(p_1)}_{2} \, \left( \frac{Y -
Y^\prime}{\sqrt{\gamma}} \, s \right) \right) \right]
\end{equation}

\noindent
According to the definition of the projector

\medskip
\( P_{\tilde{\kappa}^{(0)} + \frac{Y}{\sqrt{p}}} \,
\tilde{g}_{p_2} \left( \frac{Y^\prime}{\sqrt{1 - \gamma}} , s
\right) \, = \, \tilde{g}_{p_2} \, \left( \frac{Y^\prime}{\sqrt{1
- \gamma}} , s \right) \)

\begin{equation}
- \, \displaystyle{\frac{ < \tilde{g}_{p_2} \, \left(
\frac{Y^\prime}{\sqrt{1 - \gamma}} , s \right) , \,
\tilde{\kappa}^{(0)} \, + \, \frac{Y}{\sqrt{p}} > \, \cdot \, (
\tilde{\kappa}^{(0)} + \frac{Y}{\sqrt{p}} ) } { < \tilde{\kappa}^{(0)} \, + \,
\frac{Y}{\sqrt{p}} , \tilde{\kappa}^{(0)} \, + \, \frac{Y}{\sqrt{p}} > } }
\, = \, \tilde{g}_{p_2} \left( \frac{Y^\prime}{\sqrt{1 - \gamma}}
, s \right) \, + \, O \left( \frac{1}{\sqrt{p_2}} \right) \, .
\end{equation}

\noindent This shows that in the main order of magnitude the
projector is the identity operator and we come to a simpler
recurrent relation instead of (8):

\( \tilde{g}_p ( Y , s ) \, = \, i \, \lambda_1 \, \cdot\,
\displaystyle{\sum\limits_{p_1 , p_2 \geq p^{1/2} \atop {p_1 + p_2
= p}}} \, \frac{p^2}{p^2_1 \, \cdot \, p^2_2} \,
\displaystyle{\int_{R^3}} \, \left[ \dfrac{\gamma -
1}{\sqrt{\gamma}} \, \left( \dfrac{Y_1 - Y^\prime_1}{\sqrt{\gamma}}
\, \cdot \, \right. \right. \)

\( \left. G^{(p_1)}_{1} \, \left( \dfrac{Y -
Y^\prime}{\sqrt{\gamma}}, s \right) \, + \, \dfrac{Y_2 -
Y^\prime_2}{\sqrt{\gamma}} \, G^{(p_1)}_{2} \, \left( \dfrac{Y -
Y^\prime}{\sqrt{\gamma}} , s \right) \right)  \, + \left. \left.
\sqrt{1 - \gamma} \right. \right. \)
\begin{equation}
\left. \left( \dfrac{Y^\prime_1}{\sqrt{1 - \gamma}} \,
G^{(p_1)}_{1} \, \left( \frac{Y - Y^\prime}{\sqrt{\gamma}} , s
\right) \, + \, \frac{Y^\prime_2}{\sqrt{1 - \gamma}} \,
G^{(p_1)}_{2} \, \left( \frac{Y - Y^\prime}{\sqrt{\gamma}} , s
\right) \right) \right]  \, \cdot \, \tilde{g}_{p_2} \left(
\frac{Y^\prime}{\sqrt{1 - \gamma}} , s \right) \, d^3 Y^\prime \,
.
\end{equation}

\noindent The main assumption which we shall check below in the next
sections concerns the asymptotic form of $\tilde{g}_p ( Y , s )$ as
$p \longrightarrow \infty$: for some interval $S^{(p)} \, = \, [
S^{(p)}_- , \, S^{(p)}_+ ]$ on the time axis and some
$\Lambda$, positive  $\sigma^{(1)}
$, $\sigma^{(2)}$ and for all $r < p$

\begin{equation}
\begin{array}{l}
\tilde{g}_r ( Y , s ) \, = \,
\Lambda^{r-1}  r \, \cdot \, \nonumber
\displaystyle{\frac{\sigma^{(1)} }{2 \pi}} \, \exp \, \left\{ -
\displaystyle{\frac{\sigma^{(1)} }{2}} \, \left(| Y_1 |^2 \, +
\, | Y^2_2 | \right) \, \right\} \, \cdot \, \sqrt{\frac
{\sigma^{(2)} } {{2 \pi}} }\, \exp \, \left\{ -
\frac{\sigma^{(2)} }{2} \, | Y_3 |^2 \right\} \, \cdot
\nonumber \\
\nonumber \\
(H_1  ( Y_1 , Y_2 , Y_3 ) \, + \,
\delta^{(r)}_1 ( Y ,s ) , \,
H_2 ( Y_1 , Y_2 , Y_3 ) \, + \,
\delta^{(r)}_2 ( Y , s ) \, , \delta^{(r)}_3 ( Y , s ))
\end{array}
\end{equation}

\vspace{-.40in}
\begin{equation}
\end{equation}

\noindent
where $\delta^{(r)}_j ( Y , s ) \longrightarrow 0$ as $r \longrightarrow
\infty$, $j = 1 , 2 , 3$.
Later we shall explain in more detail in what sense
the convergence to zero takes place. The substitution of (18) into
(17) gives

\(
\tilde{g}_p ( Y , s ) \, = \,
\displaystyle{\frac{i}{| \tilde{k}^{(p)}|^4}} \, \cdot \,
p \, \cdot \, \Lambda^{p - 2}  \, \cdot
\)

\medskip

\medskip
\( \displaystyle{\sum\limits_{\gamma = \frac{p_1}{p}}} \,
\frac{1}{p} \cdot \,  \gamma^{\frac 12} (1 - \gamma)^{\frac 12} \, \cdot
\displaystyle{\int\limits_{R^3}} \, \left[ \frac{\gamma -
1}{\sqrt{\gamma}} \, \cdot \, \left( \frac{Y_1 -
Y^\prime_1}{\sqrt{\gamma}} \, \cdot \, H_1 \, \left( \frac{Y -
Y^\prime}{\sqrt{\gamma}} \right) \, + \right. \right. \)

\medskip

\medskip
\( \left. + \, \displaystyle{\frac{Y_2 -
Y^\prime_2}{\sqrt{\gamma}} } \, H_2 \left( \frac{Y -
Y^\prime}{\sqrt{\gamma}} \right) \right) \, + \, \sqrt{1 - \gamma}
\, \left( \displaystyle{\frac{Y^\prime_1}{\sqrt{1 - \gamma}}} \,
H_1 \left( \frac{Y - Y^\prime}{\sqrt{\gamma}} \right) \, + \right.
\)

\medskip
\( \left. \left. + \, \displaystyle{\frac{Y^\prime_2}{\sqrt{1 -
\gamma}}} \, H_2 \, \left( \frac{Y - Y^\prime}{\sqrt{\gamma}}
\right) \right) \right] \, H \left( \dfrac{Y^\prime}{\sqrt{1 -
\gamma}} \right) \, \cdot \)

\medskip
\(
\displaystyle{\frac{\sigma^{(1)} }{2 \pi \gamma}} \, \cdot \,
\exp \, \left\{ - \frac{\sigma^{(1)} }{2} \, \left( \frac{| Y_1
- Y_1^\prime |^2 \, + \, | Y_2 - Y^\prime_2 |^2}{\gamma} \right)
\right\} \, \cdot \)

\medskip
\( \displaystyle{\frac{\sigma^{(1)} }{2 \pi ( 1 - \gamma)}} \,
\exp \, \left\{ - \, \frac{\sigma^{(1)} }{2} \, \frac{|
Y^\prime_1|^2 \, + \, | Y^\prime_2 |^2}{1 - \gamma} \right\} \,
\cdot \)
\begin{equation}
\sqrt{ \frac{\sigma^{(2)} } {2 \pi \gamma}} \, \exp \, \left\{
- \frac{\sigma^{(2)} }{2} \, \frac{| Y_3 - Y^\prime_3
|^2}{\gamma} \right\} \, \cdot \sqrt{ \frac{\sigma^{(2)}}{2
\pi ( 1 - \gamma)}} \, \exp \, \left\{ - \frac{\sigma^{(2)}
}{2} \, \frac{| Y^\prime_3 |^2}{1 - \gamma} \right\} d^3 Y'\, .
\end{equation}

\noindent
Here

\[
H \left( \frac{Y^\prime}{\sqrt{1 - \gamma}} \right) \, = \, \left(
H_1 \, \left( \frac{Y^\prime_1}{\sqrt{1 - \gamma}} , \,
\frac{Y_2^\prime}{\sqrt{1 - \gamma}} , \,
\frac{Y^\prime_3}{\sqrt{1- \gamma}} \right) \, , \right.
\]

\[
\left. \left. H_2 \left( \frac{Y^\prime_1}{\sqrt{1 - \gamma}} , \,
\frac{Y^\prime_2}{\sqrt{1 - \gamma}} , \,
\frac{Y^\prime_3}{\sqrt{1 - \gamma }} \right) , \, 0 \right)
\right. \, .
\]

\medskip
\noindent
We do not mention explicitly the dependence of $H$ on $s$.

\medskip
\noindent
The last sum looks like a Riemannian integral sum whose limit takes the
form as $p \longrightarrow \infty$:

\vspace{.10em} \( \Lambda \, \exp \left\{ - \, \frac{\sigma^{(1)}
}{2} \: (
 | Y_1 |^2 \, + \, | Y_2 |^2 )
  \right\} \, \cdot \,
\dfrac{\sigma^{(1)} }{2 \pi} \, \cdot \, \exp \left\{ -
\dfrac{\sigma^{(2)}  | Y_3 |^2}{2} \right\} \;
\sqrt{\dfrac{\sigma^{(2)} }{2 \pi} } \: H ( Y ) \,  \)

\( = \quad\, \displaystyle{\frac{i}{| \tilde{k}^{(0)} |^4}} \,
\displaystyle{\int_{0}^1} \, \gamma^{\frac 1 2} \, (1 -
\gamma)^{\frac 1 2} \, d \gamma \, \displaystyle{\int_{R^3}} \,
\frac{\sigma^{(1)} } {2 \pi \gamma} \, \exp \, \left\{ -
\frac{\sigma^{(1)} ( | Y_1 - Y^\prime_1 |^2 \, + \, | Y_2 -
  Y^\prime_2 |^2 )}{2 \gamma}
\right\}
\)

\( \displaystyle{\frac{\sigma^{(1)}}{2 \pi ( 1 - \gamma )}} \,
\exp \left\{ - \, \frac{\sigma^{(1)}  ( | Y^\prime_1 |^2 \,
+ \, | Y^\prime_2 |^2
  )} {2(1-\gamma)} \right\} \, \cdot \,
 \sqrt{\frac{\sigma^{(2)} }{2 \pi \gamma}} \,
\exp \left\{ - \frac{\sigma^{(2)}  | Y_3 - Y^\prime_3 |^2}{2
\gamma} \right\} \, \cdot \, \)

\( \sqrt{\dfrac{\sigma^{(2)} }{2 \pi ( 1 - \gamma )}} \, \exp
\left\{ - \, \dfrac{\sigma^{(2)}  | Y^\prime_3 |^2}{2 ( 1 -
\gamma )}
  \right\} \,
\left[
- \dfrac {\gamma-1} {\sqrt {\gamma}}
\left(
\dfrac{
Y_1 - Y^\prime_1}{\sqrt{\gamma}} \,
H_1
\left(
\dfrac{Y - Y^\prime}{\sqrt{\gamma}} \right)
\, + \,
\dfrac{Y_2 - Y_2^\prime}{\sqrt{\gamma}} \,
H_2 \, \left(
\dfrac{Y - Y^\prime}{\sqrt{\gamma}}
\right) \right)
\right. \,
\)

\begin{equation}
\, + \, \gamma^{\frac{1}{2}} ( 1 - \gamma )
\left.
\left( \frac{Y^\prime_1}{\sqrt{1 - \gamma}} \,
H_1 \left( \frac{Y - Y^\prime}{\sqrt{\gamma}} \right) \, +
\,
\frac{Y^\prime_2}{\sqrt{1 - \gamma}} \,
H_2
\left( \frac{Y - Y^\prime}{\sqrt{1 - \gamma}} \right) \right) \right]
\cdot \, H
\left( \frac{Y^\prime}{\sqrt{1 - \gamma}} \right) \, d^3 Y^\prime
\, .
\end{equation}

\vspace{.10in} \noindent

The integral over $Y_3$ is the usual convolution.
Therefore we can look for functions $H_1 , H_2$ depending only on
$Y_1 , Y_2$, i.e. $H_1 ( Y ) =H_1 ( Y_1 , Y_2 )$, $H_2 ( Y ) = H_2 (
Y_1 , Y_2)$.  Write down the equation for $H_1 , H_2$ which does not
contain $Y_3$:

\medskip
\( \exp \, \left\{ - \frac{\sigma^{(1)} }{2} | Y |^2 \right\} \,
\cdot \, \dfrac{\sigma^{(1)} }{2 \pi} \, \cdot H ( Y ) \, = \,
\displaystyle{\int^{1}_{0}} \, d \gamma \,
\displaystyle{\int\limits_{R^2}} \, \frac{\sigma^{(1)} }{2 \pi
\gamma} \, \cdot \, \exp \left\{ - \frac{\sigma^{(1)}  | Y -
Y^\prime|^2}{2 \gamma} \right\} \, \cdot \, \frac{\sigma^{(1)} }{2
\pi ( 1 - \gamma )} \, \cdot \)

\medskip
\( \exp \left\{ - \frac{\sigma^{(1)} }{2 ( 1 - \gamma)} \, \cdot
\, | Y^\prime |^2
  \right\} \,
\left[
- ( 1 - \gamma)^{3/2} \,
\left(
\frac{Y_1 - Y^\prime_1}{\sqrt{\gamma}} \, \cdot \, H_1 \left(
\frac{Y - Y^\prime}{\sqrt{\gamma}}
\right) \, + \,
\frac{Y_2 - Y^\prime_2}{\sqrt{\gamma}} \,
H_2 \left(
\frac{Y - Y^\prime}{\sqrt{\gamma}} \right) \right)
\right.
\)

\begin{equation}
\left.
+ \, \gamma^{\frac{1}{2}} \, ( 1 - \gamma ) \,
\left(
\frac{Y^\prime_1}{\sqrt{1 - \gamma}} \,
H_1 \left(
\frac{Y - Y^\prime}{\sqrt{\gamma}} \right) \, + \,
\frac{Y^\prime_2}
{\sqrt{1 - \gamma}} \, H_2
\left( \frac{Y -Y^\prime}{\sqrt{\gamma}} \right)
\right)
\right] \,
\cdot \,
H \left(
\frac{Y^\prime}{\sqrt{1 - \gamma}} \right) \, d^2 Y^\prime \, .
\end{equation}

\medskip
\noindent Here 
$Y \, = \, ( Y_1 , Y_2 )$, $Y^\prime = (
Y^\prime_1, Y^\prime_2)$, $H ( Y ) = (H_1 ( Y_1 , Y_2 )$, $H_2 ( Y_1
, Y_2 )$. This is our main equation for the fixed point of the
renormalization group which we shall analyze in the next section
(see also \S 7).

\begin{center}
{\large \S4. \sf Analysis of the Equation $(21)$}
\end{center}

The solutions to the equation (21) have a natural scaling with
respect to the parameter\break $\sigma = \sigma^{(1)}$.  Namely,
if we solve the equation (21) for $\sigma = 1$ and let the
corresponding solution be $H ( Y )$, then the general solution for
arbitrary $\sigma$ is given by the formula
\begin{equation}
H_\sigma ( Y ) \, = \, \sqrt{\sigma} \, H ( \sqrt{\sigma} Y ) \, .
\end{equation}

\noindent This is analogous to the usual scaling of the Gaussian
fixed point in probability theory.  Thus, it is enough to consider
the equation (21) for $\sigma = 1$. We shall show that there exists
a three-parameter family of solutions to the equation (21) for
$\sigma = 1$. The equation (21) takes a
simpler form if we use expansions over Hermite polynomials.  All
necessary facts about Hermite polynomials are collected in the
Appendix 1.  For $ H(Y_1, Y_2) = ( H_1(Y_1, Y_2), H_2(Y_1,Y_2) )$,
we write
\begin{equation}
H_j ( Y_1 , Y_2 ) \, = \, \displaystyle{\sum\limits_{m_1 , m_2 \geq 0}} \,
h^{(j)} ( m_1 , m_2 ) \, He_{m_1} ( Y_1 ) \, He_{m_2} ( Y_2),
\quad j=1,2
\end{equation}
where $He_m ( z )$ are the Hermite polynomials of degree $m$ with respect
to the Gaussian density $\frac{1}{\sqrt{2 \pi}} \, \exp \left\{ -
\frac{1}{2} z^2 \right\}$.
We have (see \eqref{eq_A1_1}):
\begin{equation}
z He_m ( z ) \, = \, He_{m + 1} ( z ) \, + \, m He_{m-1} ( z ) \, , \, m
> 0
\end{equation}
and
\[
He_0(z)=1,\qquad \qquad z He_0 ( z ) \, =z= \, He_1 ( z ) \, .
\]

\noindent
Also we use the formula (see \eqref{eq_A1_2})

\(
\displaystyle{\int\limits_{\bR^1}} \, He_{m_1} \,
\left( \frac{Y - Y^\prime}{\sqrt{\gamma}} \right) \,
\frac{1}{\sqrt{2 \pi}} \,
\exp \left\{ - \frac{| Y - Y^\prime|^2}{2 \gamma} \right\} \,
He_{m_2} \, \left( \frac{Y^\prime}{\sqrt{1 - \gamma}} \right) \,
\frac{1}{\sqrt{2 \pi}} \,
\)
\begin{equation}
\exp \left\{
- \frac{| Y^\prime|^2}{2 ( 1 - \gamma)} \right\} \, d Y^\prime
= \, \gamma{\displaystyle^{\frac{m_1 + 1}{2}}} \,
( 1 - \gamma){\displaystyle^{\frac{m_2 + 1}{2}}} \,
He_{m_1 + m_2} ( Y ) \,
\frac{1}{\sqrt{2 \pi}} \,
\exp \left\{ - \frac{| Y |^2}{2} \right\} \, .
\end{equation}
Substituting (23) into (21) and using (24), (25), we come to the system
of equations for the coefficients $h ( m_1, m_2)$ which is equivalent to
(21):

\(
h^{(j)} ( m_1 , m_2 ) \, = \,
\displaystyle{
\sum\limits_{m^\prime_1 + m^{\prime\prime}_1 = m_1
\atop{m^\prime_2 + m^{\prime\prime}_2 = m_2}
}} \,
J^{(1)}_{m^\prime m^{\prime\prime}} \cdot
\left\{
( B_1 h^{(1)} ) (m^\prime_1, m^\prime_2) \, + \,
( B_2 h^{(2)} )(m^\prime_1, m^\prime_2) \right\}
h^{(j)}(m^{\prime\prime}_1, m^{\prime\prime}_2)
\)
\begin{equation}
+ J^{(2)}_{m^\prime m^{\prime\prime}} \cdot
\left\{
h^{(1)} (m^\prime_1, m^\prime_2) \,
( B_1 h ^{(j)} ) (m^{\prime\prime}_1, m^{\prime\prime}_2) \,
+ h^{(2)} (m^\prime_1, m^\prime_2) \,
( B_2 h ^{(j)} ) (m^{\prime\prime}_1, m^{\prime\prime}_2) \,
\right\}
\end{equation}
where
\(
m^\prime \, = \, m^\prime_1 + m^\prime_2 \, , \,
m^{\prime\prime} \, = \, m^{\prime\prime}_1 \, + \, m^{\prime\prime}_2 \
\mbox{and}
\)

\begin{equation}
\left\{
\begin{array}{l}
J^{(1)}_{m^\prime m^{\prime\prime}} \, = \,
-\displaystyle{\int\limits_{0}^{1}} \,
\gamma^{\frac {m^\prime} 2}
(1-\gamma)^{\frac {m^{\prime\prime}+3} 2} d\gamma \\
J^{(2)}_{m^\prime m^{\prime\prime}} \, = \,
\displaystyle{\int\limits_{0}^{1}} \,
\gamma^{\frac {m^\prime+1} 2}
(1-\gamma)^{\frac {m^{\prime\prime}+2} 2} d\gamma
\end{array}
\right.
\end{equation}
\begin{align*}
( B_1 \, h^{(j)} ) \,
( m^\prime_1 , m^\prime_2 ) \, = \, h^{(j)} \, (m^\prime_1 - 1 ,
m^\prime_2) \, + \,
(m^\prime_1 \, + \, 1) \,
h^{(j)} ( m^\prime_1 \, + \, 1 , m^\prime_2 ) \\
\\
(B_2 \, h^{(j)} ) \,
( m^\prime_1 , m^\prime_2 ) \, = \,
h^{(j)} \, ( m^\prime_1 , m^\prime_2 - 1) \, + \,
(m^\prime_2 \, + \, 1 ) \, h^{(j)} \, ( m^\prime_1 , m^\prime_2 \,
+ \, 1 )
\end{align*}

To simplify the system (26), we shall look for solutions with
$h^{(j)}(0,0)\,=0,\, j=1,2$.
Below we sometimes write $h^{(j)} ( m_1 , m_2)$ as $h^{(j)}_{m_1 m_2}$
or $h^{(j)}_{m_1 , m_2}$ when there is no confusion.  Similar conventions will
be applied to $J^{(j)} ( m_1, m_2)$.  For $m_1 + m_2 = 1$, we have

\[
\left\{
\begin{array}{lll}
h^{(1)}_{10} & = & J^{(1)}_{01}\cdot ( h^{(1)}_{10} + h^{(2)}_{01} ) \cdot h^{(1)}_{10}
     + J^{(2)}_{10} \cdot ( h^{(1)}_{10} h^{(1)}_{10} + h^{(2)}_{10}h^{(1)}_{01} ) \\
\\
h^{(1)}_{01} & = & J^{(1)}_{01}\cdot ( h^{(1)}_{10} + h^{(2)}_{01} ) \cdot h^{(1)}_{01}
     + J^{(2)}_{10} \cdot ( h^{(1)}_{01} h^{(1)}_{10} + h^{(2)}_{01}h^{(1)}_{01} ) \\
\\
h^{(2)}_{10} & = & J^{(1)}_{01}\cdot ( h^{(1)}_{10} + h^{(2)}_{01} ) \cdot h^{(2)}_{10}
     + J^{(2)}_{10} \cdot ( h^{(1)}_{10} h^{(2)}_{10} + h^{(2)}_{10}h^{(2)}_{01} ) \\
\\
h^{(2)}_{01} & = & J^{(1)}_{01}\cdot ( h^{(1)}_{10} + h^{(2)}_{01} ) \cdot h^{(2)}_{01}
     + J^{(2)}_{10} \cdot ( h^{(1)}_{01} h^{(2)}_{10} + h^{(2)}_{01}h^{(2)}_{01} ) \\
\end{array}
\right.
\]

\noindent
where $J^{(1)}_{01} = -1/3$ and $J^{(2)}_{10} = 1/6$. There are two cases:
\begin{namelist}{xxxxxx}
\item[\underline{\sf Case 1.}] $ h^{(1)}_{10} + h^{(2)}_{01} = -6$. In this case
($h^{(1)}_{10}$, $h^{(1)}_{01}$, $h^{(2)}_{10}$, $h^{(2)}_{01}$) only needs to satisfy:
\[
(h^{(1)}_{10}+3)^2 = 9 - h^{(1)}_{01} h^{(2)}_{10}
\]
This is a two parameter family of solutions.
\item[\underline{\sf Case 2.}] $ h^{(1)}_{10} + h^{(2)}_{01} \neq -6$. In this case
($h^{(1)}_{10}$, $h^{(1)}_{01}$, $h^{(2)}_{10}$, $h^{(2)}_{01}$)
can be uniquely determined and
we have $h^{(1)}_{10}= h^{(2)}_{01} = -2 $, $ h^{(1)}_{01} = h^{(2)}_{10}=0$.
\end{namelist}
For the rest of this paper we shall consider only the case $2$ for which
$h^{(1)}_{10}= h^{(2)}_{01} = -2 $, $ h^{(1)}_{01} = h^{(2)}_{10}=0$.
Let us write down the recurrent relations for $m_1 + m_2 = 2$, $j=1,2$ :
\[
\left\{
\begin{array}{lll}
h^{(j)}_{20} & = &  -( 2 J^{(2)}_{20} + 4 J^{(1)}_{02} +4 J^{(2)}_{11} ) h^{(j)}_{20}
 + 2 J^{(1)}_{11}\cdot h^{(j)}_{10}\cdot h^{(1)}_{20}
 + h^{(j)}_{10} \cdot J^{(1)}_{11} \cdot h^{(2)}_{11} \\
\\
h^{(j)}_{11} & = &  -( 2 J^{(2)}_{20} + 4 J^{(1)}_{02} +4 J^{(2)}_{11} ) h^{(j)}_{11}
 + J^{(1)}_{11}\dot h^{(j)}_{01}\cdot ( 2 h^{(1)}_{20} + h^{(2)}_{11} )
 + J^{(1)}_{11}\cdot h^{(j)}_{10} \cdot (h^{(1)}_{11} + 2 h^{(2)}_{02} ) \\
\\
h^{(j)}_{02} & = &  -( 2 J^{(2)}_{20} + 4 J^{(1)}_{02} +4 J^{(2)}_{11} ) h^{(j)}_{02}
 + 2 J^{(1)}_{11}\cdot h^{(j)}_{01}\cdot h^{(2)}_{02}
 + h^{(j)}_{01} \cdot J^{(1)}_{11} \cdot h^{(1)}_{11} \\
\end{array}
\right.
\]

It is not difficult to check that the only solution to the above system
is $h^{(j)}_{20}=h^{(j)}_{02}=h^{(j)}_{11}=\,0$. Solving the recurrent relations for
$m_1 + m_2 =3$ gives us:
\[
\left\{
\begin{array}{l}
h^{(1)}_{03}\,  = \, h^{(2)}_{30} =  0 \\
h^{(1)}_{12} = h^{(2)}_{03} \\
h^{(1)}_{21} = h^{(2)}_{12} \\
h^{(1)}_{30} = h^{(2)}_{21}
\end{array}
\right.
\]

\medskip
This shows that ($h^{(1)}_{12} ,
h^{(1)}_{21} , h^{(1)}_{30} $) can be considered as free parameters.
For any $p \geq 4$, the recurrent relations for $m_1 + m_2 = p$ form a
linear system of equations for the variables $\{ h^{(j)}_{m_1 , p - m_1}
\}^p_{m_1 = 0}$ with coefficients depending on $h^{(j)}_{01}$ and
$h^{(j)}_{10}$ only.  In principle, they can be solved and an explicit
expression for the solutions can be found.  We emphasize here that if the free
parameters take real values then the whole solution is also real.

It is not difficult to check that for any
values of ($h^{(1)}_{12} , h^{(1)}_{21} , h^{(1)}_{30} $),
one can find all $h^{(j)}_{m_1 , m_2}$ $(m_1 + m_2 \geq 4)$ by using (26).
The solution we obtain is formal in the sense that it satisfies (26) but
$h_{m_1 , m_2}$ with $m_1 + m_2 = p$ may not decay as $p \longrightarrow \infty$.
We are now ready to formulate the theorem concerning the existence
of formal solutions to (26).

\setcounter{section}{+4}
\begin{Thm} For any values of ($h^{(1)}_{12} ,
h^{(1)}_{21} , h^{(1)}_{30} $), there exists a unique
formal solution to the recurrent equation (26).
\end{Thm}

Thus, theorem 4.1 claims the existence of a three-parameter
family of solutions of (21) parameterized by $h^{(1)}_{12}$, $h^{(1)}_{21}$ and
$h^{(1)}_{30}$. It turns out that if  $h^{(1)}_{12}$, $h^{(1)}_{21}$ and
$h^{(1)}_{30}$ are sufficiently small, then $h^{(j)}_{m_1 , m_2}$ decay as $m_1 + m_2 = p$
tends to infinity.  Let us say that $h^{(j)}_{m_1 , m_2}$ has degree $d$
if $m_1 + m_2 = d$.  For each $d \geq 4$, introduce the vector
$h^{(d)} = ( h^{(1)}_{0 , d}, h^{(1)}_{1, d-1} , \ldots , h^{(1)}_{d, 0},
h^{(2)}_{0, d}, \ldots, h^{(2)}_{d, 0})^T$.  The
vector $h^{(d)}$ contains all terms of degree $d$.  By the
recurrent relation (26)
\begin{equation}
C^{(d)} \, h^{(d)} \, = \, b^{(d)}
\end{equation}
where the vector $b^{(d)}$ contains terms of degree $\leq d - 1$.  Also
$C^{(d)} \in \bR^{(2d + 2 ) \times (2d + 2)}$ is a matrix:

\[
C^{(d)}_{k \ell} \, = \,
\left\{
\begin{array}{l l l l}
1 - \displaystyle{\frac {16d-16+32k} {(d+1)(d+3)(d+5)}} , &&\mbox{if} \ 1\le k \, = \, l \le d+1 \\
\\
1 - \dfrac {80d+80-32k} {(d+1)(d+3)(d+5)} , &&\mbox{if} \ d+2\le k \, = \, l \le 2d+2 \\
\\
-\dfrac {32(d-k+2)} {(d+1)(d+3)(d+5)}, && \mbox{if} \ 2\le k \le d+1, l=d+k  \\
\\
-\dfrac {32(k-d-1)} {(d+1)(d+3)(d+5)}, && \mbox{if} \ d+2\le k \le 2d+1, l=k-d  \\
\\
0,  && \mbox{all other cases}
\end{array}
\right.
\]

\noindent
It is easy to check that if $d \geq 4$, then $C^{(d)}$ is nonsingular
and as $d \longrightarrow \infty$, $C^{(d)}$ converges to the identity
matrix.  This observation immediately implies the following lemma:

\setcounter{Lem}{1}
\begin{Lem}
Let $(C^{(d)} )^{-1}$ be the inverse matrix of $C^{(d)}$ for $d
\geq 4$. There exists an absolute constant $C_1 > 0$ such that for
all $d \geq 4$
\[
\parallel ( C^{(d)} )^{-1} \parallel \, \leq \, C_1 \, .
\]
\end{Lem}
We are now ready to derive an estimate which gives
the decay of solutions of the recurrent relation (26).

\setcounter{Thm}{+2}
\begin{Thm}
If $| h^{(1)}_{12} | \leq \delta$, $| h^{(1)}_{21} | \leq \delta$,
$| h^{(1)}_{30} | \leq \delta$ and $\delta$ is sufficiently small, then
for some $C_2 > 0$, $0 < \rho < \frac 1 4$, we have
\[
\abs{ h^{(j)}_{m_1, m_2} }   \le C_2 \,
 \frac { \rho^{m_1+m_2} }       {\Gamma \left( \frac {m_1+m_2+7} 2\right)}
 \quad
 \forall \, m_1 \ge 0 , \, m_2 \, \ge \, 0, \; j=1,2 .
\]
\end{Thm}

\noindent {\bf Proof}. \ We begin by noting that $h^{(j)}_{m_1 m_2}=\,0$
if $m_1+m_2$ is even. This can be easily proven by using the
recurrent relation (26) and the fact that $h^{(j)}_{00}=0$ and
$h^{(j)}_{m_1,m_2}=0$ for $m_1+m_2=2$. Let $0<\rho_1<1$, $\rho_1$ will
be chosen  sufficiently small. We shall use induction on $m_1 +
m_2$ where $m_1 + m_2$ is odd.  According to the induction
hypothesis
\begin{equation}
| h^{(j)}_{m_1 , m_2} | \, \leq \,
\displaystyle \frac {\rho_1^{m_1+m_2+2}}
              { \Gamma{\left(\frac {m_1+m_2+7} 2\right)} }
 \, g(m_1+m_2)
\end{equation}
for every $3 \leq m_1 + m_2 \leq d-2$ where $d \geq L$ is an odd
number and $L$ will be chosen later to be sufficiently large. Also
$g$ is a function to be specified later. We shall comment on the
choice of $L$ and verify the induction hypothesis for $3 \leq m_1
+ m_2 \leq L$ later. Let us show that the same inequality holds
for $m_1 + m_2 = d$.  Without any loss of generality, let us consider
$j=1$. The case $j=2$ is similar. Fix $m_1$ and let $b^{(d)}_{m_1}$ be the
$(m_1 + 1)^{\sf th}$ component of the vector $b^{(d)}$ in the
equation (28).  We now estimate $b^{(d)}_{m_1}$ using the
induction hypothesis (29) and the equation (26):
\begin{align*}
\abs{ b^{(d)}_{m_1} }
\le & \sum_{m'=2}^{d-3} \abs{J^{(1)}_{m',m''}} \cdot 2\cdot
     \frac { \rho_1^{m'+3} } {\Gamma \left( \frac {m'+8} 2   \right)}
   \cdot \frac { \rho_1^{m''+2} } {\Gamma \left( \frac {m''+7} 2   \right)}
   \cdot (m'+1) g(m'+1) g(m'')  \\
 &+ \sum_{m'=4}^{d-3} \abs{J^{(1)}_{m',m''}} \cdot 2\cdot
     \frac { \rho_1^{m'+1} } {\Gamma \left( \frac {m'+6} 2   \right)}
   \cdot \frac { \rho_1^{m''+2} } {\Gamma \left( \frac {m''+7} 2   \right)}
   \cdot (m'+1)\cdot g(m'-1) g(m'')  \\
&+ \sum_{m'=3}^{d-2} \abs{J^{(2)}_{m',m''}} \cdot 2\cdot
     \frac { \rho_1^{m'+2} } {\Gamma \left( \frac {m'+7} 2   \right)}
   \cdot \frac { \rho_1^{m''+3} } {\Gamma \left( \frac {m''+8} 2   \right)}
   \cdot (m'+1)\cdot (m''+1)\cdot g(m') g(m''+1)  \\
&+ \sum_{m'=3}^{d-4} \abs{J^{(2)}_{m',m''}} \cdot 2\cdot
     \frac { \rho_1^{m'+2} } {\Gamma \left( \frac {m'+7} 2   \right)}
   \cdot \frac { \rho_1^{m''+1} } {\Gamma \left( \frac {m''+6} 2   \right)}
   \cdot (m'+1)\cdot g(m') g(m''-1)  \\
 & + 12 \left( \,\abs{ J^{(1)}_{2,d-2} } + \abs{ J^{(1)}_{d-1,1}}
       + \abs{ J^{(2)}_{d-2,2} } + \abs{ J^{(2)}_{1,d-1}} \,\right) \cdot
     \frac { \rho_1^{d} } {\Gamma \left( \frac {d+5} 2   \right)}
     \cdot g(d-2)  \\
\end{align*}
The last term in the {\sf rhs} of the above inequality comes from
the case where $h_{m^\prime_1 m^\prime_2}$ or
$h_{m^{\prime\prime}_1 m^{\prime\prime}_2}$ is of degree one since
the induction hypothesis holds only for $3 \leq m_1 + m_2 \, \leq
\, d - 2$.  Also in the estimation of the first four terms we use
the fact that for fixed $(m',m_1)$, there are at most $\min
\{m'+1, m''+1\} $ tuples of $(m',m_1^{''},m_2', m^{''}_2)$ such
that $m_1'+m_1^{''}=m_1$, $m_2'+m_2^{''}=m_2$, $m_1'+m_2'=m'$ and
$m_1^{''}+ m_2^{''}=m^{''}$. By (27), we have
\[
\abs{ J^{(1)}_{m',m^{\prime\prime}} }= \dfrac
     { \Gamma\left( \frac {m'+2} 2 \right)   \Gamma \left( \frac {m''+5} 2\right) }
     { \Gamma \left( \frac {m'+m''+7} 2   \right) }
\]
\[
 \abs{ J^{(2)}_{m',m''} } =\dfrac
    { \Gamma\left( \frac {m'+3} 2\right)    \Gamma\left( \frac {m''+4} 2\right) }
    { \Gamma \left( \frac {m'+m''+7} 2   \right) }
\]
and for some constant $C_3 >0$
\[
\abs{ J^{(1)}_{2,d-2}} + \abs{ J^{(1)}_{d-1,1} }
+ \abs{ J^{(2)}_{d-2,2} } + \abs{ J^{(2)}_{1, d-1} } \le \frac {C_3} {d^2}
\]
Therefore
\begin{align*}
\abs{ b_{m_1}^{(d)} } \le &
\frac {2 \rho_1^{d+5}} {\Gamma\left(\frac {d+7}2\right)}
\cdot \sum_{m'=2}^{d-3}
\dfrac { \Gamma \left( \frac {m'+2} 2  \right) \cdot (m'+1)  }
       { \Gamma \left( \frac {m'+8} 2 \right) }
\cdot
\dfrac { \Gamma \left( \frac {m^{\prime\prime}+5} 2  \right) }
       { \Gamma \left( \frac {m^{\prime\prime}+7} 2 \right) }
\cdot g(m^\prime +1 ) g( m^{\prime\prime} ) \\
& + \frac {2 \rho_1^{d+3}} {\Gamma\left(\frac {d+7}2\right)}
\cdot \sum_{m'=4}^{d-3}
\dfrac { \Gamma \left( \frac {m'+2} 2  \right) \cdot (m'+1)  }
       { \Gamma \left( \frac {m'+6} 2 \right) }
\cdot
\dfrac { \Gamma \left( \frac {m^{\prime\prime}+5} 2  \right) }
       { \Gamma \left( \frac {m^{\prime\prime}+7} 2 \right) }
\cdot g(m^\prime - 1 ) g( m^{\prime\prime} ) \\
& + \frac {2 \rho_1^{d+5}} {\Gamma\left(\frac {d+7}2\right)}
\cdot \sum_{m'=3}^{d-2}
\dfrac { \Gamma \left( \frac {m'+3} 2  \right) \cdot (m'+1)  }
       { \Gamma \left( \frac {m'+7} 2 \right) }
\cdot
\dfrac { \Gamma \left( \frac {m^{\prime\prime}+4} 2  \right) \cdot (m^{\prime\prime}+1)}
       { \Gamma \left( \frac {m^{\prime\prime}+8} 2 \right) }
\cdot g(m^\prime ) g( m^{\prime\prime} +1 ) \\
& + \frac {2 \rho_1^{d+3}} {\Gamma\left(\frac {d+7}2\right)}
\cdot \sum_{m'=3}^{d-4}
\dfrac { \Gamma \left( \frac {m'+3} 2  \right) \cdot (m'+1)  }
       { \Gamma \left( \frac {m'+7} 2 \right) }
\cdot
\dfrac { \Gamma \left( \frac {m^{\prime\prime}+4} 2  \right) }
       { \Gamma \left( \frac {m^{\prime\prime}+6} 2 \right) }
\cdot g(m^\prime ) g( m^{\prime\prime} -1 ) \\
&+ \frac {\rho_1^{d+2}} {\Gamma \left( \frac {d+7} 2\right)} \cdot
\frac {C_3} {d^2} \cdot \frac {\Gamma\left( \frac {d+7} 2 \right)}
{\Gamma\left( \frac {d+5} 2 \right)}
\cdot \frac {12} {\rho_1^2} \cdot g(d-2)\\
\le &
 \frac {\rho_1^{d+2}} {\Gamma\left(\frac{d+7}2\right)} \cdot \rho_1\cdot C_4\cdot
\left( \sum_{m'=2}^{d-3} g(m'+1)g(m'') + \sum_{m'=4}^{d-3} g(m'-1)g(m'') \right. \\
&\left.
\sum_{m'=3}^{d-2} g(m')g(m''+1) + \sum_{m'=3}^{d-4} g(m')g(m''-1)
\right)
 \, + \frac {\rho_1^{d+2}} {\Gamma\left(\frac{d+7}2\right)} \cdot
C_5 \cdot \frac {g(d-2)} {d\cdot \rho_1}
\end{align*}
where $C_4,\, C_5$ are some constants. Now we specify the choice
of the function $g$. Let $g(m)$ be such that $g_1=\alpha$ and
\[
g(m) = \sum_{p=1}^{m-1} g(p) g(m-p)     \qquad \text{for} \,\, m>1
\]
By the method of formal power series it is not difficult to show
that
\[
g(m) = \frac 1 2 \cdot \frac {(2m-1)!!} {m!} \cdot {(2\alpha)}^m
\]
Clearly, we have $const \le \frac {g(m+1)} {g(m)} \le const$, and
this immediately gives us
\begin{align*}
\abs{ b_{m_1}^{(d)} } \le &\frac {\rho_1^{d+2}}
                    {\Gamma\left( \frac {d+7} 2 \right)} \cdot C_6\cdot
      \sum_{m'=1}^d g(m') g(d-m') + \frac {\rho_1^{d+2}}
                    {\Gamma\left( \frac {d+7} 2 \right)}
        \cdot \frac {C_6} {d\cdot \rho_1} g(d) \\
\le & \frac {\rho_1^{d+2}}
                    {\Gamma\left( \frac {d+7} 2 \right)}
     \cdot g(d) \cdot \left( C_6 \rho_1 + \frac {C_6} {d\cdot \rho_1} \right)
\end{align*}
where $C_6>0$ is some constant. Now by Lemma 4.2, we obtain that
\[
 \abs{ h_{m_1m_2} } \le
 \displaystyle{ \frac {\rho_1^{d+2}} {\Gamma\left(\frac{d+7} 2\right)}
 g(d) \cdot C_1\cdot \left( C_6 \rho_1 + \frac {C_6} {d\cdot \rho_1} \right)}
\]
Choose $\rho_1$ so small that $C_1C_6 \rho_1 <\frac 1 2$ and
$\rho_1\cdot 4\alpha < \frac 1 4$.  Then take $L$
so large
that $\frac {C_1 C_6} {\rho_1 L} < \frac 12$. This clearly implies
\[
| h_{m_1 , m_2} | \, \leq \, \displaystyle{\frac
{\rho_1^{d+2}} {\Gamma\left( \frac {d+7} 2\right)}    g(d)}
\]
We now justify the induction hypothesis (29).  Recall
that our free parameters are $h^{(1)}_{12}$, $h^{(1)}_{21}$ and $h^{(1)}_{30}$.
It is easy to check that if we set $h^{(1)}_{12} = h^{(1)}_{21} = h^{(1)}_{30} = 0$,
then $h_{m_1 m_2} = 0$ for any
$m_1 + m_2 \geq 2$.  Since $L$ is fixed, and $0 < | h^{(1)}_{12} | <
\delta$, $0 < | h^{(1)}_{21} | < \delta$, $ 0< | h^{(1)}_{30} | < \delta $
with sufficiently small $\delta$,
then the induction hypothesis is satisfied. A simple estimate on $g$ gives that
\[
g(m) \le (4\alpha)^m
\]
Thus the theorem is proven if one takes $\rho=4\alpha \rho_1$.

As it is stated our solutions of (20) are determined by
five parameters $\sigma^{(1)} , h^{(1)}_{12} , h^{(1)}_{21}, h^{(1)}_{30}, \sigma^{(2)}$.
However, it turns out that these parameters are not independent and $\sigma_1$ can
be expressed through ( $h^{(1)}_{12}, h^{(1)}_{21} , h^{(1)}_{30}$ ).  Namely, let
$G^{\sigma^{(1)} , h^{(1)}_{12} , h^{(1)}_{21} , h^{(1)}_{30}, \sigma^{(2)} }$
$(Y)$ be the
solution of (20).  Then
\[
G^{\displaystyle{( \sigma^{(1)} , h^{(1)}_{12} , h^{(1)}_{21} , h^{(1)}_{30}, \sigma^{(2)} )}} \,
(Y) \, = \, G^{\displaystyle{( 1 , \sigma^{(1)} (
h^{(1)}_{12}-1)+1  \, ,  \, \sigma^{(1)} h^{(1)}_{21}, \,
\sigma^{(1)} (h^{(1)}_{30} -1)+1,\,  \sigma^{(2)} )}} (Y)\, .
\]
This equality is proved at the end of \S6.
We formulate now the final result concerning the existence of solutions of (21).

\setcounter{Thm}{1}
\begin{Thm}
Let $\sigma^{(1)} > 0$, $\sigma^{(2)}> 0$ and $h^{(1)}_{12}, h^{(1)}_{21}, h^{(1)}_{30} $
be sufficiently small. Then there exists a solution of (20) which
has the following form
\[
G^{\displaystyle{( \sigma^{(1)} , h^{(1)}_{12} , h^{(1)}_{21} , h^{(1)}_{30}, \sigma^{(2)})}} \,
( Y_1 , Y_2 , Y_3) \, = \,
\exp \, \left\{
- \, \frac{\sigma^{(1)}}{2} \,\left( | Y_1 |^2 \, + \, | Y_2 |^2 \right) \right\}
\, \cdot
\]
\[
\cdot \frac{\sigma^{(1)}}{2 \pi} \, \cdot \,
\exp \left\{
- \frac{\sigma^{(2)}}{2} | Y_3 |^2 \right\}
\:
\sqrt{\frac{\sigma^{(2)}}{2 \pi}} \, \cdot \,
\sqrt{\sigma^{(1)}} \,
H^{(h^{(1)}_{12} , h^{(1)}_{21}, h^{(1)}_{30} )} \,
( \sqrt{\sigma^{(1)}} \, Y_1 , \sqrt{\sigma^{(1)}} \, Y_2 ) \, .
\]
Here $H^{( h^{(1)}_{12} , h^{(1)}_{21}, h^{(1)}_{30} )}$ is the solution of (21) with the given
$h^{(1)}_{12} , h^{(1)}_{21}, h^{(1)}_{30} $.
\end{Thm}

As it was already mentioned the parameters $\sigma_1$, $h_{12}^{(1)}$,
$h_{21}^{(1)}$, $h_{30}^{(1)}$, $\sigma_2$ are not independent and
actually the set of solutions depends on four independent parameters (see Lemma 6.2).

From the estimate in Theorem 4.3 and from known asymptotic
formulas for the Hermite polynomials it follows that the series
giving $H^{( h^{(1)}_{12} , h^{(1)}_{21}, h^{(1)}_{30} )}$
converges for every $Y=(Y_1, Y_2)$. Better estimates are also easily available.

\medskip
\begin{center}
{\large \sf \S5. \ The Linearization Near Fixed Point}
\end{center}

Denote $h^{(1)}_{12} = x^{(1)}$, $h^{(1)}_{21} = x^{(2)}$,
$ h^{(1)}_{30} = x^{(3)} $. Our fixed points have the following form
\begin{equation}
\begin{split}
&G^{(\sigma^{(1)}, x^{(1)} , x^{(2)}, x^{(3)}, \sigma^{(2)})} = \,
\frac{\sigma^{(1)}}{2 \pi } \exp \left\{ - \frac{\sigma^{(1)}(
Y^2_1 + Y^2_2 )}{2}\right\}\,
\cdot
\\
&\cdot \sqrt{\frac{\sigma^{(2)}}{2 \pi }} \, \exp \left\{ -
\frac{\sigma^{(2)}Y^2_3}{2}\right\}\left( H_1^{(\sigma^{(1)} ,\,
x^{(1)},\, x^{(2)}, \, x^{(3)} )} ( Y_1 , Y_2 ),\, H_2^{(\sigma^{(1)} ,\,
x^{(1)}, \, x^{(2)}, \, x^{(3)} )} ( Y_1 , Y_2 ), \, 0 \right)
\end{split}
\end{equation}

Recall that $H^{(\sigma^{(1)} ,\, x^{(1)}, \, x^{(2)}, \, x^{(3)}
)}=\sqrt{\sigma^{(1)}}H^{(1 ,\, x^{(1)}, \, x^{(2)},\, x^{(3)}
)}(\sqrt{\sigma^{(1)}} Y_1,\sqrt{\sigma^{(1)}}Y_2)$ and $H^{(1 ,\,
x^{(1)}, \, x^{(2)}, \, x^{(3)} )}$ are described in \S 4.

 The strategy of the proof of the main result is based on the method of
renormalization group.  At the $p$-th step of our procedure, we
consider an interval on the time axis $S^{(p)} = \left[ S^{(p)}_-,
S^{(p)}_+ \right]$ such that $S^{(p+1)} \subseteq S^{(p)}$. From
our estimates it will follow that $\bigcap\limits_{p} \, S^{(p)} =
[ S_- , S_+ ]$ is an interval of positive length.  We want to find
conditions under which $\tilde{g}_r ( Y , s )$, $s \in S^{(p)}$,
have a representation
\[
\tilde{g}_r ( Y , s ) \, = \, \Lambda^{r-1}  r
\frac{\sigma^{(1)} }{2 \pi} \exp \left\{-
\displaystyle{\frac{\sigma^{(1)}(Y_1^2+Y_2^2)}{2} } \right\}
\sqrt{\displaystyle{\frac{\sigma^{(2)} }{2 \pi}}} \exp \left \{
- \frac{\sigma^{(2)} Y^2_3}{2} \right\} \cdot
\]
\[
\left( H_1^{(\sigma^{(1)},x^{(1)},x^{(2)}, x^{(3)}
)}(Y)+ \delta^{(r)}_1(Y,
s),\,H_2^{(\sigma^{(1)},x^{(1)},x^{(2)}, x^{(3)}
)}(Y) + \delta^{(r)}_2 ( Y , s ), \, \delta^{(r)}_3 ( Y , s )
\right)
\]
where 
$\delta^{(r)}_1$, $\delta^{(r)}_2$, $\delta^{(r)}_3$ tend to zero
as $r \to \infty$.
The renormalization is based on the crucial observation (see above)
that for large $p$, the sum over $p_1$ is a Riemannian integral sum
for an integral over $\gamma$ changing from $0$ to $1$. Let us write
\begin{equation}
\begin{split}
\tilde{g}_r ( Y , s ) \, \Lambda^{- r + 1} ( r^{-1} &\exp
\left\{ \frac{\sigma^{(1)} ( Y_1^2 + Y^2_2 )}{2} +
\frac{\sigma^{(2)}  Y^2_3}{2} \right\} \left( \frac{2
\pi}{\sigma^{(1)}} \right) \left( \frac{2 \pi}{\sigma^{(2)}
} \right)^{\frac{1}{2}}=
\\
&= \; H^{(\sigma^{(1)} , x^{(1)} , x^{(2)} ,
x^{(3)} )} \, ( Y_1, Y_2 ) \, + \, \delta^{(r)} ( \gamma, Y , s )
\end{split}
\end{equation}
where $\delta^{(r)}( \gamma, Y , s ) \, = \, \Bigl\{ \delta^{(r)}_j (
\gamma, Y , s ), \, 1 \leq j \leq 3 \Bigr\}$ = $\delta^{(p)}(\gamma,
Y, s),\:\gamma \, = \, \dfrac{r}{p}$. It is natural to
consider the set of functions $\{ \delta^{(p)} ( \gamma , Y , s ) \}$ as
a small perturbation of our fixed point (30). Recall that the third
component of $H^{(\sigma^{(1)} , x^{(1)}, x^{(2)},
x^{(3)})}$ is zero because of incompressibility and
$\delta^{(p)}_3$ can be found from the incompressibility condition.
Clearly,
\[
\delta^{(p + 1)} ( \gamma, Y , s ) \, = \, \delta^{(p)}
\left( \frac{p+1}{p } \, \gamma, Y , s \right) \, , \hspace{.5em} \gamma \, \le
\, \frac {p} {p+1} \, .
\]
The formula for $\delta^{(p+1)} ( 1 , Y , s )$ follows from (21):
\medskip
\medskip
\noindent

\( \exp \, \left\{ -
{\displaystyle{\frac{\sigma^{(1)}}{2}} \, ( | Y_1 |^2 \, + \,
| Y_2 |^2 ) \, - \, \displaystyle{\frac{\sigma^{(2)} }{2}}
| Y_3 |^2} \right\}
 \, \cdot \, \displaystyle{\frac{\sigma^{(1)} }{2 \pi }  \,
\sqrt{ \frac{\sigma^{(2)} }{2 \pi }} } \, \cdot \,  \delta^{(p+1)}_{j} ( 1 , Y , s )
\)

\(
\, = \,
\displaystyle{\int\limits_{0}^{1}} d \gamma \,
\displaystyle{\int\limits_{\bR^3}} \, \frac{\sigma^{(1)}
}{2 \pi \gamma} \, \cdot \, \sqrt{ \frac{ \sigma^{(1)} }{2
\pi \gamma}} \, \cdot \, \frac{\sigma^{( 2 )} }{2 \pi ( 1 -
\gamma )} \, \cdot \, \sqrt{ \frac{ \sigma^{(2)} }{2 \pi (
1 - \gamma )} } \, \cdot \, \)

\medskip
\( \exp \left\{ \displaystyle{-\frac{\sigma^{(1)}  ( | Y_1 -
Y^\prime_1 |^2 + | Y_2 - Y^\prime_2 |^2 )}{2 \pi \gamma} } \right.
\left. - \, \displaystyle{\frac{ \sigma^{(2)}  | Y_3 -
Y^\prime_3 |^2}{2 \pi \gamma} \, - \,
\displaystyle{\frac{\sigma^{(1)}  ( | Y^\prime_1 |^2 + |
Y^\prime_2 |^2 )}{2 \pi ( 1 - \gamma )} } } \, - \, \right. \)

\medskip
\( \left. - \; \displaystyle{\frac{\sigma^{(2)}  |
Y^\prime_3 |^2}{2 \pi ( 1 - \gamma )} } \right\} \; \left\{ \left[
- ( 1 - \gamma )^{\frac{3}{2}} \, \left( \dfrac{Y_1 -
Y^\prime_1}{\sqrt{\gamma}}  \, \right. H_1 \left(
\displaystyle{\frac{Y - Y^\prime}{\sqrt{\gamma}} } \right) \, + \,
\displaystyle{\frac{Y_2 - Y^\prime_2}{\sqrt{\gamma}}} \, H_2 \left(
\displaystyle{\frac{Y -Y^\prime}{\sqrt{\gamma}} } \right) \right)
\right. \)

\medskip
\(
\left.
+ \, \gamma^{\frac{1}{2}} ( 1 - \gamma ) \,
\left(
\displaystyle{\frac{Y^\prime_1}{\sqrt{1 - \gamma}} } \,
H_1 \left(
\frac{Y - Y^\prime}{\sqrt{\gamma}} \right) \, + \,
\frac{Y^\prime_2}{\sqrt{1 - \gamma}} \, H_2
\left(
\frac{Y - Y^\prime}{\sqrt{\gamma}}
\right)
\right]
\, \delta_{j}^{(p+1)}
\left(
1 - \gamma , \,
\displaystyle{\frac{Y^\prime}{\sqrt{1 - \gamma}}} , s \right)
\right.
\)

\medskip
\(
+ \,
\left[
- \left( 1 - \gamma )^{\frac{3}{2}} \,
\left(
\displaystyle{\frac{Y_1 - Y^\prime_1}{\sqrt{\gamma}} } \,
\delta_{1}^{(p+1)}
\left(
\gamma ,
\displaystyle{\frac{Y - Y^\prime}{\sqrt{\gamma}} } \,, s
\right) \, + \,
\displaystyle{\frac{Y_2- Y^\prime_2}{\sqrt{\gamma}} } \,
\delta_{2}^{(p+1)}
\left(
\gamma ,
\displaystyle{\frac{Y - Y^\prime}{\sqrt{\gamma}} } \,, s
\right)
\right)
\right.
\right.
\)

\medskip
\(
\left.
\left.
+ \,
\gamma^{\frac{1}{2}} \,
( 1 - \gamma ) \,
\left(
\displaystyle{\frac{Y^\prime_1}{\sqrt{1 - \gamma} } } \,
\delta_{1}^{(p+1)}
\left(
\gamma, \displaystyle{\frac{Y - Y^\prime}{\sqrt{\gamma}} } \, , s
\right)
\, + \,
\displaystyle{\frac{Y^\prime_2}{\sqrt{1 - \gamma}} } \,
\delta_{2}^{(p+1)}
\left(
\gamma , \displaystyle{\frac{Y - Y^\prime}{\sqrt{\gamma}} } , s
\right)
\right)
\right)
\right]
\)
\begin{equation}
\left.
\left.
\hspace{-3.85in}H_j \left(
\displaystyle{\frac{Y^\prime}{\sqrt{1-\gamma}} } , s
\right)
\right\} \; d^3 Y^\prime,
\qquad\qquad j\,=\, 1,2
\right.
\end{equation}
We did not include in the last expression terms which are quadratic
in $\delta$ because in this section we consider only the linearized
map.

Another way to introduce the semi-group of linearized maps
is the following. Take $\theta>0$ which later will tend to zero.
Denote $\gamma_j=(1+\theta)^{-j}$, $j=0, 1, 2, \dots$ Our
semigroup will act on the space $\Delta$ of functions
$\delta(\gamma, Y)$ with values in $C^3$ such that

\begin{enumerate}
\item for each $\gamma$, $0\le\gamma\le 1$, the function
$\delta(\gamma, Y)$ belongs to the Hilbert space
$L^2=L^2\left(R^3\right)$ of square-integrable
functions with respect to the weight $\left(\frac{\sigma^{(1)}}{2\pi}\right)^{\frac 32}
\exp\{-\frac{\sigma^{(1)}Y^2}{2}\}$, $Y=(Y_1,Y_2, Y_3)$;
\item As a function of $\gamma$ it is a continuous curve in this Hilbert space and
$\max\limits_{0\le\gamma\le 1}\|\delta(\gamma, Y)\|_{L^2}<\infty$.
\end{enumerate}

Define the linearized map $L_{\theta}$ corresponding to $\theta$
as follows:

\begin{enumerate}
\item for $\gamma_{j+1}\leq \gamma\leq \gamma_{j}$, $j=1,2,\dots$
\[L_\theta(\delta(\gamma,Y))=\delta(\gamma(1+\theta),Y);\]
\item for $\dfrac{1}{1+\theta}\leq \gamma \leq 1$ the function $L_\theta(\delta(\gamma,
Y))$ is given by the formula
\[L_\theta(\delta(\gamma, Y))=\delta_{p_1}(1,Y,s)\]
where $p_1$ is found from the relation $\dfrac{p_1}{p}=\gamma$.
\end{enumerate}

In other words at $\gamma=1$ we use (32) to find the new
$\delta^{(p+1)}(1,Y,s)$. After that we apply 1.

It is easy to see that there exist the limits
$\lim\limits_{\theta\to 0\atop{n\theta\to
t}}L_{\theta}^n=A^t$ and the operators $A^t$ constitute a
semi-group. For $\gamma<1$, $t>0$ such that $\gamma e^t<1$
\[A^t\delta(\gamma, Y)=\delta(\gamma e^t, Y)\]

Let $\mathcal{A}$ be the infinitesimal generator of the semi-group
$A^t$. In \S6 we study in more detail the spectrum and
eigenfunctions of $\mathcal{A}$.

\setcounter{section}{5}
\setcounter{Lem}{0}
\begin{Lem}
The eigenfunctions of the group $A^t$ have the form
\[\delta(\gamma, Y)=\gamma^\alpha\tilde{\Phi}_\alpha(Y)\]
where $\tilde{\Phi}_\alpha$ is a function with values in $C^3$
satisfying (32).
\end{Lem}

In more detail, if we take
$\delta(\gamma,Y)=\gamma^\alpha\tilde{\Phi}_\alpha(Y)$ and
substitute it into the {\sf rhs} of (32) we get in the {\sf lhs}
$\delta^{(p+1)}(Y)=\tilde{\Phi}_\alpha(Y)$.

{\bf Proof.} If $\delta(\gamma, Y)$ is an eigenfunction then from
the formula for $A^t$
\[A^t\delta(\gamma, Y)=\delta(\gamma e^{-t}, Y)=e^{-\alpha t}\delta(\gamma, Y)\]

Let $\gamma\to 1$. Then
\[\delta(e^{-t},Y)=e^{-\alpha t}\Phi(Y)=\gamma^\alpha\Phi(Y)\]
Lemma is proven.
The space $\Delta$ is spanned by the eigenfunctions of $\{A^t\}$ in the sense
that for any $h\in \Delta$ we have the expansion

\[h(\gamma, Y)=\sum\limits_{\alpha\in {\sf spec}\mathcal{A}}C^{(\alpha)}
\gamma^\alpha\Phi_\alpha(Y)\]

The coefficients $C^{(\alpha)}$ are found with the help
of the eigenfunctions of the conjugate system $\{(A^\ast)^t\}$.
The form of the conjugate semi-group and its eigenfunctions can be
investigated using the described above discrete approximation.
We do not dwell more on this.

%
%
%
\begin{center}
{\large \S6. \ \sf The Spectrum of the Group of Linearized Maps }
\end{center}

In this section we show that the solutions of (21) studied in \S4 have $l^{(u)}=4$
unstable eigenvalues and $l^{(n)}=6$ neutral eigenvalues. Therefore in the
renormalization group approach we consider $10$-- parameter families of initial conditions
(see below).

As was already mentioned, in the limit
$p \longrightarrow \infty$ the linearized maps generate a
semi-group of operators acting in the space $\Delta$ of functions $f^{(j)} (
\gamma , Y )$, $0 \leq \gamma \leq 1$, $Y \in \bR^3$, $j=1,2$ which are
continuous as functions of $\gamma$ in the Hilbert space $L^2$.  At
$\gamma = 1$, the functions $f^{(j)} ( \gamma , Y )$ satisfy the boundary
condition which follows from (32):

\medskip
\( \exp \, \left\{ -
{\displaystyle{\frac{\sigma^{(1)}}{2}} \, ( | Y_1 |^2 \, + \,
| Y_2 |^2 ) \, - \, \displaystyle{\frac{\sigma^{(2)}}{2}}
| Y_3 |^2} \right\}
 \, \cdot \, \displaystyle{\frac{\sigma^{(1)}}{2 \pi }  \,
\sqrt{ \frac{\sigma^{(2)} }{2 \pi }} } \, \cdot \,  f^{(j)} ( 1 , Y )
\)

\(
\, = \,
\displaystyle{\int\limits_{0}^{1}} d \gamma \,
\displaystyle{\int\limits_{\bR^3}} \, \frac{\sigma^{(1)}}
{2 \pi \gamma} \, \cdot \, \sqrt{ \frac{ \sigma^{(2)}}{2
\pi \gamma}} \, \cdot \, \frac{\sigma^{( 1 )}}{2 \pi ( 1 -
\gamma )} \, \cdot \, \sqrt{ \frac{ \sigma^{(2)}}{2 \pi (
1 - \gamma )} } \, \cdot \, \)

\medskip
\( \exp \left\{ \displaystyle{-\frac{\sigma^{(1)} ( | Y_1 -
Y^\prime_1 |^2 + | Y_2 - Y^\prime_2 |^2 )}{2 \pi \gamma} } \right.
\left. - \, \displaystyle{\frac{ \sigma^{(2)} | Y_3 -
Y^\prime_3 |^2}{2 \pi \gamma} \, - \,
\displaystyle{\frac{\sigma^{(1)} ( | Y^\prime_1 |^2 + |
Y^\prime_2 |^2 )}{2 \pi ( 1 - \gamma )} } } \, - \, \right. \)

\medskip
\( \left. - \; \displaystyle{\frac{\sigma^{(2)} |
Y^\prime_3 |^2}{2 \pi ( 1 - \gamma )} } \right\} \; \left\{ \left[
- ( 1 - \gamma )^{\frac{3}{2}} \, \left( \dfrac{Y_1 -
Y^\prime_1}{\sqrt{\gamma}}  \, \right. H_1 \left(
\displaystyle{\frac{Y - Y^\prime}{\sqrt{\gamma}} } \right) \, + \,
\displaystyle{\frac{Y_2 - Y^\prime_2}{\sqrt{\gamma}}} \, H_2 \left(
\displaystyle{\frac{Y -Y^\prime}{\sqrt{\gamma}} } \right) \right)
\right. \)

\medskip
\(
\left.
+ \, \gamma^{\frac{1}{2}} ( 1 - \gamma ) \,
\left(
\displaystyle{\frac{Y^\prime_1}{\sqrt{1 - \gamma}} } \,
H_1 \left(
\frac{Y - Y^\prime}{\sqrt{\gamma}} \right) \, + \,
\frac{Y^\prime_2}{\sqrt{1 - \gamma}} \, H_2
\left(
\frac{Y - Y^\prime}{\sqrt{\gamma}}
\right)
\right]
\, f^{(j)}
\left(
1 - \gamma , \,
\displaystyle{\frac{Y^\prime}{\sqrt{1 - \gamma}}} \right)
\right.
\)

\medskip
\(
+ \,
\left[
- \left( 1 - \gamma )^{\frac{3}{2}} \,
\left(
\displaystyle{\frac{Y_1 - Y^\prime_1}{\sqrt{\gamma}} } \,
f^{(1)}
\left(
\gamma ,
\displaystyle{\frac{Y - Y^\prime}{\sqrt{\gamma}} }
\right) \, + \,
\displaystyle{\frac{Y_2- Y^\prime_2}{\sqrt{\gamma}} } \,
f^{(2)}
\left(
\gamma ,
\displaystyle{\frac{Y - Y^\prime}{\sqrt{\gamma}} }
\right)
\right)
\right.
\right.
\)

\medskip
\(
\left.
\left.
+ \,
\gamma^{\frac{1}{2}} \,
( 1 - \gamma ) \,
\left(
\displaystyle{\frac{Y^\prime_1}{\sqrt{1 - \gamma} } } \,
f^{(1)}
\left(
\gamma, \displaystyle{\frac{Y - Y^\prime}{\sqrt{\gamma}} }
\right)
\, + \,
\displaystyle{\frac{Y^\prime_2}{\sqrt{1 - \gamma}} } \,
f^{(2)}
\left(
\gamma , \displaystyle{\frac{Y - Y^\prime}{\sqrt{\gamma}} }
\right)
\right)
\right)
\right]
\)
\begin{equation}
\left.
\left.
\hspace{-3.85in} H_j \left(
\displaystyle{\frac{Y^\prime}{\sqrt{1-\gamma}} }
\right)
\right\} \; d^3 Y^\prime,
\qquad\qquad j\,=\, 1,2
\right.
\end{equation}

\noindent
Denote by $\cR_p$ the linear operator which transforms $\{
\delta^{(p)} ( \gamma , Y , s ) \}$ into $\{ \delta^{(p + 1)} ( \gamma , Y , s
) \}$.  Here $s$ is a parameter which plays no role in this section.
As it was explained in \S 5, for each $t$ there exists the limit
$\lim\limits_{p \longrightarrow \infty} \, \cR^{tp}_p = A^t$ so that the
operators $A^t$ constitute a semi-group having an infinitesimal
generator $\cA = \lim\limits_{t \downarrow 0} \, \frac{A^t - I}{t}$.
In our case $\cA \delta ( \gamma , Y , s ) \, = \, \gamma \,
\frac{\partial \delta ( \gamma , Y , s )}{\partial \gamma}$, $0 < \gamma <
1$ and for $\gamma = 1$ the function $\delta ( 1 , Y , s )$ satisfies
the boundary condition (33) in which $f ( \gamma , Y ) = \delta^{(p+1)} (
1 , Y, s )$.

If $\alpha$ is an eigenvalue of $\cA$, then the corresponding
eigenfunction has the form $\gamma^\alpha \Phi_{\alpha, \sigma^{(1)}, \sigma^{(2)} } ( Y )$
(see Lemma 5.1), where
$\Phi_{\alpha, \sigma^{(1)}, \sigma^{(2)} } ( Y )$ satisfies the equation (33) with
$f ( \gamma, Y ) = \gamma^\alpha \Phi_{\alpha, \sigma^{(1)}, \sigma^{(2)} } ( Y )$.
If $\Re(\alpha)>0$ ($\Re(\alpha)=0$) then the corresponding eigenvalue is called unstable (neutral).
All other eigenvalues are called stable. The subspaces generated by unstable,
neutral, stable eigenvalues are denoted by $\Gamma^{(u)}$, $\Gamma^{(n)}$, $\Gamma^{(s)}$ respectively.

As before, for $\Phi^{(j)}_{\alpha , \sigma^{(1)} , \sigma^{(2)} } ( Y )$ the
following scaling relation with respect to $\sigma^{(1)} , \sigma^{(2)}$ is
valid:

\[
\Phi^{(j)}_{\alpha , \sigma^{(1)} , \sigma^{(2)} } ( Y ) \, \propto \,
\Phi^{(j)}_{\alpha , 1 , 1} \,
( \sqrt{\sigma^{(1)} } Y_1 , \sqrt{\sigma^{(1)} } Y_2 , \sqrt{\sigma^{(2)} } Y_3 )
\]

\noindent
Therefore it is enough to consider the above equation (33) for $\sigma^{(1)}
= \sigma^{(2)} = 1$.  We again use the expansion over {\it Hermite}
polynomials:

\[
\Phi^{(j)}_{\alpha , 1 , 1} ( Y ) \, = \,
\Phi^{(j)}_{\alpha} ( Y ) \, = \,
\displaystyle{\sum\limits_{m_1 , \, m_2 , \, m_3}} \,
f^{(j)}_\alpha ( m_1 , m_2 , m_3) \,
He_{m_1} (Y_1) \,
He_{m_2} ( Y_2) He_{m_3} (Y_3)
\]

\noindent
Here $j$ takes values $1,2,3$. Since in $m_3$ it is the usual convolution and $H$ does not depend on
$Y_3$, it is enough to look for solutions of (33) having the form
$f_{ m_1 , m_2} \delta_{m_3}$.  Put $\beta \, = \, \alpha \, + \,
\frac{m_3}{2}$ and
$f^{(j)}_{\beta} (m_1, m_2) = f^{(j)}_{\alpha}(m_1, m_2)\delta_{m_3}$.
We come to the linear system of recurrent relations

\[
f^{(j)}_{\beta}
(m_1 , m_2)  =
\displaystyle{\sum\limits_{m_1^\prime \, + \, m^{\prime\prime}_1 \, = \,
m_1 \atop{m^\prime_2 \, + \, m^{\prime\prime}_2 \, = \, m_2}}} \
J^{(1)}_{m^\prime, m^{\prime\prime}+2\beta}
\left( ( B_1 h^{(1)} ) \, (m^\prime_1 , m^\prime_2 ) \,
  + ( B_2 h^{(2)} ) \, (m^\prime_1 , m^\prime_2 ) \, \right)
f^{(j)}_\beta ( m^{\prime\prime}_1 , m^{\prime\prime}_2 )
\]
\[
\begin{array}{lll}
+ & J^{(2)}_{m^\prime, m^{\prime\prime}+2\beta} \cdot
h^{(1)} ( m^\prime_1 , m^\prime_2)\cdot (B_1 f^{(j)}_{\beta})(m_1^{\prime\prime}, m_2^{\prime\prime}) \\
\\
+ & J^{(2)}_{m^\prime, m^{\prime\prime}+2\beta} \cdot
h^{(2)} ( m^\prime_1 , m^\prime_2)\cdot (B_2 f^{(j)}_{\beta})(m_1^{\prime\prime}, m_2^{\prime\prime}) \\
\\
+ & J^{(1)}_{m^\prime+2\beta, m^{\prime\prime}} \cdot
\Bigl(
(B_1 f^{(1)}_{\beta})(m_1^\prime,m_2^\prime) + (B_2 f^{(2)}_{\beta})(m_1^\prime,m_2^\prime)
\Bigr) h^{(j)}(m_1^{\prime\prime}, m_2^{\prime\prime}) \\
\\
+ & J^{(2)}_{m^\prime+2\beta, m^{\prime\prime}} \cdot
f_{\beta}^{(1)} ( m^\prime_1 , m^\prime_2)\cdot (B_1 h^{(j)})(m_1^{\prime\prime}, m_2^{\prime\prime}) \\
\\
+ & J^{(2)}_{m^\prime+2\beta, m^{\prime\prime}} \cdot
f_{\beta}^{(2)} ( m^\prime_1 , m^\prime_2)\cdot (B_2 h^{(j)})(m_1^{\prime\prime}, m_2^{\prime\prime}) \\
\end{array}
\eqno(34)
\]

\noindent
Introduce the vector
\[
f^{(d)}_\beta \, = \,
\left(
f^{(1)}_\beta \, ( 0 , d ) , \,
f^{(1)}_{\beta} \, ( 1 , d - 1) \, , \ldots \, ,
f^{(1)}_{\beta} \, ( d , 0 ) \,
f^{(2)}_{\beta} \, ( 0 , d ) , \,
f^{(2)}_{\beta} \, ( 1 , d - 1) \, , \ldots \, ,
f^{(2)}_\beta \, ( d , 0 ) \right)^T
\]

\noindent
The vector $f^{(d)}_\beta$ contains all terms of degree $d$.  The
recurrent relation (34) can be written as
\[
C^{(d)}_\beta \, f^{(d)}_\beta \, = \, b^{(d)}_{\beta}
\]

\noindent
where the vector $b^{(d)}_\beta$ contains terms of degree $\leq d - 1$.
Also $C^{(d)}_\beta \, \in \, \bR^{2 ( d + 1 ) \times 2 ( d + 1)}$ is a
matrix.  Let $C^{(d)}_\beta ( k , \ell )$ be its $(k , \ell)$-entry.  Then

\[
C_{\beta}^{(d)} (k, \ell) \, = \,
\left\{
\begin{array}{l l l l}
1 - \displaystyle{\frac {16d+32\beta-16+32k}
 {(d+2\beta+1)(d+2\beta+3)(d+2\beta+5)}} , &&\mbox{if} \ 1\le k \, = \, \ell \le d+1 \\
\\
1 - \dfrac {80d+160\beta+80-32k}
{(d+2\beta+1)(d+2\beta+3)(d+2\beta+5)} , &&\mbox{if} \ d+2\le k \, = \, \ell \le 2d+2 \\
\\
-\dfrac {32(d+2\beta-k+2)}
{(d+2\beta+1)(d+2\beta+3)(d+2\beta+5)}, && \mbox{if} \ 2\le k \le d+1, \ell=d+k  \\
\\
-\dfrac {32(k-d-2\beta-1)} {(d+2\beta+1)(d+2\beta+3)(d+2\beta+5)},
 && \mbox{if} \ d+2\le k \le 2d+1, \ell=k-d  \\
\\
0,  && \mbox{all other cases}
\end{array}
\right.
\]
Note that $ d + 2\Re (\beta) > -1 $.
\setcounter{section}{6}
\setcounter{Lem}{0}
\begin{Lem}
Assume $\Re ( \beta ) \geq 0$. There exists an integer $d_\ast > 0$,
independent of $\beta$, such that for all $d \geq d_\ast$, the matrix
$C^{(d)}_\beta$ is invertible.
\end{Lem}
\noindent
{\bf Proof}. \
As $d$ tends to infinity, $C^{(d)}_\beta$ tends
to the identity matrix if $\Re ( \beta ) \geq 0$.  A simple estimate on
the diagonal and off-diagonal entries shows that for all $\beta$ such
that $\Re ( \beta ) \geq 0$ and sufficiently large $d$, the matrix
$C^{(d)}_\beta$ becomes diagonally dominant.  This implies the existence
of the needed $d_\ast$ and its independence of $\beta$.

A similar statement holds if we assume that $\Re ( \beta ) \geq - A$ for
any given $A \leq 0$.  We formulate it as the following lemma.

\noindent
{\bf Lemma 6.1$^\prime$}. \ {\it For any $A \geq 0$, there exists an
integer $d_\ast ( A ) > 0$ which depends only on $A$, such that for all
$d \geq d_\ast ( A )$ and all $\beta$ with $\Re ( \beta ) \geq - A$, the
matrix $C^{(d)}_\beta$ is invertible.}

By Lemma 6.1, to find all eigen-values of $\cA$ it amounts to solve the
equation $\det ( C^{(d)}_\beta ) = 0$.  The eigenvalue $\alpha$ is then
found from the relation $\beta = \alpha + \frac {m_3}2 $. Let
\[
a_1= \left.\left( 1 - \frac {16} {(d+2\beta+3)(d+2\beta+5)} \right)
 \right /
 \left(  \frac {32} { (d+2\beta+1)(d+2\beta+3) (d+2\beta+5) } \right)
\]
Then $a_1$ is the eigen-value of the matrix
 $\tilde{C}^{(d)} \, \in \, \bR^{2 ( d + 1 ) \times 2 ( d + 1)}$ given by:

\[
\tilde{C}^{(d)} (k, \ell) \, = \,
\left\{
\begin{array}{l l l l}
k-1, &&\mbox{if} \ 1\le k \, = \, \ell \le d+1 \\
\\
2d+2-k , &&\mbox{if} \ d+2\le k \, = \, \ell \le 2d+2 \\
\\
d+2-k, && \mbox{if} \ 2\le k \le d+1, \ell=d+k  \\
\\
k-d-1
 && \mbox{if} \ d+2\le k \le 2d+1, \ell=k-d  \\
\\
0,  && \mbox{all other cases}
\end{array}
\right.
\]
It is not difficult to find that the eigen-values of $ {\tilde C}^{(d)}$
are $0$ and $d+1$ with algebraic multiplicity $d+2$ and $d$
respectively. Solve the equations $a_1\, =\, 0$ or $ a_1\, = \, d+1$
and use the condition $d+2\Re(\beta)>-1$. The possible values of $\beta$
are then given by
\[
\beta = \frac {3-d} 2 \quad \mbox{or} \quad
 \frac { \sqrt{17}-4-d }2, \quad d=\, 1,\, 2,\, 3,\, \cdots
\]
This fact immediately gives the following lemma.

\begin{Lem}
Let $( \tilde{C}^{(d)}_{\beta} )^{-1}$ be the inverse matrix of
$\tilde{C}^{(d)}_{\beta}$ for $d \geq d^{*}(\beta)$, where
$ d^{*}(\beta) = 3-2\beta $ or $\sqrt{17}-4-2\beta$ is an integer.
Then there exists an absolute constant $C_2 > 0$ such
that for all $d \geq d^{*}(\beta)$
\[
\parallel ( \tilde{C}^{(d)}_{\beta} )^{-1} \parallel \, \leq \, C_2 \, .
\]
\end{Lem}

We now state our theorem about the properties of the solutions to the
recurrent relation (34).
\begin{Thm}
The only possible values of $\ \beta$ for which (34) have nonzero solutions
$f^{(j)}_{\beta} (m_1,m_2)$ is given by:
\[
\beta = \frac {3-m} 2 \quad \mbox{or} \quad
 \frac { \sqrt{17}-4-m }2, \quad m=\, 1,\, 2,\, 3,\, \cdots
\]
The corresponding solutions $f^{(j)}_{\beta} (m_1,m_2)$ have the following
property:
\begin{enumerate}
\item[a)] $\beta= (\sqrt{17}-4-m)/ 2$. In this case $f^{(j)}_{\beta}(m_1,m_2) = 0$ for
any $0\le m_1+m_2 < m$. For $d=m$, we have
\[
f^{(1)}_{\beta}(r,d-r) =  - (d-r+1) f^{(2)}_{\beta}(r-1,d-r+1), \quad r=1,2,\cdots, d
\]
$f^{(1)}_{\beta}(0,d)$, $ f^{(2)}_{\beta}(d,0)$ are free parameters.
$f^{(j)}_{\beta}(m_1,m_2)$ for $m_1+m_2\ge m+1$ are uniquely determined if the values of
the $m+2$ free parameters $ f^{(1)}_{\beta}(r,m-r), r=0,1,\cdots, m$,
and $f^{(2)}_{\beta}(m,0)$ are specified.

\item[b)] $\beta= (3-m)/ 2$. In this case $f^{(j)}_{\beta}(m_1,m_2) = 0$ for
any $0\le m_1+m_2 < m$. For $d=m$, we have
$f^{(1)}_{\beta}(0,d) = f^{(2)}_{\beta}(d,0) = 0$, and
\[
f^{(1)}_{\beta}(r,d-r) = f^{(2)}_{\beta}(r-1,d-r+1), \quad r=1,\cdots, d
\]
are free parameters.
$f^{(j)}_{\beta}(m_1,m_2)$ for $m_1+m_2\ge m+1$ are uniquely determined if the values of
the $m$ free parameters $ f^{(1)}_{\beta}(r,m-r), r=1,\cdots, m$ are specified.

\end{enumerate}
In both case a) and b), the solutions $f^{(j)}_{\beta}(m_1,m_2)$ is zero for
$m_1+m_2=m+1,m+3,\ldots$.  Since $f^{(j)}_{\beta}$ depends linearly on the free
parameters, we have for some $C_3 > 0$, $0 < \rho < \frac 1 {4000}$
\[
\abs{ f^{(j)}_{\beta} (m_1, m_2) }   \le C_3 \,
 \frac { \rho^{m_1+m_2+2\beta} }       {\Gamma \left( \frac {m_1+m_2+2\beta+3} 2\right)}
 \, ,
 \quad
 \forall \, m_1 \ge 0 , \, m_2 \, \ge \, 0, \; j=1,2 .
\]
\end{Thm}
\medskip
\noindent

{\bf Proof}. \ Property a) and b) are straightforward computations. From recurrent
relation (34), by parity it is obvious that $f^{(j)}_{\beta}(m_1,m_2)=0$ for $m_1+m_2=m+1$.
An easy induction shows that $f^{(j)}_{\beta}(m_1,m_2)=0$ for $m_1+m_2=m+3,m+5,\ldots$
We now prove the decay estimate. The strategy of the proof is the same as in theorem
4.3. From the proof of theorem 4.3, it is clear that by choosing the parameters
$(x^{(1)}, x^{(2)}, x^{(3)})$ sufficiently small, we have
\[
\abs{ h^{(j)}_{m_1, m_2} } \le
 \frac { \rho^{m_1+m_2+2} }       {\Gamma \left( \frac {m_1+m_2+7} 2\right)}
 \quad
 \forall \, m_1 \ge 0 , \, m_2 \, \ge \, 0, m_1+m_2\ge 3,\, j=1,2 .
\]
Our induction hypothesis for $f^{(j)}_{\beta}(m_1,m_2)$ is
\[
\abs{ f^{(j)}_{\beta} (m_1, m_2) }   \le \,
 \frac { \rho^{m_1+m_2+2\beta} }       {\Gamma \left( \frac {m_1+m_2+2\beta+3} 2\right)}
 \quad
 \forall \,  m \le m_1+ m_2 < d, \; j=1,2 .
\]
where $d\ge L$ and $d-m$ is an even number (note that $f^{(j)}_{\beta}(m_1,m_2) = 0 $ for
$m_1+m_2=m+1, m+3,\ldots$).  We assume that $L$ is a sufficiently large number and will
verify the induction assumption for $m\le d \le L$ later. Now for $m_1+m_2 = d$,
by lemma 6.2, we have
\begin{align*}
\abs{ f^{(j)}_{\beta}(m_1,m_2) }
 \le &
C_2 \cdot \sum_{m'=2}^{d-m} (m'+1) \cdot \abs{ J^{(1)}_{m',m^{\prime\prime}+2\beta} }
\cdot 2 (m^\prime +1) \cdot
\dfrac { \rho^{m^\prime+3} } {\Gamma\left( \frac {m^\prime+8} 2\right) }
\cdot
\dfrac { \rho^{m^{\prime\prime}+2\beta} }
   {\Gamma\left( \frac {m^{\prime\prime}+2\beta+3} 2\right) }
\\
+ &
C_2 \cdot \sum_{m'=4}^{d-m} (m'+1) \cdot \abs{ J^{(1)}_{m',m^{\prime\prime}+2\beta} }
\cdot 2 \cdot
\dfrac { \rho^{m^\prime+1} } {\Gamma\left( \frac {m^\prime+6} 2\right) }
\cdot
\dfrac { \rho^{m^{\prime\prime}+2\beta} }
   {\Gamma\left( \frac {m^{\prime\prime}+2\beta+3} 2\right) }
\\ + &
C_2 \cdot \abs{ J^{(1)}_{2,d-2+2\beta}} \cdot 4\cdot
\dfrac { \rho^{{d-2+2\beta} }}
   {\Gamma\left( \frac {d+2\beta+1} 2\right) }
\\
+ &
C_2 \cdot \sum_{m'=3}^{d-m+1} (m'+1) \cdot \abs{ J^{(2)}_{m',m^{\prime\prime}+2\beta} }
\cdot 2 \cdot
\dfrac { \rho^{m^\prime+2} } {\Gamma\left( \frac {m^\prime+7} 2\right) }
\cdot (m^{\prime\prime}+1) \cdot
\dfrac { \rho^{m^{\prime\prime}+2\beta+1} }
   {\Gamma\left( \frac {m^{\prime\prime}+2\beta+4} 2\right) }
\\
+ &
C_2 \cdot \sum_{m'=3}^{d-m-1} (m'+1) \cdot \abs{ J^{(2)}_{m',m^{\prime\prime}+2\beta} }
\cdot 2 \cdot
\dfrac { \rho^{m^\prime+2} } {\Gamma\left( \frac {m^\prime+7} 2\right) }
 \cdot
\dfrac { \rho^{m^{\prime\prime}+2\beta-1} }
   {\Gamma\left( \frac {m^{\prime\prime}+2\beta+2} 2\right) }
\\ + &
C_2 \cdot \abs{ J^{(2)}_{1,d-1+2\beta}} \cdot 4\cdot
\dfrac { \rho^{{d-1+2\beta} }}
   {\Gamma\left( \frac {d+2\beta+2} 2\right) }
\\
+ &
C_2 \cdot \sum_{m'=m-1}^{d-3} (m^{\prime\prime}+1) \cdot \abs{ J^{(1)}_{m'+2\beta,m^{\prime\prime}} }
\cdot 2 \cdot (m'+1) \cdot
\dfrac { \rho^{m^\prime+2\beta+1} } {\Gamma\left( \frac {m^\prime+2\beta+4} 2\right) }
\cdot
\dfrac { \rho^{m^{\prime\prime}+2} }
   {\Gamma\left( \frac {m^{\prime\prime}+7} 2\right) }
\\
+ &
C_2 \cdot \sum_{m'=m+1}^{d-3} (m^{\prime\prime}+1) \cdot \abs{ J^{(1)}_{m'+2\beta,m^{\prime\prime}} }
\cdot 2 \cdot
\dfrac { \rho^{m^\prime+2\beta-1} } {\Gamma\left( \frac {m^\prime+2\beta+3} 2\right) }
\cdot
\dfrac { \rho^{m^{\prime\prime}+2} }
   {\Gamma\left( \frac {m^{\prime\prime}+7} 2\right) }
\\ + &
C_2 \cdot \abs{ J^{(1)}_{d-1+2\beta,1}} \cdot 4\cdot
\dfrac { \rho^{{d-2+2\beta} } }
   {\Gamma\left( \frac {d+2\beta+1} 2\right) } \\
+ &
C_2 \cdot \sum_{m'=m}^{d-2} (m^{\prime\prime}+1) \cdot \abs{ J^{(2)}_{m'+2\beta,m^{\prime\prime}} }
\cdot 2 \cdot
\dfrac { \rho^{m^\prime+2\beta} } {\Gamma\left( \frac {m^\prime+2\beta+3} 2\right) }
\cdot
\dfrac { \rho^{m^{\prime\prime}+3} }
   {\Gamma\left( \frac {m^{\prime\prime}+8} 2\right) } \cdot (m^{\prime\prime}+1)
\\
+ &
C_2 \cdot \sum_{m'=m}^{d-4} (m^{\prime\prime}+1) \cdot \abs{ J^{(2)}_{m'+2\beta,m^{\prime\prime}} }
\cdot 2 \cdot
\dfrac { \rho^{m^\prime+2\beta} } {\Gamma\left( \frac {m^\prime+2\beta+3} 2\right) }
\cdot
\dfrac { \rho^{m^{\prime\prime}+1} }
   {\Gamma\left( \frac {m^{\prime\prime}+6} 2\right) }
\\ + &
C_2 \cdot \abs{ J^{(2)}_{d-2+2\beta,2}} \cdot 4\cdot
\dfrac { \rho^{{d-2+2\beta} }}
   {\Gamma\left( \frac {d+2\beta+1} 2\right) }
 \end{align*}
\begin{align*}
\\
\le & C_2 \cdot
  \dfrac { \rho^{d+2\beta} } {\Gamma\left( \frac{d+2\beta+3}2\right)}
\cdot
  \left(
     \sum_{m^\prime=2}^{d-m} \frac {8\rho^3} {d+2\beta+5}
     +\sum_{m^\prime=4}^{d-m} \frac {8\rho} {d+2\beta+5}
     + \frac 8 {\rho^2\cdot (d+2\beta+5)}
 \right.
\\
&+ \sum_{m^\prime=3}^{d-m+1} \frac {8\rho^3} {d+2\beta+5}
   \cdot \frac {2(m^{\prime\prime}+1)} {d+2\beta+3} +
   \sum_{m^\prime=3}^{d-m-1} \frac{8\rho}{d+2\beta+5}
   +\frac 8 {\rho(d+2\beta+5)}
\\
&+ \sum_{m^\prime=m-1}^{d-3} \frac {8\rho^3} {d+2\beta+5}
   \cdot \frac {2(m^{\prime}+1)} {d+2\beta+3} +
+ \sum_{m^\prime=m+1}^{d-3} \frac{8\rho}{d+2\beta+5}
   +\frac {16} {\rho^2(d+2\beta+5)}
\\
& + \left. \sum_{m^\prime=m}^{d-2}
\frac {16\rho^3}{d+2\beta+5}
+\sum_{m^\prime=m}^{d-4} \frac {8\rho}{d+2\beta+5}
+ \frac {16} {\rho^2 (d+2\beta+5)} \right)
\\
& \le C_2 \cdot \frac {\rho^{d+2\beta}}
{\Gamma{ \left( \frac {d+2\beta+3} 2 \right) } }
\cdot 2000\rho
\, \le\, \frac {\rho^{d+2\beta}}
{\Gamma{ \left( \frac {d+2\beta+3} 2 \right) } }
\end{align*}
where in the second last inequality above we have used the fact that $d\ge L$ and
$L$ is sufficiently large such that $d/(d+2\beta+3)\le 2$. It is clear that
it suffices for us to take $L=2m$. To check the inductive assumption for
$m\le d\le 2m$, we recall that $f^{(j)}_{\beta}(m_1,m_2)$ depends linearly
on several free parameters. If we let them be sufficiently small, then it
is clear that the inductive assumption is satisfied for $m\le d\le 2m$.
Our theorem is proved.

We now formulate our main theorem about the spectrum of the linearized
operator.

\begin{Thm}
The spectrum of the operator $\cA$ consists of the following eigen-values

\[
{\sf spec} \, ( \cA) \, = \, \left\{
 1 , \frac{1}{2} ,
0 , \lambda^{(1)}_m , \lambda^{(2)}_m ,
m \geq 1 \right\} \, .
\]
where $\lambda^{(1)}_m \, = \, - \frac{m}{2}$,
$ \lambda^{(2)}_m  = \frac {\sqrt{17}-4-m} 2, \, m\ge 1$.

The first eigen-values have multiplicities $\nu_1 = 1$,
$\nu_{\frac{1}{2}} = 3$,
$\nu_0 = 6$. The eigen-values $\lambda^{(1)}_m$, $\lambda^{(2)}_m$ correspond
to the stable part of the spectrum and also have finite multiplicities given by:
$\nu_{\lambda^{(1)}_m} = \frac {(m+3)(m+4)} 2, \qquad
  \nu_{\lambda^{(2)}_m} = \frac {m(m+5)} 2$.

For each $\alpha\in {\sf spec}\,(\cA)\,$, the eigenfunctions
$f^{(j)}_{\alpha} (m_1,m_2,m_3)$ have the following property:
\begin{enumerate}
\item[a)] $f^{(j)}_{\alpha} (m_1,m_2,m_3)$ is compactly supported in the $m_3$
variable, i.e., there exists an integer $m_3^*=m_3^*(\alpha)$ such that
\[
f^{(j)}_{\alpha}(m_1,m_2,m_3) = 0 \quad\text{\ if\, } m_3>m_3^*
\]
\item[b)] $f^{(j)}_{\alpha}(m_1,m_2,m_3)$ decays faster than exponentially, more
precisely, there exist constants $C_3=C_3(\alpha)>0$ and $0<\rho<\frac {1}{4000}$,
such that
\[
\abs{ f^{(j)}_{\alpha}(m_1,m_2,m_3) } \le C_3
\frac {\rho^{m_1+m_2+m_3+2\alpha}}
{\Gamma \left( \frac {m_1+m_2+m_3+2\alpha+3} 2\right)},
\quad\forall\ m_1,m_2,m_3\ge 0
\]
\end{enumerate}

The system of eigenfunctions is complete in the following sense.  Let
$\Gamma^{(s)}$ be the stable linear subspace of $\Delta$ generated by all
eigenfunctions with $\Re ( \lambda ) < 0$, $\Gamma^{(u)}$ be the
unstable subspace generated by all eigenfunctions with eigenvalues
$\lambda > 0$,  and $\Gamma^{(n)}$ be the
neutral subspace generated by all eigenfunctions with eigenvalue
$\lambda = 0$.  Then $\dim \Gamma^{(u)} = 4$, $\dim \Gamma^{(n)} = 6$ and

\[
\Delta\, = \, \Gamma^{(u)} \, + \, \Gamma^{(n)} \, + \, \Gamma^{(s)} \,
.
\]
\end{Thm}

\medskip
\noindent

{\bf Proof}. \ By Lemma 6.1, we only need to examine $\beta$ for which
$\det ( C^{(d)}_\beta ) = 0$.  From previous arguments, we have that for
$d \geq 1$, $\beta = - \frac{d-3}{2}$ or $\frac{\sqrt{17} -4 -d }  2$.
\noindent
We discuss the spectrum separately in the following three cases.

\begin{namelist}{xxxxxxxx}
\item[\underline{1$^\circ$ unstable spectrum:}] \hspace{.9em} $\alpha \, =
\, 1 , 1/2 $.
\end{namelist}
\begin{enumerate}
\item[a)] $\alpha \, = 1$.  Since $\beta = \alpha +
\frac{m_3}{2}$, the only possibility is that $\beta = 1$,
$d=1$ and $m_3 = 0$.  The eigenspace is one-dimensional with
$f^{(1)}_{000}=f^{(2)}_{000} = f^{(1)}_{010} = f^{(2)}_{100}=0$,
$f^{(1)}_{100}=f^{(2)}_{010}$ is a free parameter and the remaining part of all
higher degree terms ( $f^{(j)}_{m_1,m_2,0}$ with $m_1+m_2\ge 2$ )
is uniquely determined once we specify $f^{(1)}_{100}$.
\item[b)] $\alpha = 1/2$.  Possible cases are $m_3 = 0$, $\beta = 1/2$,
$d=0,\, 2$ or $m_3 = 1$, $\beta = 1$, $d=1$. In the first case we have
$f^{(j)}_{m_1,m_2,0} = 0$ for $m_1+m_2 \le 1$,
$f^{(1)}_{110}=f^{(2)}_{020}$, $f^{(1)}_{200} = f^{(2)}_{110}$ are two free parameters,
all other terms of higher degee ( $f^{(j)}_{m_1,m_2,0}$ with $m_1+m_2 \ge 3$ ) are uniquely
determined once we specify the above four parameters. In the second case
we have  $f^{(1)}_{001}=f^{(2)}_{001} = f^{(1)}_{011} = f^{(2)}_{101}=0$,
$f^{(1)}_{101}=f^{(2)}_{011}$ is a free parameter and the remaining part of all
higher degree terms ( $f^{(j)}_{m_1,m_2,1}$ with $m_1+m_2\ge 2$ )
is uniquely determined once we specify $f^{(1)}_{101}$. Putting two cases
together, we see that the dimension of the eigenspace is $3$.

\end{enumerate}
This gives $\dim \Gamma^{(u)} = 4$.
\begin{namelist}{xxxxxxxx}
\item[\underline{2$^\circ$ neutral spectrum:}] \hspace{.9em} Here we
have $\alpha = 0$, and three cases.
\end{namelist}

\begin{enumerate}
\item[a)] $m_3= 2$. Then $\beta=1$. The eigenspace is one-dimensional with
$f^{(1)}_{002}=f^{(2)}_{002} = f^{(1)}_{012} = f^{(2)}_{102}=0$,
$f^{(1)}_{102}=f^{(2)}_{012}$ is a free parameter and the remaining part of all
higher degree terms ( $f^{(j)}_{m_1,m_2,2}$ with $m_1+m_2\ge 2$ )
is uniquely determined once we specify $f^{(1)}_{102}$. This eigenvector is connected with
$\frac {\partial} {\partial \sigma^{(2)}}$ which corresponds to the variation
 of the parameter $\sigma^{(2)}$
of the fixed point.

\item[b)] $m_3= 1$.  Then $\beta = 1/2$. We have
$f^{(j)}_{m_1,m_2,1} = 0$ for $m_1+m_2 \le 1$,
$f^{(1)}_{111}=f^{(2)}_{021}$, $f^{(1)}_{201} = f^{(2)}_{111}$ are two free parameters,
all other terms of higher degee ( $f^{(j)}_{m_1,m_2,1}$ with $m_1+m_2 \ge 3)$ are uniquely
determined once we specify the above two parameters. Clearly the eigenspace
is two-dimensional. This eigenspace does not correspond to any change of parameters
of the fixed point.

\item[c)] $m_3 = 0$. Then $\beta = 0 $.  We have $f^{(j)}_{m_1,m_2,0} = 0 $ for $m_1+m_2\le 2$,
$f^{(1)}_{030}=f^{(2)}_{300}=0$,
$f^{(1)}_{120}=f^{(2)}_{030}$, $f^{(1)}_{210}=f^{(2)}_{120}$, $f^{(1)}_{300}=f^{(2)}_{210}$ are three
free parameters. All other terms of higher degee ( $f^{(j)}_{m_1,m_2,0}$ with $m_1+m_2 \ge 4$ ) are uniquely
determined once we specify the above three parameters. This eigenspace corresponds to
($\frac {\partial} {\partial x^{(1)}}$, $\frac {\partial} {\partial x^{(2)}}$,  $\frac {\partial} {\partial x^{(3)}}$).
\end{enumerate}
Putting all three cases together, we see that $\dim \Gamma^{(n)} = 6$.
 \begin{namelist}{xxxxxxxx}
 \item[\underline{3$^\circ$ stable spectrum:}] \hspace{.9em} $\Re ( \alpha ) < 0$.
 \end{namelist}
 There are two cases.

 \begin{namelist}{xxxxxx}
 \item[\underline{\sf Case 1}:]  $\alpha = - \frac{m}{2} ,
 \, m \geq 1$. Recall that $\beta = \alpha + \frac{m_3}{2}$, and $m_3$
 satisfies $0\le m_3 \le m+2$. By theorem 6.3, for each such $\beta$,
 the number of free parameters is $3-2\beta $. Then the total multiplicity
 $\nu_{\alpha}$ is given by
 \[
 \nu_\alpha \, = \,
 \displaystyle{\sum\limits_{m_3 = 0}^{m +2}} \,
 3-(-m+m_3) = \frac {(m+3)(m+4)} 2
 \]

 \item[\underline{\sf Case 2}:] $\alpha =\frac {\sqrt{17}-4-m} 2, \, m\ge 1$.
 $\beta = \alpha + \frac{m_3}{2}$, and $m_3$
 satisfies $0\le m_3 \le m-1$.
 By theorem 6.3, we have
\[
 \nu_\alpha \, = \,
 \displaystyle{\sum\limits_{m_3 = 0}^{m-1}} (m-m_3 +2)
  = \frac {m(m+5)} 2
 \]
 \end{namelist}
It follows easily  that the eigenfunctions
$f^{(j)}_{\alpha}(m_1,m_2,m_3)$ is compactly supported in the $m_3$
variable. By theorem 6.3, the decay estimate on $f^{(j)}_{\alpha}(m_1,m_2,m_3)$
is obvious.
\medskip

It turns out that the eigenvector corresponding to
$\frac {\partial} {\partial \sigma^{(1)}}$ is in the eigenspace spanned by
the eigenvectors ($\frac {\partial} {\partial x^{(1)}}$, $\frac {\partial} {\partial x^{(2)}}$,
 $\frac {\partial} {\partial x^{(3)}}$).  More precisely we have the following:
\begin{Lem}
Let $t_1 = x^{(1)} - 1$, $t_2= x^{(2)}$, $t_3=x^{(3)}-1$.  Then

\begin{equation}
\tilde{G}^{(\sigma^{(1)}, t_1 , t_2, t_3, \sigma^{(2)})} ( Y ) \, = \, G^{(\sigma^{(1)},
x^{(1)}, x^{(2)}, x^{(3)}, \sigma^{(2)} )} ( Y )
\end{equation}
where $G^{(\sigma^{(1)}, x^{(1)}, x^{(2)}, x^{(3)}, \sigma^{(2)} )}$ is defined in (30).  The function
$\tilde{G}$ satisfies the following scaling relation:

\begin{equation}
\tilde{G}^{(\sigma^{(1)} , t_1 , t_2,t_3, \sigma^{(2)})} ( Y ) \, = \,
\tilde{G}^{(1 , \, \sigma^{(1)} t_1, \, \sigma^{(1)} t_2 ,\, \sigma^{(1)} t_3,\, \sigma^{(2)})} ( Y )
\end{equation}
\end{Lem}

\medskip
\noindent
{\bf Proof}. \ Let $f^{(j), 0}_{m_1, m_2, 0}$ correspond to the
eigenvector $\frac{\partial}{\partial \sigma^{(1)}}$, then
a simple calculation shows that
\[
f^{(j), 0}_{m_1, m_2, 0} \, = \, ( m_1+ m_2 - 1) \, h^{(j)}_{m_1 m_2} \, +
\, h^{(j)}_{m_1- 2 , m_2} \, + \, h^{(j)}_{m_1, m_2 - 2} \, .
\]

If $f^{(j), 1}_{m_1, m_2,0 }$,  $f^{(j),2}_{m_1, m_2, 0}$ and
$f^{(j),3}_{m_1,m_2,0}$ correspond to the eigenvectors
$\frac{\partial}{\partial x^{(1)}}$,
$\frac{\partial}{\partial x^{(1)}}$,
and $\frac{\partial}{\partial x^{(3)}}$
respectively, then clearly we have
\[
f^{(j),0}_{m_1,m_2, 0} \, = \, \left( x^{(1)} - 1 \right) \,
f^{(j),1}_{m_1, m_2, 0} \,
+\,
 x^{(2)}  f^{(j),2 }_{m_1, m_2, 0} \,
+ \, \left( x^{(3)} -  1\right) \,
f^{(j),3}_{m_1, m_2, 0} \,
\]
This immediately gives

\[
\left[
\sigma^{(1)} \, \frac{\partial}{\partial \sigma^{(1)}} \, -
\,
\left( x^{(1)} - 1 \right) \,
\frac{\partial}{\partial x^{(1)}} \, -
 x^{(2)}   \,
\frac{\partial}{\partial x^{(2)}} \, -
\left( x^{(3)} - 1 \right) \,
\frac{\partial}{\partial x^{(3)}} \,
\right]
\, G^{(\sigma^{(1)} , x^{(1)} , x^{(2)} , x^{(3)},  \sigma^{(2)})} \, ( Y ) \, = \, 0 \, .
\]
Regarding this as a transport equation in the variables $( \sigma^{(1)}, t_1 ,
t_2, t_3)$, we can easily find that $\tilde{G}$ satisfies the scaling (35).
Lemma is proved.

This lemma actually shows in what sense the parameters  $\sigma^{(1)} , x^{(1)}
, x^{(2)} , x^{(3)}$ are dependent.

\medskip
As was shown in \S4, we have the five-parameter family of fixed points
$G^{(\sigma^{(1)}, x^{(1)}, x^{(2)}, x^{(3)}, \sigma^{(2)} )}$.
We use the notation
$\pi = (\sigma^{(1)}, x^{(1)}, x^{(2)}, x^{(3)}, \sigma^{(2)} )$
and write $G^{(\pi)}$ instead of
$G^{(\sigma^{(1)}, x^{(1)}, x^{(2)}, x^{(3)}, \sigma^{(2)} )}$.  The
spectrum of the linearization of the equation
for the fixed point does not depend on $\pi$ (see \S5) and has $\ell^{(u)}=4$
unstable eigenvectors $\Phi^{(u)}_j ( Y_1 , Y_2 , Y_3 )$, $1 \leq j \leq
\ell^{(u)} =4$ and $\ell^{(n)}=6 $ neutral eigenvectors $\Phi^{(n)}_{j^\prime} ( Y_1
, Y_2 , Y_3 )$, $1 \leq j^\prime \leq \ell^{(n)}=6$.

\begin{center}
{\large \S7. \ The Choice of Initial Conditions and the
Initial Part of the Inductive Procedure}
\end{center}

\setcounter{section}{7}

The equation (21) for the fixed point which was derived in \S 3 is
non-typical from the point of view of the renormalization group
theory because it contains the integration over $\gamma$,
$0\le\gamma\le 1$. On the other hand, since we consider the Cauchy
problem for (1) we are given only the initial condition $v(k,0)$ which
produces through the recurrent relations $(4)$, $(5)$, $(6)$ or
$(4')$, $(5')$, $(6')$ the whole set of functions $h_r(k,s)$ or
$g_r(\tilde{k},s)$. For large $p$ and $r\le p$ they can be
considered as depending on $\gamma=\dfrac rp$ and our procedure is
organized in such a way that for $\gamma$ which are away from zero
$\tilde{g}_r$ are close to their limits. Therefore the initial part of
our process should be discussed in more detail. This is done in this
section.

We take ${k}^{(0)}$ which will be assumed to be sufficiently large,
introduce the neighborhood
\[
A_1\, =\, \{ k:\, \abs{ k- \kappa^{(0)} } \le D_1
     \sqrt { k^{(0)} ln k^{(0)} } \}
\]
where $\kappa^{(0)}= ( 0, 0, k^{(0)})$ and $D_1$ is also sufficiently
large. Our initial conditions will be zero outside $A_1$. Inside
$A_1$ they have the form
\begin{align*}
v(k,0) = \frac 1 {2\pi} \exp\left\{ - \frac {Y_1^2+Y_2^2} 2 \right\}
\left( H^{(0)}(Y_1,Y_2) + \sum_{j=1}^4 b_j^{(u)} \Phi_j^{(u)}(Y_1,Y_2,Y_3)+ \right. \\
\left.
\sum_{j'=1}^6 b_{j'}^{(n)} \Phi_{j'}^{(n)} (Y_1,Y_2,Y_3)
+\Phi( Y_1, Y_2, Y_3 ; b^{(u)}, b^{(n)} ) \right)
\frac 1{\sqrt{2\pi}} \exp \left\{ - \frac {Y_3^2} 2 \right\}
\end{align*}
In this expression $k = k^{(0)} + \sqrt{ k^{(0)}} Y$,
$H^{(0)}(Y_1,Y_2) = ( H^{(0)}_1(Y_1,Y_2), H^{(0)}_2(Y_1,Y_2), 0 ) $ is the
fixed point of our renormalization group (see \S 4) corresponding to
the parameters $\sigma_1^{(1)}=\sigma_1^{(2)}=1$, $x_1=x_2=x_3=0$.
Also $\Phi_j^{(u)}$, $\Phi_{j'}^{(n)}$
are unstable and neutral eigen-functions for $H^{(0)}$ described in $\S 6$,
$b_j^{(u)}$ and $b_{j'}^{(n)}$ are our main parameters,
$-\rho_1 \le b_j^{(u)},\, b_{j'}^{(n)} \le \rho_1$ where $\rho_1$ is another constant
which depends on $k^{(0)}$. Its value will also be specified later. Each function
$\Phi(Y_1, Y_2, Y_3; b^{(u)}, b^{(n)} )$, $b^{(u)} =\{ b_j^{(u)} \}$,
$b^{(n)} = \{ b_{j'}^{(n)} \}$ is small in the sense that they satisfy
\[
\sup_{Y,b} \abs{ \Phi(Y_1,Y_2,Y_3; b^{(u)}, b^{(n)}) } \le D_2,
\]
\[
\sup \norm{ \Phi(Y_1,Y_2,Y_3; \bar{b}^{(u)}, \bar{b}^{(n)} )
  -\Phi(Y_1,Y_2,Y_3; b^{(u)}, b^{(n)} ) }
\le D_2 ( \abs{ \bar{b}^{(u)} - b^{(u)} }
 + \abs{ \bar{b}^{(n)} -b^{(n)} } ).
\]
Due to the presence of $b^{(u)}$, $b^{(n)}$, we have $l=l^{(u)} + l^{(n)}=10$-parameter
families of initial conditions, due to the presence of $\Phi$  we have an open
set in the space of such families.



Let
\[A_r=\{k:\:|k-r\kappa^{(0)}|\le D_1\sqrt{r k^{(0)}\ln
rk^{(0)}}\}\] and the variable $Y$ be such that
$k=r\kappa^{(0)}+\sqrt{rk^{(0)}} Y$. Assume that for $r<p$,
$|Y|\le D_1\sqrt{\ln rk^{(0)}}$
\[h_r(r\kappa^{(0)}+\sqrt{r k^{(0)}}Y,s)=Z_p(s) \Lambda^{r-1}_p(s)r\tilde{g}_r(Y,s)\] and

\begin{equation*}\begin{split}\tilde{g}_r(Y,s)=&\frac{1}{2\pi}\exp\left\{-\frac{Y_1^2+Y_2^2}{2}\right\}
\frac{1}{\sqrt{2\pi}}\exp\left\{-\frac{Y_3^2}{2}\right\}\cdot r
\\
&\begin{split}\cdot
\left(H_1^{(0)}(Y_1,Y_2)+\delta_1^{(r)}(Y_1,Y_2,Y_3),H_2^{(0)}(Y_1,Y_2)+\delta_2^{(r)}(Y_1,Y_2,Y_3),
\phantom{\frac11}\right.
\\
\left.\frac{1}{\sqrt{rk^{(0)}}}(F^{(r)}(Y_1,Y_2)+\delta_3^{(r)}(Y_1,Y_2,Y_3))\right)\end{split}
\end{split}
\end{equation*}
where in view of incompressibility
\begin{equation}
\label{eq: sec8_renormalised_incompressibility}
H_1^{(0)}Y_1+H_2^{(0)}Y_2+F^{(r)}=0
\end{equation}

We shall derive a system of recurrent relations for $Z_p(s)$ and $\Lambda_p(s)$ for $p<p_0$.
All $\delta_j^{(r)}$ will be considered as remainder terms.

Outside $A_r$ we assume that

\[|h_r(r\kappa^{(0)}+\sqrt{rk^{(0)}}Y,s)|\le\frac{1}{(rk^{(0)})^{\lambda_1}}\]
where $\lambda_1$ is another constant which depends on $C_1$.

Returning back to $(6)$ take the term with some $p_1$, $p_2$,
$p_1+p_2=p$ and introduce the new variable of integration $Y'$
where $k'=p_2\kappa^{(0)}+\sqrt{pk^{(0)}}Y'$. Introduce also the
variables $\theta_1$, $\theta_2$, $0\le\theta_1\le(p_1k^{(0)})^2$,
$0\le\theta_2\le(p_2k^{(0)})^2$ where
$s_1=s\left(1-\dfrac{\theta_1}{(p_1k^{(0)})^2}\right)$,
$s_2=s\left(1-\dfrac{\theta_2}{(p_2k^{(0)})^2}\right)$.

Then from (6)

\begin{equation}
\label{eq:sec8_recurrent relation_wide}
\begin{split}
    & h_p(p\kappa^{(0)}+\sqrt{pk^{(0)}}Y,s)=Z_{p+1}(s) \Lambda_{p+1}^p(s)p \tilde{g}_p(Y,s)\\
    &\begin{split}
        &\begin{split}
            &=(pk^{(0)})^{\frac
             52}i\int\limits_0^{((p-1)k^{(0)})^2}d\theta_2\int\limits_{\mathbb{R}^3}\exp\left\{-\theta_2|\kappa^{(0,0)}
             +\frac{\scriptstyle{Y'}}{\scriptstyle{\sqrt{(p-1)k^{(0)}}}}|^2\right\} \cdot \\
             &\qquad Z_{p}( s(1-\frac{\scriptstyle{\theta_2}}{\scriptstyle{((p-1)k^{(0)})^2}})\cdot
            \Lambda_{p}^{p-1}(s(1-\frac{\scriptstyle{\theta_2}}{\scriptstyle{((p-1)k^{(0)})^2}} ))
            \cdot (p-1) \cdot Z_{p}(s) \Lambda_{p}(s) \\
            &\left\langle\tilde{g}_1((Y-Y')\sqrt{pk^{(0)}},0),\kappa^{(0,0)}
             +\frac{\scriptstyle{Y}}{\scriptstyle{\sqrt{pk^{(0)}}}}\right\rangle
             P_{\kappa^{(0,0)}+\frac{\scriptstyle{Y}}{\scriptstyle{\sqrt{pk^{(0)}}}}}\tilde{g}_{p-1}
             \left(
             Y'\sqrt{\frac{\scriptstyle{p}}{\scriptstyle{p-1}}},
             s(1-\frac{\scriptstyle{\theta_2}}{\scriptstyle{((p-1)k^{(0)})^2}})\right)
             d^3Y'+
        \end{split}
        \\
        &\begin{split}
            &+ip\sum\limits_{p_1+p_2=p\atop{p_1,p_2>1}}\frac 1p \frac{(pk^{(0)})^{\frac 52} p_1
             p_2}{(p_1k^{(0)})^2(p_2k^{(0)})^2}\int\limits_{0}^{(p_1k^{(0)})^2}
             d\theta_1\int\limits_{0}^{(p_2k^{(0)})^2}
             d\theta_2\int\limits_{\mathbb{R}^3}\left(\frac{1}{2\pi}\right)^{\frac
             32}\exp\left\{-\frac{\scriptstyle{(Y_1-Y_1')^2+(Y_2-Y_2')^2+(Y_3-Y_3')^2}}
             {\scriptstyle{2\gamma}}\right\}
            \\
            &\left\langle\tilde{g}_{p_1}\left(\frac{\scriptstyle{Y-Y'}}{\scriptstyle{\sqrt{\gamma}}},
             s\left(1-\frac{\scriptstyle{\theta_1}}{\scriptstyle{(p_1k^{(0)})^2}}\right)\right),
             \kappa^{(0,0)}+\frac{\scriptstyle{Y}}
             {\scriptstyle{\sqrt{pk^{(0)}}}}\right\rangle
             P_{\kappa^{(0,0)}+\frac{\scriptstyle{Y}}{\scriptstyle{\sqrt{pk^{(0)}}}}}
             \tilde{g}_{p_2}\left(\frac{\scriptstyle{Y'}}{\scriptstyle{\sqrt{1-\gamma}}},
             s\left(1-\frac{\scriptstyle{\theta_2}}{\scriptstyle{(p_2k^{(0)})^2}}\right)\right)
        \end{split}
        \\
&\qquad Z_{p}( s(1-\frac{\scriptstyle{\theta_1}}{\scriptstyle{(p_1k^{(0)})^2}}))\cdot
            \Lambda_{p}^{p_1-1}(s(1-\frac{\scriptstyle{\theta_1}}{\scriptstyle{({p_1}k^{(0)})^2}} ))
            \cdot
 Z_{p}( s(1-\frac{\scriptstyle{\theta_2}}{\scriptstyle{(p_2k^{(0)})^2}}))\cdot
            \Lambda_{p}^{p_2-1}(s(1-\frac{\scriptstyle{\theta_2}}{\scriptstyle{({p_2}k^{(0)})^2}} ))
            \cdot \\
        &\left(\frac{1}{2\pi}\right)^{\frac
         32}\exp\left\{-\frac{\scriptstyle{(Y_1')^2+(Y_2')^2+(Y_3')^2}}{\scriptstyle{2(1-\gamma)}}\right\}
         \exp\left\{-\theta_1|\kappa^{(0,0)}
         +\frac{\scriptstyle{Y-Y'}}{\scriptstyle{\gamma\sqrt{pk^{(0)}}}}|^2\right\}
         \exp\left\{-\theta_2|\kappa^{(0,0)}
         +\frac{\scriptstyle{Y'}}{\scriptstyle{(1-\gamma)\sqrt{pk^{(0)}}}}|^2\right\}+
        \\
        &\begin{split}
            &+\frac{i(pk^{(0)})^{\frac
             52}(p-1)}{((p-1)k^{(0)})^2}\int\limits_0^{((p-1)k^{(0)})^2}d\theta_1
             \int\limits_{\mathbb{R}^3}\exp\left\{-\theta_1|\kappa^{(0,0)}
             +\frac{\scriptstyle{Y-Y'}}{\scriptstyle{\sqrt{(p-1)k^{(0)}}}}|^2\right\}
            \\
&\qquad Z_{p}( s(1-\frac{\scriptstyle{\theta_1}}{\scriptstyle{((p-1)k^{(0)})^2}})\cdot
            \Lambda_{p}^{p-1}(s(1-\frac{\scriptstyle{\theta_1}}{\scriptstyle{((p-1)k^{(0)})^2}} ))
            \cdot  Z_{p}(s) \Lambda_p(s) \\
            &\left\langle\tilde{g}_{p-1}((Y-Y')\sqrt{\frac{\scriptstyle{p}}
             {\scriptstyle{p-1}}},s(1-\frac{\scriptstyle{\theta_1}}{\scriptstyle{((p-1)k^{(0)})^2}}),\kappa^{(0,0)}
             +\frac{\scriptstyle{Y-Y'}}{\scriptstyle{\sqrt{pk^{(0)}}}}\right\rangle
             P_{\kappa^{(0,0)}+\frac{\scriptstyle{Y}}{\scriptstyle{\sqrt{pk^{(0)}}}}}\tilde{g}_1
             \left(Y'\sqrt{\scriptstyle{p}},
             s)\right)
             d^3Y'
        \end{split}
    \end{split}
\end{split}
\end{equation}

Here $\gamma=\dfrac{p_1}{p}$ and $\kappa^{(0,0)}=(0,0,1)$. Now we shall modify
\eqref{eq:sec8_recurrent relation_wide}
for $p_1>1$, $p_2>1$ similar to what we did in \S 3. Later
we discuss the terms with $p_1=1$ and $p_2=1$. The modification
consists of four steps.

\underline{\textbf{Step 1.}} All terms
$s\left(1-\frac{\theta_1}{(p_1k^{(0)})^2}\right)$,$s\left(1-\frac{\theta_2}{(p_2k^{(0)})^2}\right)$
are replaced by $s$.

\underline{\textbf{Step 2.}} Write
\[\frac{(pk^{(0)})^{\frac{5}{2}}p_1p_2}{(p_1k^{(0)})^2(p_2k^{(0)})^2}=\frac{(pk^{(0)})^
{\frac{1}{2}}}{(k^{(0)})^2\gamma(1-\gamma)}\]

\underline{\textbf{Step 3.}} Consider the inner product
\[(pk^{(0)})^{\frac{1}{2}}\left\langle
\tilde{g}_{p_1}\left(\frac{Y-Y'}{\sqrt{\gamma}},s\right),\kappa^{(0,0)}+\frac{Y}{\sqrt{pk^{(0)}}}\right\rangle
\] Up to
remainders and from
\eqref{eq: sec8_renormalised_incompressibility}
it equals to
\[
\left(\frac{1}{2\pi}\right)^{\frac{3}{2}}\exp\left\{-\frac{(Y_1-Y_1')^2+(Y_2-Y_2')^2+(Y_3-Y_3')^2}{2\gamma}\right\}
\]
\[
\left[H_{1}^{(0)}\left(\frac{Y_1-Y_1'}{\sqrt{\gamma}},\frac{Y_2-Y_2'}{\sqrt{\gamma}}\right)Y_1+
H_{2}^{(0)}\left(\frac{Y_1-Y_1'}{\sqrt{\gamma}},\frac{Y_2-Y_2'}{\sqrt{\gamma}}\right)Y_2+
\frac{1}{\sqrt{\gamma}}F^{(p_1)}\left(\frac{Y-Y'}{\sqrt{\gamma}},s\right)\right]=
\]
\[
=\left(\frac{1}{2\pi}\right)^{\frac{3}{2}}\exp\left\{-\frac{(Y_1-Y_1')^2+(Y_2-Y_2')^2+(Y_3-Y_3')^2}{2\gamma}\right\}\]
\[
\left[
H_{1}^{(0)}\left(\frac{Y_1-Y_1'}{\sqrt{\gamma}},\frac{Y_2-Y_2'}{\sqrt{\gamma}}\right)Y_1+
H_{2}^{(0)}\left(\frac{Y_1-Y_1'}{\sqrt{\gamma}},\frac{Y_2-Y_2'}{\sqrt{\gamma}}\right)Y_2-\right.
\]
\[
\left.
-H_{1}^{(0)}\left(\frac{Y_1-Y_1'}{\sqrt{\gamma}},\frac{Y_2-Y_2'}{\sqrt{\gamma}}\right)\frac{Y_1-Y_1'}{\gamma}-
H_{2}^{(0)}\left(\frac{Y_1-Y_1'}{\sqrt{\gamma}},\frac{Y_2-Y_2'}{\sqrt{\gamma}}\right)\frac{Y_2-Y_2'}{\gamma}\right]=
\]
\[
=\left(\frac{1}{2\pi}\right)^{\frac{3}{2}}\exp\left\{-\frac{(Y_1-Y_1')^2+(Y_2-Y_2')^2+(Y_3-Y_3')^2}{2\gamma}\right\}
\]
\[
\left\{-\frac{\gamma-1}{\sqrt{\gamma}}\left[
H_{1}^{(0)}\left(\frac{Y_1-Y_1'}{\sqrt{\gamma}},\frac{Y_2-Y_2'}{\sqrt{\gamma}}\right)\frac{Y_1-Y_1'}{\sqrt{\gamma}}+
H_{2}^{(0)}\left(\frac{Y_1-Y_1'}{\sqrt{\gamma}},\frac{Y_2-Y_2'}{\sqrt{\gamma}}\right)\frac{Y_2-Y_2'}{\sqrt{\gamma}}
\right]+\right.
\]
\[\left.
+\sqrt{1-\gamma}\left[
H_{1}^{(0)}\left(\frac{Y_1-Y_1'}{\sqrt{\gamma}},\frac{Y_2-Y_2'}{\sqrt{\gamma}}\right)\frac{Y_1'}{\sqrt{1-\gamma}}+
H_{2}^{(0)}\left(\frac{Y_1-Y_1'}{\sqrt{\gamma}},\frac{Y_2-Y_2'}{\sqrt{\gamma}}\right)\frac{Y_2'}{\sqrt{1-\gamma}}
\right]\right\}
\]
Let us stress again that $H_{j}^{(0)}(Y,s)$ depend only on
$Y_1$, $Y_2$ and $s$. With respect
to $Y_3$ we have the usual convolution.

\underline{\textbf{Step 4.}} Replace the projection operator by
the identity operator. It is not the reduction to the Burgers
system because the incompressibility condition is preserved.

Now we shall modify the first and the last terms in \eqref{eq:sec8_recurrent relation_wide}. For
the first one we can write
\begin{equation}
\label{eq: first_term_of_recurrent_relation}
\begin{split}
     &\frac{(pk^{(0)})^{\frac
     52}(p-1)}{((p-1)k^{(0)})^2}
     \int\limits_0^{((p-1)k^{(0)})^2}d\theta_2\int\limits_{\mathbb{R}^3}\exp\left\{-\theta_2|\kappa^{(0,0)}
     +\frac{\scriptstyle{Y'}}{\scriptstyle{\sqrt{(p-1)k^{(0)}}}}|^2\right\}\cdot
    \\
     &\begin{split}
         \cdot
         \exp\left\{-s|\kappa^{(0)}+(Y-Y')\sqrt{pk^{(0)}}|^2\right\}
         \left\langle v(\kappa^{(0)}+(Y-Y')\sqrt{pk^{(0)}},0),\kappa^{(0,0)}
         +\frac{\scriptstyle{Y}}{\scriptstyle{\sqrt{pk^{(0)}}}}\right\rangle\cdot
        \\
        \cdot
         P_{\kappa^{(0)}+\frac{\scriptstyle{Y}}{\scriptstyle{\sqrt{pk^{(0)}}}}}\tilde{g}_{p-1}
         \left(Y'\sqrt{\frac{\scriptstyle{p}}{\scriptstyle{p-1}}},
         s(1-\frac{\scriptstyle{\theta_2}}{\scriptstyle{((p-1)k^{(0)})^2}})\right)
         d^3Y'
     \end{split}
\end{split}
\end{equation}

The factor $(p-1)$ comes from the inductive assumption concerning
$h_{p-1}$. As before, we replace
$\exp\left\{-\theta_2|\kappa^{(0,0)}
    +\frac{\scriptstyle{Y'}}{\scriptstyle{\sqrt{(p-1)k^{(0)}}}}|^2\right\}$ by $\exp\{-\theta_2\}$,
$P_{\kappa^{(0)}+\frac{\scriptstyle{Y}}{\scriptstyle{\sqrt{pk^{(0)}}}}}$
by the identity operator and $\tilde{g}_{p-1}\left(
    Y'\sqrt{\frac{\scriptstyle{p}}{\scriptstyle{p-1}}},
    s(1-\frac{\scriptstyle{\theta_2}}{\scriptstyle{((p-1)k^{(0)})^2}})\right)$ by
    $\tilde{g}_{p-1}(Y'\sqrt{\frac{p}{p-1}},s)$. All corrections are
included in the remainder terms.

For the Gaussian term in $v(\kappa^{(0)}+(Y-Y')\sqrt{pk^{(0)}},0)$
we can write $\dfrac{1}{(2\pi)^{\frac
32}}\exp\left\{\frac{|Y-Y'|^2p}{2}\right\} $. This shows that
typically $Y-Y'=O(\frac{1}{\sqrt{p}}) $. For the third component
$F^{(1)}$ of $v(\kappa^{(0)}+(Y-Y')\sqrt{pk^{(0)}},0)$ using the
incompressibility condition we can write
\[F^{(1)}(\kappa^{(0)}+(Y-Y')\sqrt{pk^{(0)}},0)=\]
\[-\frac{1}{\sqrt{k^{(0)}}}\left((Y_1-Y_1')\sqrt{p}H_1^{(0)}
((Y-Y')\sqrt{p})+(Y_2-Y_2')\sqrt{p}H_1^{(0)}
((Y-Y')\sqrt{p})+O(\frac{1}{\sqrt{k^{(0)}}})\right)\cdot\]
\[\cdot\exp\left\{-\frac{p|Y-Y'|^2}{2}\right\}\]
For the inner product in \eqref{eq: first_term_of_recurrent_relation}
\[\sqrt{pk^{(0)}}\left\langle v(\kappa^{(0)}+(Y-Y')\sqrt{pk^{(0)}},0),\kappa^{(0,0)}
    +\frac{\scriptstyle{Y}}{\scriptstyle{\sqrt{pk^{(0)}}}}\right\rangle=\exp\left\{-\frac{p|Y-Y'|^2}{2}\right\}\cdot\]
\[\cdot\left[
H_1^{(0)} ((Y-Y')\sqrt{p})Y_1+H_1^{(0)} ((Y-Y')\sqrt{p})Y_2
-\right.
\]\[-\left.\sqrt{p}\left((Y_1-Y_1')\sqrt{p}H_1^{(0)}
((Y-Y')\sqrt{p})+(Y_2-Y_2')\sqrt{p}H_1^{(0)}
((Y-Y')\sqrt{p})\right)+O(\frac{1}{\sqrt{k^{(0)}}}) \right]\]

The expression in the square brackets grows as $\sqrt{p}$ and therefore
\[\sqrt{pk^{(0)}}\left\langle v(\kappa^{(0)}+(Y-Y')\sqrt{pk^{(0)}},0),\kappa^{(0,0)}
    +\frac{\scriptstyle{Y}}{\scriptstyle{\sqrt{pk^{(0)}}}}\right\rangle=\]
can be replaced by
\[-\sqrt{p}\left[(Y_1-Y_1')\sqrt{p}H_1^{(0)}
((Y-Y')\sqrt{p})+(Y_2-Y_2')\sqrt{p}H_1^{(0)}
((Y-Y')\sqrt{p})-\right. \]\[\left.
-\frac{1}{\sqrt{p}}\left(H_1^{(0)} ((Y-Y')\sqrt{p})Y_1+H_1^{(0)}
((Y-Y')\sqrt{p})Y_2\right) \right]\]
Further,
\[\exp\{-s|\kappa^{(0)}+(Y-Y')\sqrt{pk^{(0)}}|^2\}=\exp\{-s|k^{(0)}|^2\}\cdot\]\[\cdot\exp\{-2sk^{(0)}\langle\kappa^{(0,0)},
(Y-Y')\sqrt{p}\sqrt{k^{(0)}}\rangle\}\exp\{-s|Y-Y'|^2pk^{(0)}\}\]
The first factor takes values $O(1)$, the others can be written as
$1+O(\frac{1}{\sqrt{k^{(0)}}})$. The main order of magnitude of
\eqref{eq: first_term_of_recurrent_relation} takes the form
\[p\exp\{-s(k^{(0)})^2\}\frac{(p-1)}{p}\frac{1}{p}\left[
-\int\limits_{\mathbb{R}^3}\left[
(Y_1-Y_1')\sqrt{p}H_1^{(0)}((Y-Y')\sqrt{p})+(Y_2-Y_2')\sqrt{p}H_1^{(0)}((Y-Y')\sqrt{p})
\right]+\right.\]\[+\frac{1}{\sqrt{p}}\left[H_1^{(0)}((Y-Y')\sqrt{p})Y_1+H_1^{(0)}((Y-Y')\sqrt{p})Y_2\right]
\left(\frac{p}{2\pi}\right)^{\frac
32}\exp\left\{-\frac{\scriptstyle{|Y-Y'|^2
p}}{\scriptstyle{2}}\right\}\cdot\]
\[\cdot\left.\left(\frac{1}{2\pi}\right)^{\frac
32}\exp\left\{-\frac{\scriptstyle{|Y'|^2
p}}{\scriptstyle{2(p-1)}}\right\}H^{(0)}\left(Y'\sqrt{\frac{\scriptstyle{p}}{\scriptstyle{p-1}}}\right)d^3Y'\right]\]
A similar expression can be written for the last term in \eqref{eq:sec8_recurrent relation_wide}.
Remark that due to our choice of the interval $S^{(1)}$ the product $s(k^{(0)})^2=O(1)$.

\medskip

Now we derive the recurrent formula for $Z_p(s)$ and $\Lambda_p(s)$.
Since our special fixed point $H^{(0)}$ is a Hermite polynomial of
first degree, the convolution of $H^{(0)}$ over $Y'$ (see
\eqref{eq:sec8_recurrent relation_wide}) gives us simply the product of $H^{(0)}$ and the Gaussian term
and a polynomial in $\gamma$. The function $H^{(0)}$ and the Gaussian term can then
be taken out of the summation in $\gamma$ and this gives us the following recurrent
system for $Z_p(s)$ and $\Lambda_p(s)$:
\begin{align}
&Z_{p+1}(s) \Lambda_{p+1}^p(s) \nonumber \\
=&\sum_{p_1+p_2=p} \frac 1p \cdot \frac {i} {(k^{(0)})^2}\cdot (6\gamma^2-10\gamma+4) \cdot
Z_p(s)^2 \cdot \Lambda_p^{p}(s)\cdot ( 1- e^{-s(p_1 k^{(0)})^2} )
\cdot ( 1- e^{-s(p_2 k^{(0)})^2} ) \label{eq:ZpLambda}
\end{align}
where the factor $(6\gamma^2-10\gamma+4)$ comes from the convolution of $H^{(0)}$ with
itself. Now if we take $Z_{p}(s) = -i (k^{(0)})^2$ and write
$\frac {\Lambda_{p+1}(s)} {\Lambda_p(s)} = 1 + \frac {\xi_{p+1}} {p^2}$, then we have
\[
\left( 1+ \frac {\xi_{p+1}} {p^2} \right)^p = \sum_{\gamma} \frac 1p\cdot(6\gamma^2 -10\gamma+4)
\cdot (1 - e^{-s(p_1 k^{(0)})^2}) ( 1- e^{-s(p_2 k^{(0)})^2} )
\]
then it is not difficult to see that there exists bounded $\xi_{p+1}$ (with an bound independent
of $p$) such that the equality holds. It is an elementary fact that the limit
\[
\Lambda(s) = \lim_{p\rightarrow\infty} \Lambda_{p+1}(s) = \Lambda_1 \prod_{k=1}^{\infty}
 \left(1 + \frac {\xi_{k+1}} {k^2} \right)^k
\]
exists.

Now we discuss the behavior of all remainders for $p<\left( k^{(0)} \right)^{\lambda_2}$.

\ By $\Phi_j^{(u)}$, $\Phi_{j'}^{(n)}$ we denote the
eigen-vectors of the linearized renormalization group corresponding to
the fixed point $H^{(0)}$. For each $p$ we make the following inductive
assumption for $\delta^{(r)}(Y,s),\, r<p:$
\[
\delta^{(r)}(Y,s) = \sum_{j=1}^4 \left( b^{(u)}_{j,r} + \beta_{j,r}^{(u)}\right) \Phi_j^{(u)}
+ \sum_{j'=1}^6 \left( b_{j',r}^{(n)} + \beta_{j',r}^{(n)} \right) \Phi_{j'}^{(n)}
+ \Phi_r^{(st)}, \quad \gamma = \frac r {p-1}
\]
where $b_{j,r}^{(u)} = (p-1)^{\alpha_j} b_j^{(u)} \gamma^{\alpha_j^{(u)}}$,
$b_{j',r}^{(n)} = b_{j'}^{(n)}$, $\Phi_r^{(st)}$ is a function which belongs
to the stable subspace of the linearized renormalization group,
$\gamma= \frac r{p-1}$.

As we go from $p-1$ to $p$, the variable $\gamma=\frac {r} {p-1}$ is replaced by
$\gamma' = \frac r p = \gamma\cdot \frac {p-1} p$. Therefore
\begin{align*}
\left ( b_{j,r}^{(u)} + \beta_{j,r}^{(u)} \right) \gamma^{\alpha_j} \Phi_j^{(u)}
=&
\left( (p-1)^{\alpha_j} b_j ^{(u)} + \beta_{j,r}^{(u)} \right)
 \cdot \left( \frac p {p-1} \right)^{\alpha_j} \cdot
 (\gamma')^{\alpha_j} \Phi_j^{(u)} \\
 =& \left( p^{\alpha_j} b_j^{(u)} +
  \left( \frac p{p-1} \right)^{\alpha_j} \cdot \beta_{j,r}^{(u)}
  \right) \cdot (\gamma')^{\alpha_j} \Phi_j^{(u)} .
\end{align*}
In the same way for the neutral eigen-functions we have
\[
\left( b_{j'}^{(n)} + \beta_{j', r}^{(n)} \right) \Phi_{j'}^{(n)}
\]
because $\alpha_{j'} = 0$. In the same way one can transform $\Phi^{(st)}$.The
coefficients $\beta_{j,r}^{(u)},\, \beta_{j',r}^{(n)}$ are small compared
to the first term . An important conclusion is that the projections to the unstable directions
increase, projections to the neutral directions remain the same and projections
to the stable directions decrease. As was already said, in the case of unstable and neutral
directions the term containing $b_j^{(u)}$ or $b_{j'}^{(n)}$ is the main term.

Now we discuss the form of $\delta^{(p)}(Y,s)$. It is the sum of three types of terms.
\begin{enumerate}
\item [ $a_1$).] The term which depends linearly on all $\delta^{(r)}(Y,s)$. Especially important
is the part which contains all $p^{\alpha_j} b_j^{(u)} (\gamma')^{\alpha_j} \Phi_j^{(u)}$,
$b_{j'}^{(n)} \Phi_{j'}^{(n)}$. If we were to have and be in the limiting regime $H^{(0)}$ then
the integral will give $p^{\alpha_j} b_j^{(u)} \left( 1 + \frac 1p\right)^{\alpha_j} \Phi_j^{(u)}
= (p+1)^{\alpha_j} b_j^{(u)} \Phi_j^{(u)}$ since $\gamma'=1$. However, $H^{(r)}$ are slightly
different from $H^{(0)}$. Therefor we shall have a small correction which is included in
all $\beta_{j, p}^{(u)}$, $\beta_{j',p}^{(n)}$ and in $\Phi_p^{(st)}$. We denote it as
$\beta_{pj1}^{(u)}$, $\beta_{pj'1}^{(n)}$, $\Phi_{p,1}^{(st)}$. The we have terms which are linear
functions of all $\beta_{j, r}^{(u)}$, $\beta_{j',r}^{(n)}$ and $\Phi_r^{(st)}$. They will give us
$\beta_{pj2}^{(u)}$, $\beta_{pj'2}^{(n)}$, $\Phi_{p,2}^{(st)}$.

\item [ $a_2$). ] The term which is the sum of all quadratic expressions depending on
$\delta^{(p_1)}$, $\delta^{(p_2)}$. We expand it using our basis of $\Phi_j^{(u)}$, $\Phi_{j'}^{(n)}$ and
all stable eigen-vectors.

\item [ $a_3$). ] The term which contains all corrections which arise during the four steps described
above. We also expand it in the same way as in $a_2)$).
\end{enumerate}

The sum of all terms gives $\beta_{p,j}^{(u)}$, $\beta_{p,j'}^{(n)}$, $\Phi_{p}^{st}$.

We use this procedure till $p = p_0=(k^{(0)})^{\lambda_2}$.
%
The procedure for $p>p_0$ is discussed in \S 9.


\begin{center}
{\large \S8. \  The List of Remainders and Their Estimates}
\end{center}

In the beginning of \S 7 we described 10-parameter families of initial
conditions which we consider in this paper. We mentioned above that for
each p we have an interval $S^{(p)} = \left[ S^{(p)}_{-}, S^{(p)}_{+}
\right ]$ on the time axis. Actually these intervals are changing when
$p=p_n = (1+\epsilon)^n$ where $\epsilon>0$ is a constant. Therefore we
shall write $S^{(n)} = \left[ S^{(p_n)}_{-}, S^{(p_n)}_{+} \right]$ and
hope that there will be no confusion.

In this and the next section we consider $p > (k^{(0)})^{\lambda_2}$. Each
function $\tilde g_r(Y,s)$, $3\le r<p$, has the following representation:



in the domain $\abs{Y} \le C_1 \sqrt{ \ln rk^{(0)}}$, $Y=(Y_1,Y_2,Y_3)\in\cR^3$
\\
\[
\tilde{g}_r ( Y , s ) \, = \, \Lambda^{r-1} \, \cdot \,
r \, \cdot \, \frac{\sigma^{(1)} }{2 \pi} \, \exp \left\{
\frac{\sigma^{(1)} }{2} \, \left( | Y_1 |^2 \, + \, \right.
\right.
\]
\[
\left. \left. + \, | Y_2 |^2 \right) \right\} \, \cdot \,
\sqrt{\displaystyle{\frac{\sigma^{(2)} }{2 \pi} }} \, \exp
\left\{ - \frac{\sigma^{(2)} }{2} \, | Y_3 |^2 \right\} \,
\cdot \, \left( H^{(0)} ( Y_1 , Y_2 ) \, + \, \delta^{(r)} (
Y , s ) \right);
\]

in the domain $| Y | > C_1 \sqrt{\ln (r k^{(0)})}$ :
\begin{align*}
&\frac{\sigma^{(1)} }{2 \pi } \, \exp \left\{ - \,
\frac{\sigma^{(1)} }{2} \, (| Y_1 |^2 \, + \, | Y_2 |^2 )
\right\} \, \cdot \, \sqrt{\displaystyle{\frac{\sigma^{(2)} }{2
\pi }} } \cdot\\
&\quad \cdot\exp \left\{ - \, \frac{\sigma^{(2)} }{2} \, | Y_3 |^2 \right\}
\, \cdot \, | H^{(0 )} \, (Y_1 , Y_2 ) \, + \, \delta^{(r)}
( Y , s ) | \, \\
& \le\;  \Lambda^{r-1}\cdot r \cdot
\frac 1 {r^{\lambda_3 -1}}
\end{align*}
for some constant $\lambda_3 >0$.
We use the formula (7) to get $\tilde{g}^{(p)}(Y,s)$. New remainders appear in one of
the following ways.

\begin{namelist}{xxxxxxxx}
\item[ Type I.] The recurrent relation (7) does not coincide with the
equation for the fixed point and actually is some perturbation of this
equation. The difference produces some remainders which tend to zero
as $p\rightarrow\infty$.

\item[ Type  II.] For the limiting equation all eigen-vectors in the linear
approximation are multiplied by some constant. In the equation (7) it is no
longer true and the difference generate some remainders. (see also \S 9).

\item[ Type III.] The remainders which are quadratic functions of all
previous remainders.
\end{namelist}

\pagebreak
\begin{center}
{\large \S8A. \ The Remainders of Type I.}
\end{center}

We call the domain $A$ the set $\{ | Y | \leq D_1 \sqrt{\ln
(rk^{(0)})} \}$ and the domain $B$ the set\break
$\{ | Y | > D_1 \sqrt{\ln (r k^{(0)})} \}$.  The
estimates will be done separately in each domain.  We include the first,
the second and the last two terms in (7) in the remainders.  We shall
estimate only the first one, the others are estimated in the same way.

\begin{namelist}{xxxxxxxx}
\item[\underline{\sf Domain A}:] We have

\(
\beta^{(1)}_p ( Y , s ) \,   =   \,
(p + 1)^{\frac{5}{2}} \, \cdot \,
\dfrac{i}{s p^2} \, \cdot \,
\displaystyle{\int\limits_{0}^{p^2}} \, d \theta_2 \,
\displaystyle{\int\limits_{\bR^{3}}} \, < \,
v \left( \left( k^{(0)}  +
\frac{Y - Y^\prime}{\sqrt{s}}  \, \sqrt{p+1} \,  , \, 0 \right) \, ; \,
b \right),
\)

\medskip
\(
\hspace{.25in}
\displaystyle{\sqrt{s}} \, k^{(0)}  +
\frac{Y}{\sqrt{p + 1}}  >
P_{{\displaystyle{\sqrt{s}} \, k^{(0)}}  +
{\textstyle{\frac{Y}{\sqrt{p+1}}}}} \,
\tilde{g}_p \, \left(
Y^\prime , \left( 1 - {\displaystyle{\frac{\theta_2}{p^2}}}
\right) s \right)
\)

\medskip
\(
\hspace{.25in}
\exp \left\{
- \left|
\sqrt{s} \, k^{(0)}  +  ( Y - Y^\prime ) \;
\sqrt{p+1} \right|^2 \, -
\displaystyle{\frac{\theta_2}{p^2}} \, \left|
\sqrt{s} \, k^{(0)} \, p \, + \, Y^\prime \, \sqrt{p+1} \right|^2
\right \} \, d^3 Y^\prime
\)
\end{namelist}

\noindent
Here $b$ means the collection of all parameters of $v(k;0)$.
The the main contribution to the integral comes from $Y -
Y^\prime \, = \, O
\, \left(
\frac{1}{\sqrt{p + 1}} \right)$.  In this domain in the main order of
magnitude

\[
\langle
v ( k^{(0)} \, + \, \frac{Y - Y^\prime}{\sqrt{s}} \;
\sqrt{p+1} , \, 0 \, ; \, b ) , \, \sqrt{s} \, k^{(0)} \rangle \, =
\, O(1)
\]

\noindent
Assuming that
$v ( k^{(0)} \, + \, \frac{Y - Y^\prime}{\sqrt{s}} \sqrt{ p+1}
 , \ 0 ; \, b )$
is differentiable {\tt w.r.t} the first three variables we see that the
inner product
\[
\langle v ( k^{(0)} \, + \, \frac{Y - Y^\prime}{\sqrt{s}}\sqrt{p+1}
 , \, 0 \, ; \,
\alpha ) \, , \sqrt{s} \, k^{(0)} \, + \, \frac{Y}{\sqrt{p+1}} \rangle
\]
is of order $O ( 1 )$.  For $\tilde{g}_p$ we can write using our
inductive assumptions
\[
\tilde{g}_p \left( Y^\prime , \left(
1 - \frac{\theta_2}{p^2} \right) s \right) \, = \,
\Lambda^{p-1} \cdot p
\cdot \, \displaystyle{\frac{\sigma^{(1)}}{2 \pi} \, \cdot \,
\displaystyle{\sqrt{\frac{\sigma^{(2)}}{2 \pi} } \,
\cdot \, \exp \left\{ - \displaystyle{ \frac{
\sigma^{(1)} \, ( | Y_1|^2 \,
+ \, Y_2 |^2 )}{2} } \right\} } }
\]

\[
\cdot \, \exp \left\{ - \displaystyle{\frac{ \sigma^{(2)}
 \, ( | Y_3 |^2)}{2}} \right\} \, \cdot \,
\cH^{(p)}\, \left( Y^\prime, \left( 1 -
\frac{\theta_2}{p^2} \right) s \right) \, .
\]
Also
\[
\exp \left\{
- \frac{\theta_2}{p^2} \left|
\sqrt{s} \, k^{(0)} \, p \, + \, Y^\prime \, \sqrt{p + 1} \right|^2
\right\} \, = \,
\exp \left\{
- \theta_2 \left|
\sqrt{s} \, k^{(0)} \, + \,
\frac{Y^\prime \sqrt{p + 1}}{p} \right|^2 \right\}
\]

\noindent
and in the main order of magnitude the integration over $\theta_2$ does
not depend on $Y^\prime$.  Thus we can write
\begin{equation*}
| \beta^{(1)}_p ( Y , s ) | \, \leq \,
\Lambda^{(p - 2)}  \, \cdot \, p \, \cdot \, \exp \left\{
- \frac{\sigma^{(1)} }{2} \,
( | Y_1 |^2 \, + \, | Y_2 |^2 ) \right\}
\end{equation*}
\begin{equation} \label{eq:beta estimate}
 \cdot \, \exp \left\{
- \frac{\sigma^{(2)} }{2}  \,
| Y_3 |^2 \right\} \, \cdot \,
  \frac{D_4}{p}
\end{equation}

\noindent Here and later various constants whose exact values play
no role in the arguments will be denoted by the letter $D$ with
indices.  Since in the expression for $\tilde{g}_{p+1}$ we have
the factors $\Lambda^{p}  \,
\cdot \, (p + 1) \, \cdot \, \exp \left\{ - \frac{\sigma^{(1)}
} 2 \, ( | Y_1 |^2 \, + \, | Y_2 |^2 \right\} \, \cdot \,
\frac{\sigma^{(1)} }{2 \pi}$ $\sqrt{\frac{\sigma^{(2)} (
s )}{2 \pi}} \, \cdot \, \exp \left\{ - \frac{\sigma^{(2)}}
{2} \, | Y_3 |^2 \right\} \, $,  the estimate \eqref{eq:beta estimate}
shows that $| \beta^{(1)}_p ( Y , s ) |$ is relatively smaller than
$\tilde{g}_{p+1}$ with an order $O ( \frac{1}{p} )$.  This is good
enough for our purposes.  We did not discuss the errors which follow
from the fact that the expressions in the previous formulas depend
on $\theta_2$.

\begin{namelist}{xxxxxxxxx}
\item[\underline{\sf Domain B}:] The smallness of $\beta^{(1)}_p ( Y , s
)$ in this case follows easily from several inequalities and arguments.
\end{namelist}
\begin{namelist}{xxxx}
\item[1$^\circ$:] $| Y | \, \leq \, D_4 \, \sqrt{p k^{(0)}}$ because $| k | \,
\leq \, D_5 p k^{(0)}$.
\item[2$^\circ$:] $| Y - Y^\prime | \, \leq \, D_6 \sqrt{k^{(0)}}$ because $v ( k , 0 ;
b )$ has a compact support.
\item[3$^\circ$:] If $\left| Y - Y^\prime \, \right| \, \leq \,
\displaystyle{\frac{2
s_+}{\sqrt{p}}}$ \hspace{.3em} then
\[ \exp \left\{
- \, \left| \sqrt{s} \: k^{(0)} + ( Y - Y^\prime ) \, \sqrt{p + 1} \,
  \right|^2 \right\} \,
  \leq 1
\]
If $\left| Y - Y^\prime \right| \, \geq \,
\displaystyle{\frac{2 s_+}{\sqrt{p}}}$
\hspace{.3em} then
\[
\exp \left\{
- \left| \sqrt{s} \: k^{(0)} \, + \,
( Y - Y^\prime ) \sqrt{p+1} |^2 \right\} \, \leq \,
\exp \left\{ - \frac{s_+}{4} \, \right| Y - Y^\prime |^2 \right\} \, .
\]
\item[4$^\circ$:] If $| Y^\prime |  \, \geq \, D_7 \, \sqrt{p}$
\hspace{.3em} then
\[
\exp \left\{
- \frac{\theta_2}{p^2} \, \left|
\sqrt{s} \, k^{(0)} p \, + \,
Y^\prime \sqrt{p+1} \, \right|^2 \right\} \, \leq \,
\exp \{ - C_8 \theta_2 \}
\]
\item[5$^\circ$:] If $| Y^\prime |  \, \leq \, D_7 \, \sqrt{p}$
\hspace{.3em} then
\[
\exp \left\{ - \,
\frac{\theta_2}{p^2} \left| \, \sqrt{s} \: k^{(0)} \, p + Y^\prime \,
\sqrt{p+1} \, \right|^2 \right\} \, \leq \, 1 \, .
\]
\item[6$^\circ$:] We have
\[
\exp
\left\{
- \,
\displaystyle{
\frac{\sigma^{(1)} }{2} } \,
( | Y^\prime_1 |^2 \, + \, | Y^\prime_2 |^2 ) \,
- \, \displaystyle{\frac{\sigma^{(2)}}{2} } \, | Y^\prime_3
  |^2
\right\}
\]

\[
=  \, \exp
\left\{
-  \, \displaystyle{\frac{\sigma^{(1)} }{2}}
\left(
\left|
Y_1 - (Y_1 - Y^\prime_1 )
\right|^2 +  | Y_2 - ( Y_2 - Y^\prime_2 )
|^2 \right) \right. \]

\[ \left. - \,
\displaystyle{\frac{\sigma^{(2)} }{2 }}
| Y_3 - ( Y_3 - Y^\prime_3 ) |^2 \right\}
\]

\[
= \, \exp \left\{ - \,
\displaystyle{\frac{\sigma^{(1)} }{2}} \, ( | Y_1|^2 \, + \, | Y_2
|^2 ) \, - \,
\displaystyle{\frac{\sigma^{(2)} }{2}} \, (| Y_3 |^2 ) \right\}
\]

\[
\cdot \,
\exp
\left\{
\sigma^{(1)}  \,
( Y_1 ( Y_1 - Y^\prime_1 ) \, + \,
Y_2 ( Y_2 - Y^\prime_2 ) ) \, + \,
\sigma^{(2)} \, Y_3 ( Y_3 - Y^\prime_3 ) \right.
\]

\[ \left.
- \, \displaystyle{\frac{
\sigma^{(1)} ( s )}{2}} \,
\left(
| Y_1 - Y^\prime_1 |^2 \, + \, | Y_2 - Y^\prime_2 |^2 \right) \, - \,
\displaystyle{\frac{\sigma^{(2)} }{2}} \,
| Y_3 - Y^\prime_3 |^2 \right\} \, .
\]

If $| Y - Y^\prime| \, \leq \, \displaystyle{\frac{2 s_+}{\sqrt{p}}}$
\hspace{.3em} then
\[
\exp \left\{
\sigma^{(1)}  \, ( Y_1 (Y_1 - Y^\prime_1 ) \, + \,
Y_2 (Y_2- Y^\prime_2 )) \, + \,
\sigma^{(2)} ( s ) \, Y_3 (Y_3 - Y^\prime_3 ) \right.
\]

\[ \left. - \, \displaystyle{\frac{\sigma^{(1)}}{2}} \,
( | Y_1 -Y^\prime_1 |^2 \, + \, | Y_2 - Y^\prime_2 |^2 ) \, -
\displaystyle{\frac{\sigma^{(2)} }{2}} \, | Y_3 - Y^\prime_3
|^2 \right \} \, \leq \, C_8 \, .
\]
\end{namelist}
If $| Y - Y^\prime | \, > \, \displaystyle{\frac{2 s_+}{\sqrt{p}}}$
then we have an integral of the function which is the product of some
Gaussian factor and $| \cH^{(p)}( Y ) |$.  Direct estimate
shows as before that in this case
\begin{equation*}
| \beta^{(1)}_p ( Y , s ) | \, \leq \, \Lambda^{(p - 1)}
\, \cdot \, p \, \cdot \, e^{- \, \displaystyle{\frac{\sigma^{(1)}
}{2}} ( | Y_1 |^2 \, + \, Y_2 |^2 )} \, \cdot \, e^{ -
\displaystyle{\frac{\sigma^{(2)} }{2}} \, | Y_3 |^2} \, \cdot \,
\displaystyle{\frac{D_8}{p^{\frac{3}{2}}}}
\end{equation*}
which is also good for us.

In the same way one can estimate terms with relatively small $p_1$ and
$p - p_1$ (i.e., $p_1 \leq \sqrt{p}$ or $p_1 \geq p - \sqrt{p}$.  The
remainders will be of order $\frac{1}{\sqrt{p_1}} \, \cdot \,
\frac{1}{p}$.  The next set of remainders comes from splitting the
integration over $\theta$ and $Y^\prime$ (see (7) and beginning of \S3).
We may assume that $p_1 > \sqrt{p}$ or $p_1 < p - \sqrt{p}$ because
other terms were estimated before.  Put
\[
\tilde{\tilde{g}}_{p+1} ( Y , s ) \, =\,
i \, (p+1)^{\frac{5}{2}} \;
\displaystyle{\sum\limits_{p_1 + p_2 \, = \, p_{+1}\atop{
p_1, p_2 \, > \, \sqrt{p}}}} \;
\displaystyle{\int\limits_{0}^{p_1^2}} \,
d \theta_1 \,
\displaystyle{\int\limits_{0}^{p_2^2}} \, d \theta_2 \, \cdot \,
\displaystyle{\frac{1}{p^2_1 p^2_2}}
\]

\[
\displaystyle{\int\limits_{\bR^{3}}} \, \langle \tilde{g}_{p_1} \,
\left( ( Y - Y^\prime ) \; \displaystyle{ \frac{ ( 1 - \frac{
\theta_1}{ p^2_1} )^{ \frac{1}{2}}} {\sqrt{\gamma}} }  \, , \left(
1 - \frac{\theta_1}{p^2_1} \right) s \right) \, , \, \sqrt{s} \,
k^{(0)} \, + \, \frac{Y}{\sqrt{p + 1}} \rangle
\]

\[
P_{\displaystyle{ {\sqrt{s}} \, k^{(0)} } \, + \,
\displaystyle{\frac{Y} {\sqrt{p+1}} }}   \; \tilde{g}_{p_2} \;
\left( \displaystyle{\frac{Y^\prime ( 1 - \frac{\theta_2}{p^2_2}
)^{\frac{1}{2}}}{\sqrt{(1-\gamma)}}} \, , \left( 1 -
\frac{\theta_2}{p^2_2} \right) s \right)
\]

\[
\cdot \, \exp \left\{ - \theta_1 \, | \, \sqrt{s} \, k^{(0)} \, +
\, \displaystyle{\frac{ Y - Y^\prime}{\sqrt{p + 1} \, \cdot \,
\gamma}}|^2 \, - \, \theta_2 | \, \sqrt{s} \, k^{(0)} \, + \,
\displaystyle{\frac{Y - Y^\prime}{\sqrt{p+1} \gamma}} |^2 \right\}
\, .
\]

\noindent
Using the inductive assumption we can rewrite the last expression as
follows:

\[
\tilde{\tilde{g}}_{p+1} \, ( Y , s ) \, = \,
i \, (p+1) \;
\displaystyle{\sum\limits_{p_1 + p_2 \, = \, p + 1 \atop{p_1 , p_2 \, > \,
\sqrt{p}}}} \;
\displaystyle{\int\limits_{0}^{p^2_1}} \, d \theta_1 \;
\displaystyle{\int\limits_{0}^{p^2_2}} \, d \theta_2
\]

\[
\Lambda^{p_1-1}  \cdot
\, \Lambda^{p_2 - 1}  \, \cdot
\, \frac{1}{\gamma (1-\gamma)} \, \cdot \, \frac{1}{p + 1}
\exp \, \left\{ - \displaystyle{\frac{\sigma^{(1)}}{2}} \;
\displaystyle{\frac{| Y_1 - Y^\prime_1 |^2 \, + \, | Y_2 -
Y^\prime_2 |^2}{\gamma}} \right.
\]

\[
- \, \displaystyle{\frac{\sigma^{(2)}}{2} }
\; \displaystyle{\frac{| Y_3 - Y^\prime_3 |^2}{\gamma}}
\, - \, \displaystyle{\frac{\sigma^{(1)}}{2}} \; \displaystyle{\frac{|
Y^\prime_1 |^2 \, + \, | Y^\prime_2 |^2}{(1-\gamma)}} \]

\[ \left.
- \, \displaystyle{ \frac{\sigma^{(1)}}{2}}   \, \cdot \, \displaystyle{\frac{
| Y^\prime_3 |^2}{1-\gamma }} \right\} \, . \, p^{\frac{1}{2}} \, <
\, \cH^{(p_1)} \, \left( Y - Y^\prime , \, s \left( 1 -
\frac{\theta_1}{p^2_1} \right) \right) \, ,
\]

\[ \sqrt{s} \, k^{(0)} \, + \,
\displaystyle{\frac{Y}{\sqrt{p}} } \, > \, \, \cdot \,
P_{\displaystyle{\sqrt{s} \, k^{(0)} \, +\,
\displaystyle{\frac{Y}{\sqrt{p}}}}} \;
\cH^{(p_2)} \, \left(
Y^\prime, s \left( 1 - \frac{\theta_2}{p^2_2}
\right) \right) \, d^3 Y^\prime \, .
\]

\noindent
As was explained before, due to incompressibility in the {\sf Domain A}
, the inner product
\[
\langle
\cH^{(p_1)} \, \left( Y - Y^\prime ; \, s \left( 1 -
\frac{\theta_1}{p^2_1} \right) \right)
\, ,
\sqrt{s} \, k^{(0)} \, + \, \frac{Y}{\sqrt{p}} \rangle
\]

\noindent
takes values $O ( \frac{1}{\sqrt{p}} )$ because the first two components
of the vector $\sqrt{s} \, k^{(0)} + \frac{Y}{\sqrt{p}}$ are of order $O
( \frac{1}{\sqrt{p}} )$.  Therefore the product
\[
\sqrt{p} \, \langle \,
\cH^{(p_1)} \, \left(
Y - Y^\prime , \, s \left( 1 - \frac{\theta_1}{p^2_1} \right) \right) \,
,
\sqrt{s} \, k^{(0)} \, + \,
\displaystyle{\frac{Y}{\sqrt{p}} } \rangle
\]

\noindent
takes values of order $O ( 1 )$.

The remainder can be written in the following form:

\[
\beta^{(2)}_p \, ( Y , s ) \, = \, i \,
\displaystyle{\sum\limits_{p_1 + p_2 \, = \, p+1 \atop{ p_1 , p_2
\, > \, \sqrt{p}}}} \; \displaystyle{\frac{1}{\gamma (1-\gamma)}}
\, \cdot \, \frac{1}{p} \, \cdot \,
\displaystyle{\int\limits_{0}^{p_1^2}} \, d \theta_1 \,
\displaystyle{\int\limits_{0}^{p^2_2}} \, d \theta_2
\]

\[
\cdot \,
\Lambda^{p_1 - 1}  \cdot
\Lambda^{p_2 - 1} \,  \, \cdot
\frac{1}{\Lambda^p}
\cdot \, \displaystyle{\int_{\bR^{3}}} \, \exp \left\{ - \,
\displaystyle{ \frac{\sigma^{(1)} \;
( | Y_1 - Y^\prime_1 |^2 \, + \, | Y_2 - Y^\prime_2 |^2 )}{2
\gamma}} \right.
\]

\[
- \, \displaystyle{ \frac{\sigma^{(2)}}{2 \gamma}} \, \cdot \, \displaystyle{ \frac{| Y_3 -
Y^\prime_3 |^2}{2 \gamma}} \, - \, \displaystyle{\frac{ \sigma^{(1)} \, ( | Y^\prime_1 |^2 \, + \,
Y^\prime_2 |^2 )}{2 (1-\gamma)}}
\left. - \, \displaystyle{ \frac{\sigma^{(2)} \, | Y^\prime_3 |^2}{2 (1-\gamma)} }
\right\} \, < \, \cH^{(p_1)}  
 \; \left( Y - Y^\prime , \, s \left( 1 -
\frac{\theta_1}{p^2_1} \right) \right) \, ,
\]

\[
\sqrt{s} \, k^{(0)} \, + \,
\frac{Y}{\sqrt{p}} \, > \, \, \cdot \,
P_{\sqrt{s} \, k^{(0)} \, + \, \frac{Y}{\sqrt{p}} }
\, \cH^{(p_2)} \, \left( Y^\prime, \left( 1 - \frac{\theta_2}{p^2_2}
\right) s \right) \,
\cdot
\]

\[
\cdot \, \exp \left\{ - \theta_1 | \sqrt{s} \, k^{(0)} \, + \,
\displaystyle{\frac{Y - Y^\prime}{\sqrt{p} \gamma}} |^2 \, - \,
\theta_2 | \sqrt{s} \, k^{(0)} \, + \,
\displaystyle{\frac{Y^\prime}{\sqrt{p} \, (1-\gamma)}} |^2
\right\} \, \cdot \]

\[ - \, i \,
\displaystyle{\sum\limits_{p_1 + p_2 \, = \, p+1 \atop{ p_1 , p_2
\, > \, 1}}} \; \displaystyle{\frac{1}{\gamma (1-\gamma)}} \,
\cdot \, \frac{1}{p} \, \cdot \,
\displaystyle{\int\limits_{0}^{p_1^2}} \, \exp \left\{ - \theta_1
s \right\} \, d \theta_1 \, \displaystyle{\int\limits_{0}^{p^2_2}}
\, \exp \left\{ - \theta_2 s \right\} \, d \theta_2
\]

\[
\displaystyle{\int\limits_{\bR^{3}}} \, \exp \left\{ - \,
\displaystyle{\frac{\sigma^{(1)}  \, ( | Y_1 - Y^\prime_1 |^2
\, + \, | Y_2 -Y^\prime_2 |^2 )}{2 \gamma}} \, - \,
\displaystyle{\frac{\sigma^{(2)} ( | Y_3 - Y^\prime_3 |^2 )}{2
\gamma}} \right.
\]

\[
\left. - \, \displaystyle{\frac{\sigma^{(1)}  \, ( | Y^\prime_1
|^2 \, + \, Y^\prime_2 |^2 )}{2 (1-\gamma)}} \, - \,
\displaystyle{\frac{\sigma^{(2)} \, | Y^\prime_3 |^2}{2
(1-\gamma)}} \right\}
\]

\[
\cdot p^{\frac{1}{2}} \, \cdot \langle
\cH^{(p_1)} \,( Y - Y^\prime ) , \,
\sqrt{s} \, k^{(0)} \, + \, \frac{Y}{\sqrt{p}} \rangle \,
P_{\displaystyle{\sqrt{s} \, k^{(0)} \, + \,
\frac{Y}{\sqrt{p}}
}} \,
\cH^{(p_2)} \, ( Y^\prime , s ) \, d^3 Y^\prime
\, .
\]

\noindent
We did not include the factor $\Lambda^{p - 1}  \, \cdot \,
p$ because it is a part of the inductive assumption.  This remainder is
estimated in the following way.

First we consider

\[
R_1 \, = \, \left( \left| \sqrt{s} \, k^{(0)} \, + \,
\displaystyle{\frac{Y - Y^\prime}{\sqrt{p} \gamma}} \right|^2 - s
\right) \, + \, \left( \left| \sqrt{s} \, k^{(0)} \, + \,
\displaystyle{\frac{Y^\prime}{\sqrt{p} \, (1-\gamma)}} \right|^2 -
s \right)
\]

\noindent
As before, consider the domain where
\[
| Y - Y^\prime | \, \leq \, D_9 \,
\sqrt{\ln \, (pk^{(0)})}, \, | Y^\prime | \, \leq \, D_{10} \,
\sqrt{\ln \, (pk^{(0)})} \, .
\]

\noindent
We write

\[
R_1 \, = \, \displaystyle{\frac{| Y - Y^\prime|^2}{p \, \cdot
\gamma^2_1}} \, + \, \displaystyle{\frac{| Y^\prime|^2}{p \, \,
\cdot \gamma^2_2}} \, + \, C_{11} \left( \displaystyle{\frac{| Y -
Y^\prime|}{\sqrt{p} \, \gamma|}} \, + \, \displaystyle{\frac{|
Y^\prime|}{\sqrt{p} (1-\gamma)}} \right) \, .
\]

\noindent
In the {\sf Domain A}
\[
| R_1 | \, \leq \, \frac{C_{12} \ln ( p k^{(0)}) }{p k^{(0)}}
\, .
\]

Therefore
\[
R_2 \, = \, \exp \left\{ - \theta_1 \left| \sqrt{s} \, k^{(0)}  +
\displaystyle{\frac{Y - Y^\prime}{\sqrt{p} \, \gamma}} \right|^2
- \theta_2 \left| \sqrt{s} \, k^{(0)}  +
\displaystyle{\frac{Y^\prime}{\sqrt{p} \gamma^2}} \right|^2
\right\}
\]
\[ - \,  \exp
\{ - \, \theta_1 s \} \, \cdot \, \exp \{ - \theta_2 s \}
\]

\[
= \, \exp \{ - ( \theta_1 + \theta_2 ) s ) \} \, \cdot \, \left[
\exp \left\{ - \theta_1 \left( \left| \sqrt{s} \, k^{(0)} \, + \,
\displaystyle{\frac{Y - Y^\prime}{\sqrt{p} \gamma}} \right|^2 - s
\right) \right. \right.
\]

\[
\left. \left. \cdot \, \exp \left\{ - \theta_2 \left( \left|
\sqrt{s} \, k^{(0)} \, + \, \displaystyle{\frac{Y^\prime}{\sqrt{p}
(1-\gamma)}} \right|^2 - s \right) \right\} - 1 \right] \right.
\]

\noindent
and in the {\sf Domain A}
\[
| R_2 | \, \leq \, \exp \{ - ( \theta_1 \, + \, \theta_2 ) s \} \,
\left( \displaystyle{\frac{\theta_1 \, \cdot \, C_{13}}{\sqrt{p}
\gamma}} \, + \, \displaystyle{\frac{\theta_2 \, \ln \, p
}{\sqrt{p} (1-\gamma)}} \right) \, .
\]

\noindent
This shows that in the {\sf Domain A} we can replace the exponent

\[
\exp \left\{ - \theta_1 | \sqrt{s} \, k^{(0)} \, + \,
\displaystyle{\frac{Y -Y^\prime}{\sqrt{p} \gamma}} |^2 - \,
\theta_2 | \sqrt{s} \, k^{(0)} \, + \,
\displaystyle{\frac{Y^\prime}{\sqrt{p} (1-\gamma)}} |^2 \right\}
\]

\noindent
by $\exp \{ - ( \theta_1 + \theta_2 ) s (k^{(0)})^2 \}$ and the remainder will be
not more than $\frac{D_{14} \ln \, p}{\sqrt{p}}$.  This is enough for
our purposes.  In the {\sf Domain B} the estimates are similar because
again the main contribution to the integral comes from $| Y - Y^\prime|
\leq D_9 \sqrt{\ln \, p}$, $| Y^\prime | \leq D_{10} \sqrt{\ln \, p}$.
In other words, in the {\sf Domain B} we can replace the product of the
Gaussian factors and $\cH^{(p)}$ by

\[
\exp \left\{ - \frac{1}{2} \, \sigma^{(1)}  \, ( | Y_1 -
Y^\prime_1 |^2 \, + \, |Y_2 - Y^\prime_2 |^2 ) \, - \, \frac{1}{2}
\, \sigma^{(2)}  \, | Y_3 - Y^\prime_3 |^2 \right.
\]

\[
\left. - \, \frac{1}{2} \, \sigma^{(1)}  \, ( | Y^\prime_1 |^2
\, + \, | Y^\prime_2 |^2 ) \, - \frac{1}{2} \, \sigma^{(2)}  (
| Y^\prime_3 |^2 ) \right\} \, .
\]

\noindent
This is also enough for our purpose.

The next remainder of Type I comes from the difference between the sum
over $\gamma$ and the corresponding integral.  The remainder
$\beta^{(3)}_p ( Y , s )$ is the difference between the sum

\[
i \displaystyle{\sum\limits_{p_1 + p_2 \, = \, p+1 \atop{ p_1 , p_2
\, > \, \sqrt{p}}}} \; \sqrt{\gamma} \, \sqrt{(1-\gamma)} \, \cdot
\, \frac{1}{p} \, \cdot \, \displaystyle{\int\limits_{\bR^{3}}} \,
\exp \left\{ - \, \displaystyle{\frac{\sigma^{(1)}  \, ( | Y_1
- Y^\prime_1 |^2 \, + \, | Y_2 - Y^\prime_2 |^2 )}{2 \gamma}}
\right.
\]

\[
\left. - \, \displaystyle{\frac{ \sigma^{(2)}  ( | Y_3 -
Y^\prime_3 |)^2}{2 \gamma}} \, - \, \displaystyle{\frac{\sigma^{(1)}
 ( | Y^\prime_1 |^2 \, + \, | Y^\prime_2 |^2 )}{2 (1-\gamma)}}
\, - \, \displaystyle{\frac{\sigma^{(2)}   | Y^\prime_3 |
^2}{2 (1-\gamma)}} \right\}
\]

\[
\cdot \, \left( \frac{1}{2 \pi \gamma} \right)^{\frac{3}{2}} \,
\cdot \, \left( \frac{1}{2 \pi (1-\gamma)} \right)^{\frac{3}{2}}
\, \cdot \, p^{\frac{1}{2}} \, \cdot \, \langle \cH^{(p_1)}
( ( Y- Y^\prime)) \, , \sqrt{s} \, k^{(0)} \, + \,
\frac{Y}{\sqrt{p}} \rangle
\]

\[
P_{\displaystyle{\sqrt{s} \, k^{(0)} \, + \, \frac{Y}{\sqrt{p}}}} \,
\cH^{(p_2)} ( Y^\prime , s ) \, d^3 Y^\prime
\]

\noindent and the corresponding integral over $\gamma$ from $0$ to
$1$.  It is easy to check that this difference is not more than
$\frac{C_{14}}{\sqrt{p}}$.

\begin{center}
{\large \S8B. \ \sf The Remainders of Type II and III }
\end{center}

All remainders of Type II appear because we use the sums (over $p_1$)
instead of the integrals .  The functions $\cH \left( \frac{Y
- Y^\prime}{\sqrt{\gamma}} \right)$ are defined for all $\gamma$.  We
  use a linear interpolation to define $\delta ( \gamma , Y , s )$ for
all $\gamma$.  From our inductive assumptions it follows that $| \delta_p
( \gamma, Y , s ) | \leq \, \frac{C_{16}}{\sqrt{p}}$. Therefore, the remainders
which follow from the difference between the sum and the integral
also satisfy this estimate.

It remains to consider quadratic expressions of $\delta_p ( \gamma, Y ,
s )$.  The Gaussian density is present in all these expressions.
Therefore, all the remainders are not more than $\frac{C_{17}}{p}$.

\begin{center}
{\large \S9. \ \sf Final Steps in the Proof of the Main Result}
\end{center}

In this section we consider our procedure for $p>p_0$. Introduce the sequence $p_n$,
$p_n = (1+\epsilon)p_{n-1} = (1+\epsilon)^n p_0$, where $\epsilon>0$ is small
 (see below). These are the values of $p$ when we make the renormalization of
our parameters. For $p\neq p_n$, no renormalization is done.

In \S 7 we explained the choice of our fixed point ${H}^{(0)}$. The corresponding
eigen-functions are denoted by ${\Phi}_j^{(u)}$ and ${\Phi}_{j'}^{(n)}$.
Also we have eigen-functions of the stable part of the spectrum. Consider p,
$p_m < p <p_{m+1}$. By induction we assume that we have an interval on the time
axis $\left [ S^{(m)}_{-}, S^{(m)}_{+} \right]$ and
$s\in \left [ S^{(m)}_{-}, S^{(m)}_{+} \right]$, $r<p$, so that
\begin{align*}
\tilde g_r(Y,s) = &\Lambda^{r-1}\cdot r\cdot ( H^{(0)}(Y) + \delta^{(r)}(Y,s) ) \cdot\\
&\cdot \frac {\sigma^{(1)}} {2\pi}
\exp \left\{ - \frac {\sigma^{(1)} (Y_1^2+Y_2^2)}2 \right\} \cdot
\sqrt{ \frac{\sigma^{(2)}} {2\pi}}
\exp \left\{ - \frac {\sigma^{(2)} Y_3^2}2 \right\}
\end{align*}
If $\gamma = \frac {r} {p-1}$ then
\begin{align*}
\delta^{(r)} (Y,s) =
\sum_{j=1}^4 \left(  b_{j,p}^{(u)} + {\beta}_{j,r}^{(u)} \right)
\gamma^{\alpha_j^{(u)}} {\Phi}_j^{(u)} (Y)
+ \sum_{j'=1}^6 \left(  b_{j', p}^{(n)} + {\beta}_{j',r}^{(n)} \right)
 {\Phi}_j^{(n)} (Y) + \Phi^{(st)}_r (Y,\gamma).
\end{align*}
here ${\beta}_{j,r}^{(u)}, {\beta}_{j',r}^{(n)}$ are small corrections to the
main terms $ b_{j,p}^{(u)}$, $ b_{j',p}^{(n)}$, $\Phi_r^{(st)}$ can be
written as a series w.r.t. the stable eigen-functions. (see Appendix II).

At one step of our procedure $p-1$ is replaced by $p$, $\gamma$ is replaced by
$\gamma' = \gamma \cdot \frac {p-1} p$
and $\gamma^{\alpha_j^{(u)}}$ is replaced by $\left( 1 + \frac 1{p-1} \right)^{\alpha_j^{(u)}}
\cdot {(\gamma')}^{\alpha_j^{(u)}}$,  $ b_{j,p}^{(u)}+ \bar{\beta}_{j,r}^{(u)}$ is replaced
by $( \bar b_{j,p}^{(u)} +  \beta_{j,r}^{(u)} ) \left
( 1 + \frac 1{p-1}\right)^{\alpha_j^{(u)}}$. During the whole interval $p_m < p < p_{m+1}$ the
variable $ b_{j,p_m}^{(u)}$ acquires the factor
\[
\prod_{p_m < p < p_{m+1}}
\left( 1 + \frac 1{p-1} \right)^{\alpha_j^{(u)}} \thickapprox e^{(1+\epsilon)\alpha_j^{(u)}}.
\]
A similar statement holds for the stable part of the spectrum. The neutral part remains the
same since $\alpha_{j'}^{(n)} = 0$.

Now we shall discuss $\delta^{(p)} ( Y,s)$ using (7). As in \S 7 $\delta^{(p)}(Y,s)$ consists
of three parts.

\begin{namelist}{xxxxxxx}
\item[ Part I. ]   In all $\delta^{(r)}$ the main term is the one which contains our basic
parameters $ b_j^{(u)}$, $ b_{j'}^{(n)}$. We consider terms in (7) which are linear in
$ b_j^{(u)}$, $ b_{j'}^{(n)}$. As it follows from the definition of the linearized group
and its spectrum we get $\left( 1  + \frac 1 p \right )^{\alpha_j^{(u)}}  b_{j,p}^{(u)}$.
For the neutral part we get $1$ because $\alpha_{j'}^{(n)}=0$. We put
$ b_{j, p+1}^{(u)} =  b_{j,p}^{(u)} \cdot \left (1 + \frac 1 p \right)^{\alpha_j^{(u)}}
 b_{j,p}^{(u)}$, $ b_{j', p+1}^{(n)}=  b_{j',p}^{(n)}$. The stable part is
transformed accordingly.

\item[ Part II.] The term which is the sum of quadratic functions of all $\delta^{(r)}$. We
expand it using the basis of our functions $ \Phi_j^{(u)}$, $ \Phi_{j'}^{(n)}$ and the
functions from the stable part of the spectrum. The result is included in
$ \beta_{j,p}^{(u)}$, $ \beta_{j', p}^{(n)}$ and the stable function $\Phi_p^{(st)}(Y,s)$.

\item[ Part III.] All remainders which arise because the formulas for finite $p$ are different
from the limiting formulas. These remainders were estimated in \S 6. The result is written as
a series w.r.t. our basis and the corresponding terms are included in
$ \beta_{j,p}^{(u)}$, $ \beta_{j', p}^{(n)}$ and the stable part of the spectrum.
\end{namelist}

Finally we have
\[
 b_{j,p+1}^{(u)} =  b_{j,p}^{(u)}
\left( 1 + \frac 1 p \right)^{\alpha_j^{(u)}},
\quad
 b_{j, p+1}^{(n)} =  b_{j,p}^{(n)}
\]
and the formulas for $ \beta_{j,p}^{(u)}$, $ \beta_{j',p}^{(n)}$ and
$\Phi_p^{(st)}(Y,s)$. This works for $p<p_{m+1}$. If $p=p_{m+1}$, then
we introduce new variables (rescaling!)
\[
 b_{j,p_{m+1}}^{(u)} =
 b_{j, p_{m+1}-1}^{(u)}
\left( 1 + \frac 1{p_{m+1}}\right)^{\alpha_j^{(u)}} +
 \beta_{j,p_{m+1}}^{(u)},
\]
\[
 b_{j',p_{m+1}}^{(n)} =  b_{j',p_{m+1}-1}^{(n)}
+ \beta_{j',p_{m+1}}^{(n)} .
\]
It is our other inductive assumption that
\[
-\rho_1^m \le  b_{j,p_m}^{(u)} \le \rho_1^m,\quad
-\rho_1^m \le  b_{j,p_m}^{(n)} \le \rho_1^m
\]
where $0<\rho_1<1$ but $\rho_1$ is sufficiently close to $1$.

Let $\Delta_{m+1}^{(m+1)} = \left[ -\rho_1^{m+1}, \rho_1^{m+1}\right]$ and
$\Delta_m^{(m+1)} = \left\{
( b_{j,p_m}^{(u)},  b_{j',p_m}^{(n)} ):\;
-\rho_1^{m+1} \le  b_{j,p_{m+1}}^{(u)},\,
 b_{j',p_{m+1}}^{(n)} \le \rho_1^{m+1}
\right\}.$
It follows easily from the estimates of $ \beta_{j,p_{m+1}}^{(u)}$,
$ \beta_{j',p_{m+1}}^{(n)}$ that
$\Delta_m^{(m+1)}\subseteq \Delta_m^{(m)}$. If
$\Delta_0^{(m)} =
\left\{ (  b_j^{(u)},  b_{j'}^{(n)} ):\;
( b_{j,m}^{(u)},  b_{j',m}^{(n)} ) \in \Delta_m^{(m)}
\right\}$, then $\Delta_0^{(m)}$ is a decreasing sequence of closed sets.
The intersection $\bigcap_m \Delta_0^{(m)}$ gives us the values of parameters
for which $\delta^{(p)}\rightarrow\infty$ as $p\rightarrow\infty$.

We make also some shortening of the time interval $S^{(m)}$. In the formulas
for $\delta^{(r)}$ there are several remainders which appear when we replace in
all expressions $s'$ and $s^{\prime\prime}$ by $s$. We estimate these remainders
using the fact that our functions satisfy the Lipschitz condition and the Lipschitz
constants and the maxima of their values decay as some power of $p$. We choose
the interval $S^{(m+1)}\subset S^{(m)}$ so that when we consider
$s\in S^{(m+1)}$ these remainders do not violate the basic inclusion
$\Delta_m^{(m+1)}\subset \Delta_m^{(m)}$. It is easy to see $S^{(m+1)}$ can
be chosen so that $S^{(m)}\setminus S^{(m+1)}$ consists of two intervals whose lengths
decay exponentially. Therefore $\bigcap_m S^{(m)}$ is an interval of positive
length.

The transformation
$( b_{j,p_{m+1}}^{(u)},\,  b_{j',p_{m+1}}^{(n)}) \rightarrow
( b_{j,p_{m}}^{(u)},\,  b_{j',p_{m}}^{(n)}) $
is given by smooth functions and is sufficiently close to the identity map.
The last step in the renormalization is the replacement in all $\delta^{(r)}$,
$r<p_{m+1}$ the variables $ b_{j,p_{m}}^{(u)}$, $ b_{j',p_m}^{(n)}$ by
their expressions through $ b_{j,p_{m+1}}^{(u)}$, $ b_{j',p_m+1}^{(n)}$.
The form of $\delta^{(r)}$ in new variables is the same as before.

\textbf{{The Choice of Constants}}

The main constants which are used in the
construction are the following:
\begin{enumerate}
\item  $k^{(0)}$ which determines the position of the domain where $v(k,0)$ is
concentrated.

\item $D_1$ is the constant which determines the size of the neighborhood where
$v(k,0)$ is concentrated.

\item $\rho_1$ determines the size of the neighborhood where the main parameters
$b_j^{(u)}$, $b_{j'}^{(n)}$ vary.

\item $D_2$ is the constant which determines the possible size of perturbations
$\Phi^{(st)}$ in the form of $v(k,0)$.

\item $\lambda_1$ is the power which gives the estimation of the decay of $h_r$ in
the domain $B$.

\item $\lambda_2$ is the parameter which determines the size of the first part of
the procedure.

\item $\epsilon$ determines the values of $p$ where we make the renormalization.
\end{enumerate}

The value of $k^{(0)}$ should be sufficiently large. All estimate of the remainders
which appear during the first half of the procedure are less than
$\frac {const} { \left( k^{(0)} \right)^{\frac 12}}$. They should be so small that the
estimates of all $\beta_{jr}^{(u)},\, \beta_{j'r}^{(n)}$ are much smaller than $\rho_1$.
On the other hand, $\rho_1$ should be small but not too small. It should be small
in order to make the quadratic part of our formulas smaller than the linear part. However
$\rho$ cannot be  too small in order that we could choose the next interval
$\left [ - \rho^{m+1}, \rho^{m+1} \right]$. This can be achieved by the choice of $k^{(0)}$.
The parameter $\lambda_2$ should be small. In this case the estimates of all corrections
are easier. However, after $\lambda_2$ is chosen the value of $k^{(0)}$ can be taken
sufficiently large depending on $\lambda_2$. The parameter $\lambda_1$ can be arbitrarily
large in order to make the perturbation arbitrarily small. The value of $D_1$ determines
the estimates in the domain $B$ which decay as $\frac 1 { \left( k^{(0)} \right)^{\lambda_1}}$.
We choose $D_1$ so that $\lambda_1 > \frac 12$. The value of $\epsilon$ is chosen small so
that we can write with a good precision the action of the linearized renormalization group.

\bigskip
\begin{center}
{\large \S10. \sf Critical Value of Parameters and Behavior\\
 of Solutions
 near the Singularity Point}
\end{center}

We return back to the first formulas:

\( v_A ( k , t ) \, = \,
\exp \, \{ - t | k |^2 \} \,
A \cdot v ( k , 0 ) \, +\, \displaystyle{\int\limits_{0}^t} \,
\exp \, \{
- ( t - s ) | k |^2 \} \, \cdot \,
\displaystyle{\sum\limits_{p > 1}} \, A^p h_p ( k , s ) \, d s
\)

\medskip
\noindent
or
\begin{equation} \label{eq_S10_main}
v_A ( k , t ) \, = \, \exp \,
\{ - \, t | k |^2 \} \,  A \cdot v ( k , 0 )\, + \,
\displaystyle{\int\limits_{0}^t} \, \exp \,
\{ - \, ( t - s ) | k |^2 \} \, \cdot \,
\displaystyle{\sum\limits_{p > 1}} \, A^p g_p ( k \sqrt{s} , s ) \, ds
\, .
\end{equation}

Our construction gives us the interval $S=\bigcap_n S^{(n)}$ on the time
axis such that for each $t\in S$ we can find the values of parameters
$b_j^{(u)} = b_j^{(u)}(t)$, $1\le j\le 4$ and
$b_{j'}^{(n)} = b_{j'}^{(n)} (t)$, $1\le j'\le 6$ such that we have the
representation (31) with $\delta^{(r)}\rightarrow 0$ as $r \rightarrow\infty$.
It is easy to see that $A_{cr} (t) = \Lambda^{-1} (t)$. If so then
$A^p h_p(k,t)$ is concentrated in the domain with the center at
$\frac {\kappa^{(0)} p} {\sqrt t}$ having the size $O(\sqrt p)$ and there
it takes values $O(p)$. This immediately implies that at $t$ the energy
is infinite.

Consider $t^\prime < t$.  It is important to investigate the behavior of
$E ( t' )$ and the enstrophy $\Omega ( t^\prime )$
of the same solution with $A=A_{cr}(t)$ when $t^\prime$ is close to $t$.
Denote $\Delta t = t -t^\prime$.  It follows easily
from the proof of the main result that
$\Lambda ( t^\prime) \slash_{\Lambda ( t )} = ( 1 - C_1 \Delta t + O (
\Delta t ))$ for some constant $C_1$.  Since $A_{cr}^p  \, \cdot \, ( \Lambda
( t' ))^p $ $=A_{cr}^p \cdot (\Lambda(t))^p \cdot \left( \Lambda(t')
\slash_{\Lambda (t)} \right)^p = (1-C\Delta t + o(\Delta t) )^p$. It is
clear that the terms in \eqref{eq_S10_main} are close to each other
for $p \le O \left( \frac {ln (\Delta t)^{-1}} {\Delta t} \right)$. For
$p >> \frac {ln (\Delta t)^{-1}} {\Delta t}$ the product $A_{cr}^p (\Lambda(t'))^p$
tends exponentially to zero and dominates other terms of the expansion. Therefore
it is enough to consider $\abs k \le O \left ( \frac {ln (\Delta t)^{-1}} {\Delta t}
\right )$ and in this domain the solution grows as $\abs k^{\frac 3 2}$. The
factor $\abs k^{\frac 1 2}$ appears because for any $k$ the values for which
the terms in \eqref{eq_S10_main} give the essential contribution to the solution
belonging to an interval of the size $O(\sqrt{\abs k}) = O( \sqrt p)$. From
this argument it follows easily that
$E(t') \thicksim  \left(\frac {ln (\Delta t)^{-1}} {\Delta t} \right)^6$ and
$\Omega(t') \thicksim \left( \frac {ln (\Delta t)^{-1}} {\Delta t}
\right )^8$.

%
%
%

\bigskip
\newpage
\begin{center}
{\large Appendix I. \ Hermite Polynomials and their basic properties}
\end{center}

Take $\sigma>0$ and write
\[
 {He}^{(\sigma)}_{n}(x) = (-1)^n
   e^{ \frac {\sigma x^2} 2} \frac {d^n} {dx^n}
   e^{- \frac {\sigma x^2} 2},
   \quad n\ge 0.
\]

It is clear that $He^{(\sigma)}_n(x) = \sigma^n x^n + \cdots $, where dots
mean terms of smaller degree. We shall call $He^{(\sigma)}_n$ the n-th
Hermite polynomial. It is clear that $He^{(\sigma)}_{0}(x) = 1,\,
He^{(\sigma)}_1 (x) = \sigma x,\, He^{(\sigma)}_{2}(x)=\sigma^2x^2 - \sigma$
and so on. In general, $ He^{(\sigma)}_n (x) = \sigma^{\frac n2} He^{(1)}_n(\sqrt{\sigma} x)$.
It is easy to check that
\begin{equation} \label{eq_A1_1}
\sigma x He^{(\sigma)}_n (x) = He^{(\sigma)}_{n+1} (x) + \sigma n He^{(\sigma)}_{n-1} (x).
\end{equation}
The Fourier transform of $ He^{(\sigma)}_m (x) e^{-\frac {\sigma x^2} 2} \sqrt { \frac {\sigma}{2\pi}}$
is $ (i\lambda)^m e^{-\frac {\lambda^2} {2\lambda}}$. This implies the formula for convolution:
\begin{equation} \label{eq_A1_2}
\int_{\bR^1} He^{(\sigma)}_{m_1} (x-y) e^{-\frac {\sigma (x-y)^2} 2} \sqrt {\frac {\sigma} {2\pi}}
\cdot He^{(\sigma)}_{m_2} (y) e^{-\frac {\sigma y^2}2} \sqrt {\frac {\sigma} {2\pi}} dy
 = He^{(\sigma)}_{m_1+m_2} (x) e^{-\frac {\sigma x^2}2} \sqrt{ \frac {\sigma} {2\pi}}
\end{equation}

Take positive $\gamma_1,\, \gamma_2,\, \gamma_1 + \gamma_2 = 1$ and consider the convolution of
$He^{(\sigma)}_{m_1} ( \frac x{\sqrt{\gamma_1}}) e^{-\frac {\sigma x^2}{2\gamma_1}} \cdot
\sqrt{ \frac {\sigma} {2\pi\gamma_1}}$ and
$He^{(\sigma)}_{m_2} ( \frac x{\sqrt{\gamma_2}}) e^{-\frac {\sigma x^2}{2\gamma_2}} \cdot
\sqrt{ \frac {\sigma} {2\pi\gamma_2}}$. Their Fourier transforms are
$(i \lambda \sqrt{\gamma_1})^{m_1} e^{-\frac {\lambda^2 \gamma_1} {2\sigma}}$ and
$(i \lambda \sqrt{\gamma_2})^{m_2} e^{-\frac {\lambda^2 \gamma_2} {2\sigma}}$
respectively. The product of these two functions is
$\gamma_1^{\frac {m_1}2} \gamma_2^{\frac {m_2} 2} (i\lambda)^{m_1+m_2}
e^{ - \frac {\lambda^2} {2\sigma}} $. Therefore the convolution is
$\gamma_1^{\frac {m_1} 2} \cdot \gamma_2^{\frac {m_2} 2} He^{(\sigma)}_{m_1+m_2}
(x) e^{-\frac {\sigma x^2} 2}$.

\newpage
\noindent
\baselineskip=15pt
\begin{center}
{\large \sf \ References}
\label{S_ref}
\end{center}
\baselineskip=13.5pt
\begin{enumerate}
\item[{[}C{]}] M. Cannone. Harmonic Analysis Tools for Solving the
Incompressile Navier-Stokes Equations. Handbook of Mathematical Fluid
Dynamics, vol. 3, 2002.
\item[{[}Cl{]}] Clay Mathematical Institute. The Millennium Prize Problems,
2006.

\item[{[}F-T{]}] C. Foias and R. Temam. Gevrey Classes of Regularity for the
Solutions of the Navier-Stokes Equations. J. of Funct. Anal. 87, 1989,
359-369.
\item[{[}G{]}] Y. Giga, T. Miyakawa. Navier-Stokes Flow in $R^3$ with Measures
as Initial Vorticity and Morrey spaces. Commu. Partial Differential Equations,
14, 1989, 577-618.

\item[{[}K{]}] T. Kato. Strong $L^p$-solution of the Navier-Stokes Equation in
$R^m$, with Applications to Weak Solutions. Math. Zeitschrift, 187, 1984,
471-480.
\item[{[}La{]}] O. Ladyzenskaya.
The mathematical theory of viscous incompressible flow.
New York: Gordon and Breach Science Publishers, 1969.

\item[{[}Le{]}] J. Leray. \'Etude de diverses \'equations int\'egrales
non lin\'eaires et de quelques probl\'emes que pose l'hydrodynamique.
J. Math. Pures Appl. 12, 1993, 1-82

\item[{[}Si 1{]}] Ya. G. Sinai. Power Series for Solutions of the Navier-Stokes
System on $R^3$. Journal of Stat. Physics, vol. 121, No. 516, 2005, 779-804.

\item[{[}Si 2{]}] Ya. G. Sinai. Diagrammatic Approach to the $3D$-Navier-Stokes
System. Russian Math. Surveys, vol. 60, No.5, 2005, 47-70.

\item[{[}Y{]}] V.I. Yudovich. The Linearization Method in Hydrodynamical
Stability Theory. Trans. Math. Mon. Amer. Math. Soc. Providence, RI 74(1984).
\end{enumerate}

\vfill

\today:\/gpp
\end{document}